\RequirePackage{fix-cm}
\documentclass{svjour3}                     
\smartqed  

\usepackage{graphicx}
\usepackage[caption=false]{subfig}
\usepackage[hidelinks]{hyperref}
\usepackage{cite}
\usepackage{paralist}
\usepackage{xspace} 
\usepackage{amsfonts}
\usepackage{listings}
\usepackage{multirow}
\usepackage{amsmath}
\usepackage{todonotes} 
\presetkeys{todonotes}{inline}{} 

\newcommand{\eg}{\textit{e.g.,~}}

\newcommand{\ie}{\textit{i.e.,~}}

\newcommand{\etal}{\textit{et~al.}\xspace}
\newcommand{\etc}{\textit{etc.}}
\newcommand{\wrt}{\textit{w.r.t.~}}

\newcommand{\botsing}{\textsc{Bot\-sing}\xspace}
\newcommand{\recore}{\textsc{Re\-Core}\xspace}
\newcommand{\integ}{\textit{STDistance}\xspace}
\newcommand{\WS}{\textit{WeightedSum}\xspace}
\newcommand{\integA}{\textit{STD}\xspace}
\newcommand{\WSA}{\textit{WS}\xspace}
\newcommand{\bbc}{\textit{BBC}\xspace}

\newcommand{\WSOBB}{\textit{WS + BBC}\xspace}
\newcommand{\integOBB}{\textit{STD + BBC}\xspace}

\newcommand{\evocrash}{\textsc{Evo\-Crash}\xspace}
\newcommand{\crashpack}{\textsc{JCrash\-Pack}\xspace}
\newcommand{\evosuite}{\textsc{Evo\-Suite}\xspace}
\newcommand{\dynamosa}{\textsc{Dy\-na\-MO\-SA}\xspace}
\newcommand{\defectsforj}{\textsc{De\-fects\-4J}\xspace}
\newcommand{\sleep}{\textsc{Sleep Time}\xspace}
\newcommand{\usage}{\textsc{Usage Rate}\xspace}
\newcommand{\tet}{\textsc{TT}\xspace}

\newcommand{\ncuts}{219\xspace}
\newcommand{\nruns}{30\xspace}

\journalname{ }

\begin{document}

\title{Basic Block Coverage for Search-based Unit Testing and Crash Reproduction
}


\author{Pouria Derakhshanfar \and
  Xavier Devroey \and
  Andy Zaidman
}


\institute{P. Derakhshanfar (orcid.org/0000-0003-3549-9019),  
  Andy Zaidman (orcid.org/0000-0003-2413-3935) 
  \at 
  Delft University of Technology, Postbus 5, 2600 AA Delft, The Netherlands \\
  \email{p.derakhshanfar@tudelft.nl},\\
  \email{a.e.zaidman@tudelft.nl}\\
  X. Devroey (orcid.org/0000-0002-0831-7606)
  \at
  NADI, University of Namur, rue de Bruxelles 61, 5000 Namur, Belgium\\
  \email{xavier.devroey@unamur.be}
}

\date{}

\maketitle

\begin{abstract}
Search-based techniques have been widely used for white-box test generation. Many of these approaches rely on the \emph{approach level} and \emph{branch distance} heuristics to guide the search process and generate test cases with high line and branch coverage. 
Despite the positive results achieved by these two heuristics, they only use the information related to the coverage of explicit branches (\eg indicated by conditional and loop statements), but ignore potential implicit branchings within basic blocks of code. 
If such implicit branching happens at runtime (\eg if an exception is thrown in a branchless-method), the existing fitness functions cannot guide the search process. 
To address this issue, we introduce a new secondary objective, called Basic Block Coverage (\bbc), which takes into account the coverage level of relevant basic blocks in the control flow graph. 
We evaluated the impact of \bbc on \emph{search-based unit test generation} (using the \dynamosa algorithm) and \emph{search-based crash reproduction} (using the \integ and \WS fitness functions).
Our results show that for unit test generation, \bbc improves the branch coverage of the generated tests. Although small ($\sim$1.5\%), this improvement in the branch coverage is systematic and leads to an increase of the output domain coverage and implicit runtime exception coverage, and of the diversity of runtime states. 
In terms of crash reproduction, in the combination of \integ and \WS, \bbc helps in reproducing 3 new crashes for each fitness function. \bbc significantly decreases the time required to reproduce 43.5\% and 45.1\% of the crashes using \integ and \WS, respectively. For these crashes, \bbc reduces the consumed time by 71.7\% (for \integ) and 68.7\% (for \WS) on average.

\end{abstract}

\section{Introduction}
\label{sec:introduction}

Various search-based techniques have been introduced to automate different white-box test generation activities (\eg unit testing \cite{fraser2012whole,Fraser2011}, integration testing \cite{derakhshanfar2020integ}, or system-level testing \cite{arcuri2019restful}). Depending on the testing level, each of these approaches utilizes dedicated fitness functions to guide the search process and produce a test suite satisfying given criteria (\eg line coverage, branch coverage, \etc). 

Fitness functions typically rely on \textit{control flow graphs (CFGs)} to represent the source code of the software under test \cite{McMinn2004}. Each node in a CFG is a \textit{basic block} of code (\ie maximal linear sequence of statements with a single entry and exit point without any internal branch), and each edge represents a possible \textit{execution flow} between two blocks.
Two well-known heuristics are usually combined to achieve high line and branch coverage: the \textit{approach level} and the \textit{branch distance} \cite{McMinn2004}. The former measures the distance between the execution path of the generated test and a target basic block (\ie a basic block containing a statement to cover) in the CFG. The latter measures, using a set of rules, the distance between an execution and the coverage of a \textit{true} or \textit{false} branch of a particular predicate in a branching basic block of the CFG.

Both \textit{approach level} and \textit{branch distance} assume that only a limited number of basic blocks (\ie \textit{control dependent} basic blocks \cite{Allen:1970:CFA:800028.808479}) can change the execution path away from a target statement (\eg if a target basic block is the true branch of a conditional statement).
However, basic blocks are not atomic due to the presence of \textbf{implicit branches} \cite{borba2010testing} (\ie branches occurring due to the exceptional behavior of instructions).
As a consequence, any basic block between the entry point of the CFG and the target basic block can impact the execution of the target basic block.
For instance, a generated test case may stop its execution in the middle of a basic block with a runtime exception thrown by one of the statements of that basic block. 
In these cases, the search process does not benefit from any further guidance from the approach level and branch distance.

Fraser and Arcuri \cite{fraser20151600} introduced testability transformation for \textbf{unit testing}, which instruments the code to guide the unit test generation search to cover implicit exceptions happening in the class under test. However, this approach does not guide the search process in scenarios where an implicit branch happens in another class called by the class under test. This is due to the extra cost added to the search process stemming from the calculation and monitoring of implicit branches in all the classes coupled to the class under test.
For instance, the class under test may be heavily coupled with other classes in the project, thereby finding implicit branches in all of these classes can be expensive.

In contrast, other test case generation scenarios, like \textbf{crash reproduction}, aim to cover only a limited number of paths, and thereby we only need to analyse a limited number of basic blocks~\cite{Chen2015, Xuan2015, nayrolles2015jcharming, Rossler2013, Soltani2018a}. Current crash reproduction approaches rely on information about a reported crash (\eg a stack trace, a core dump, \etc) to generate a crash reproducing test case.
Among these approaches, \textbf{search-based crash reproduction} \cite{Rossler2013, Soltani2018a} takes as input a \textbf{stack trace} to guide the generation process. More specifically, the statements pointed to by the stack trace act as target statements for the approach level and branch distance.
Hence, current search-based crash reproduction techniques suffer from a lack of guidance in cases where the involved basic blocks contain implicit branches (which is common when trying to reproduce a crash). 

In our prior work we have introduced a novel secondary objective called \textbf{Basic Block Coverage (\bbc)} to address the guidance problem in crash reproduction~\cite{Derakhshanfar2020c}.
The secondary objective guides the search process to differentiate two generated tests with the same fitness values (here, same approach level and branch distance).
This paper extends our prior work on \bbc to the more general unit test case generation context. 
\bbc helps the search process to compare two generated test cases with the same distance (according to approach level and branch distance) to determine which one is closer to the target statement. In this comparison, \bbc analyzes the coverage level, achieved by each of these test cases, of the basic blocks in between the closest covered control dependent basic block and the target statement.

To assess the impact of \bbc on search-based unit test generation, we implemented \bbc in \evosuite \cite{Fraser2011}, the state-of-the-art tool for search-based unit test generation, and evaluate its performance against the classical \dynamosa \cite{Panichella2018} for various activation probabilities of \bbc (11 configurations in total). We applied these eleven configurations to \ncuts classes under test selected from the last version of \defectsforj v.2.0.0 \cite{Just2014b}, a collection of existing faults. We compare the performance in terms of effectiveness for branch coverage, weak mutation score, output coverage, and real fault detection capabilities.

Our results show that \bbc improves the branch coverage of the generated tests when activating \bbc as a secondary objective in \dynamosa. Utilizing this secondary objective improves the average branch coverage achieved by \dynamosa (74.5\% average branch coverage with standard deviation 28\%) to 76.1\% with standard deviation 27.5\%. 
Despite the slight improvement in the average branch coverage, this increase in branch coverage is systematic, as indicated by the static analysis performed in this study: for 59 target classes, \bbc improves the branch coverage achieved by \dynamosa significantly ($p-value < 0.01$) with a large effect size.
This improvement in the branch coverage leads to an increase of coverages and scores achieved by tests generated by the unit test generation process in terms of output domain (\ie the number of pre-defined partitions of the output values domain) coverage, implicit runtime exception coverage, and the diversity of runtime states (denoted by the weak mutation score). \bbc increases the average output domain coverage of the generated tests from 54.2\% (with standard deviation 26.6\%) up to 55.5\% (with standard deviation 26.2\%). The improvement achieved by this secondary objective is statistically significant and has a large effect size in 57 classes under test.
Moreover, \bbc improves the average implicit runtime exception coverage when using \dynamosa from 75.1\% (with standard deviation 22.8\%) up to 80.3\% (with standard deviation 21\%). Besides, this secondary objective significantly improves the implicit runtime exception coverage with large effect size in 67 classes. 
Also, \bbc improves the weak mutation score achieved by the tests generated by \dynamosa from 73.2\% (with standard deviation 30.1\%) up to 74.6\%  (with standard deviation 29.6\%). Finally, our static analysis shows that activating \bbc also significantly improves with a large effect the fault detection rate for 3 real faults out of 92.

Similarly, to assess the impact of \bbc on search-based crash reproduction, we re-implemented the existing \integ \cite{Rossler2013} and \WS \cite{Soltani2018a} fitness functions and empirically compared their performance with and without using \bbc (4 configurations in total). We applied these four crash reproduction configurations to 124 hard-to-reproduce crashes introduced in \crashpack \cite{Derakhshanfar2019}, a crash benchmark used by previous crash reproduction studies \cite{Derakhshanfar2020}. 
We compare the performance in terms of \textit{effectiveness in crash reproduction ratio} (\ie percentage of times that an approach can reproduce a crash) and \textit{efficiency} (\ie time required by for reproducing a crash).

Our results show that \bbc significantly improves the crash reproduction ratio over the 30 runs in our experiment for respectively 10 and 4 crashes when compared to use \integ and \WS without any secondary objective. Also, \bbc helps these two fitness functions to reproduce 3 (for \integ) and 3 (for \WS) crashes that could not be reproduced without the secondary objective. 
Besides, on average, \bbc increases the crash reproduction ratio of \integ and \WS from 70.5\% (with standard deviation 38.1\%) to 79.7\% (with standard deviation 37.3\%) and from 74.8\% (with standard deviation 38.1\%) to 78.1\% (with standard deviation 36.1\%), respectively.
Applying \bbc also significantly reduces the time consumed for crash reproduction guided by \integ and \WS in 56 (45.1\% of cases) and 54 (43.5\% of cases) crashes, respectively. In cases where \bbc has a significant impact on efficiency, this secondary objective improves the average efficiency of \integ and \WS by 71.7\% (with standard deviation 36\%) and 68.7\% (with standard deviation 28.9\%), respectively.

The remainder of this paper is organized as follow: Section \ref{sec:background} reports the background on CFG-based guidance. Section \ref{sec:approach} describes our novel \bbc secondary objective and how it can be used for search-based crash reproduction and search-based unit test generation. Section \ref{sec:evaluation} describes our evaluation to assess the importance of implicit branches (\textbf{RQ 0}) and the impact of \bbc on search-based unit test generation (\textbf{RQ 1}) and search-based crash reproduction (\textbf{RQ 2}). Section \ref{sec:results} presents our results on \ncuts classes under test selected from the last version of \defectsforj and 124 hard-to-reproduce crashes from \crashpack. Sections~\ref{sec:discussion} and~\ref{sec:threats} discuss our results and their implications for search-based test case generation, Section~\ref{sec:relatedwork} discusses related work, and Section~\ref{sec:conclusion} concludes the paper.

\section{Background}
\label{sec:background}

\subsection{Coverage distance heuristics}
\label{sec:background:distance}

\begin{lstlisting}[caption={Method \texttt{fromMap} from XWIKI version 8.1\cite{Derakhshanfar2019}},
    label=list:fromMap,
    language=java,
    captionpos=t,
    numbers=left,
    float=t,
    firstnumber=402]
public BaseCollection fromMap(Map<[...]> map, BaseCollection object){
    for (PropertyClass property : (Collection<[...]>) getFieldList()) {
        String name = property.getName();
        Object formvalues = map.get(name);
        if (formvalues != null) {
            BaseProperty objprop;
            if (formvalues instanceof String[]) {
                [...]
            } else if (formvalues instanceof String) {
                objprop = property.fromString(formvalues.toString());
            } else {
                objprop = property.fromValue(formvalues);
            }
            [...]
        }
    }
    return object;
}
\end{lstlisting}
 
\begin{figure}[t]
    \centering
    \includegraphics[width=0.35\linewidth]{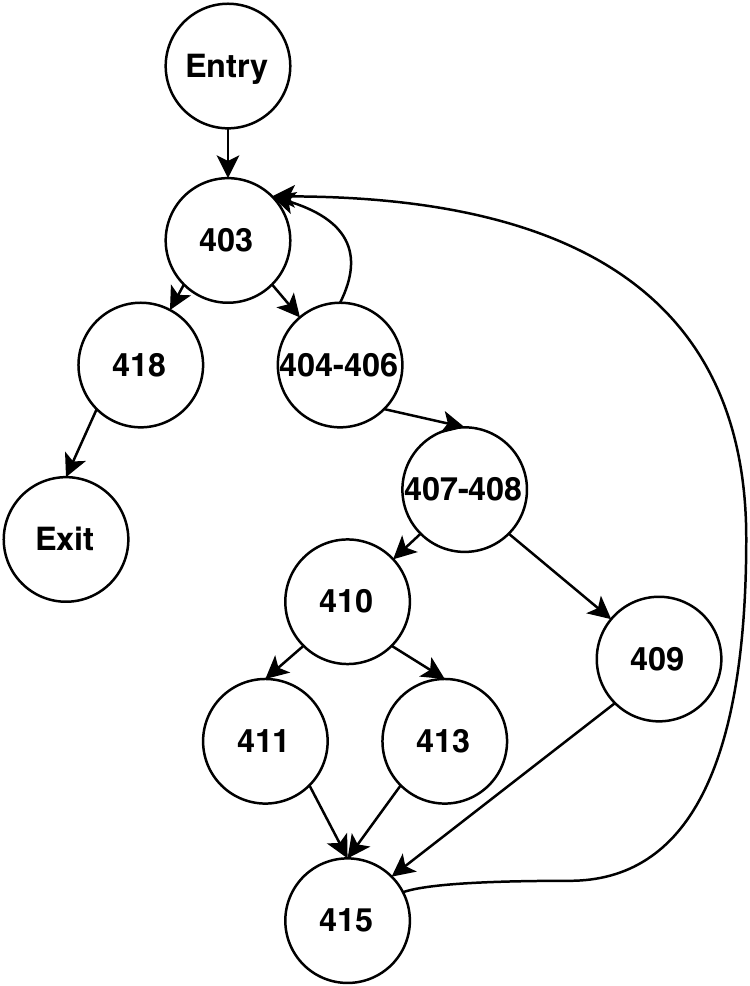}    
    \caption{CFG for method \texttt{fromMap}}
  \label{fig:CFG}
  \vspace{-2em}
\end{figure}

Many structural-based search-based test generation approaches mix the \textit{branch distance} and \textit{approach level} heuristics to achieve a high line and branch coverage~\cite{McMinn2004}. These heuristics measure the distance between a test execution path and a specific statement or a specific branch in the software under test.
For that, they rely on the coverage information of \emph{control dependent basic blocks}, \ie basic blocks that have at least one outgoing edge leading the execution path toward the \emph{target basic block} (containing the targeted statement) and at least another outgoing edge leading the execution path away from the target basic block.
As an example, Listing \ref{list:fromMap} shows the source code of the method \texttt{fromMap} from XWIKI\footnote{https://github.com/xwiki}, and Figure~\ref{fig:CFG} contains the corresponding CFG. In this graph, the basic block \texttt{409} is control dependent on the basic block \texttt{407-408} because the execution of line 409 is dependent on the  condition at line 408 (\ie line 409 will be executed only if elements of array \texttt{formvalues} are \texttt{String}).

The \textit{approach level} is the number of uncovered control dependent basic blocks for the target basic block between the closest covered control dependent basic block and the target basic block. The \textit{branch distance} is calculated from the predicate of the closest covered control dependent basic block, based on a set of predefined rules. Assuming that the test $t$ covers only line \texttt{403} and \texttt{418}, and our target line is \texttt{409}, the approach level is 2 because two control dependent basic blocks (\texttt{404-406} and \texttt{407-408}) are not covered by $t$. The branch distance for the predicate in line \texttt{403} (the closest covered control dependency of node \texttt{409}) is measured based on the rules from the establised technique \cite{McMinn2004}.

To the best of our knowledge, there is no related work studying the extra heuristics helping the combination of approach level and branch distance to improve the coverage. Most related to our work, Panichella \etal \cite{Panichella2018} and Rojas \etal \cite{rojas2015combining} introduced two heuristics called \textit{infection distance} and \textit{propagation distance}, to improve the weak mutation score of two generated test cases. However, these heuristics do not help the search process to improve the general statement coverage (\ie they are effective only after covering a mutated statement).

In this paper, we introduce a new secondary objective to improve the statement coverage achieved by fitness functions based on the approach level and branch distance, and analyze the impact of this secondary objective on \textbf{search-based unit test generation} and \textbf{search-based crash reproduction}.

\begin{lstlisting}[
        caption=XWIKI-13377 crash stack trace \cite{Derakhshanfar2019},
        label=lst:stacktrace,
        numbers=left,
        firstnumber=0
        ]
java.lang.ClassCastException: [...]
    at [...].BaseStringProperty.setValue(BaseStringProperty.java:45)
    at [...].PropertyClass.fromValue(PropertyClass.java:615)
    at [...].BaseClass.fromMap(BaseClass.java:413)
    [...] (*@\label{line:lowestframe}@*)
\end{lstlisting}

\subsection{Search-based unit test generation}
\label{sec:background:unit}

Search-based software test generation (SBST) algorithms use meta-heuristic optimization search techniques (\eg genetic algorithms) to automate the test generation tasks at different testing levels. One of these levels is unit testing, where the search algorithm tries to generate tests satisfying various criteria (such as line coverage and branch coverage) for a given class under test (CUT). SBST techniques are widely used for unit test generation. Prior studies showed that the tests generated by these techniques achieve a high code coverage~\cite{Panichella2018a, Campos2018} and real-bug detection~\cite{Almasi2017}, hence complementing the hand-written test cases.

\paragraph{Dynamic many-objective sorting algorithm (\dynamosa).}
Panichella \etal have recently introduced an evolutionary-based algorithm, called \dynamosa, for unit test generation \cite{Panichella2018}. Their study \cite{Panichella2018a}, independently confirmed by Campos \etal~\cite{Campos2018}, shows that \dynamosa outperforms other unit test generation techniques in terms of structural coverage and mutation coverage. This approach is currently used as the default algorithm in \evosuite, which is the state-of-the-art tool for search-based unit test generation.

\dynamosa relies on the hierarchy of dependencies between the coverage targets (\eg lines and branches) to perform a dynamic selection of the objectives during the search process. For instance, by applying \dynamosa to generate tests for method \texttt{fromMap} (Listing \ref{list:fromMap}), this algorithm, first, tries to cover targets that do not have any dependencies. So, first, it tries to generate test cases to cover nodes \texttt{403} and \texttt{418}. After covering node \texttt{403}, it tries to cover the node \texttt{404-406}, which is control-dependent on the covered node. \dynamosa continuously changes the search objectives up to the point that all of the targets are covered.

Since \dynamosa uses the approach level and branch distance heuristics to guide the search process towards achieving the high line, branch, and weak mutation coverage, \bbc may help this technique to cover more targets. This study performs an in-depth experiment and analysis to see whether \bbc can improve \dynamosa.

\subsection{Search-based Crash Reproduction}
\label{sec:background:crash}

After a crash is reported, one of the essential steps of software debugging is to write a \textbf{crash reproducing test case} to make the crash observable to the developer and help them in identifying the root cause of the failure \cite{Zeller2009}. Later, this crash reproducing test can be integrated into the existing test suite to prevent future regressions. Despite the usefulness of a crash reproducing test, the process of writing this test can be labor-intensive and time-taking \cite{Soltani2018a}. 
Various techniques have been introduced to automate the reproduction of a crash \cite{Chen2015, Xuan2015, nayrolles2015jcharming, Rossler2013, Soltani2018a}, and search-based approaches (\evocrash \cite{Soltani2018a} and \recore \cite{Rossler2013}) yielded the best results~\cite{Soltani2018a}.

\paragraph{\evocrash.}
This approach utilizes a single-objective genetic algorithm to generate a crash reproducing test from a given stack trace and a \textit{target frame} (\ie a frame in the stack trace that its class will be used as the class under test). The crash reproducing test generated by \evocrash throws the same stack trace as the given one up to the target frame. 
For example, by passing the stack trace in Listing~\ref{lst:stacktrace} and target frame 3 to \evocrash, it generates a test case reproducing the first three frames of this stack trace (\ie thrown stack trace is identical from line 0 to~3).

\evocrash uses a fitness function, called \WS, to evaluate the candidate test cases. \WS is the sum scalarization of three components:
\begin{inparaenum}[(i)]
    \item the \textbf{target line coverage} ($d_{s}$), which measures the distance between the execution trace and the \textit{target line} (\ie the line number pointed to by the target frame) using \textit{approach level} and \textit{branch distance};
    \item the \textbf{exception type coverage}  ($d_{e}$), determining whether the type of the triggered exception is the same as the given one; and 
   \item the \textbf{stack trace similarity}  ($d_{tr}$), which indicates whether the stack trace triggered by the generated test contains all frames (from the most in-depth frame up to the target frame) in the given stack trace.
\end{inparaenum}

\begin{definition}[\WS \cite{Soltani2018a}]
    For a given test case execution $t$, the \WS ($ws$) is defined as follows:
    \begin{equation}
    ws(t) = 
    \left\{
        \begin{array}{ll}
          3 \times d_{s}(t) + 2 \times max(d_{e}) + max(d_{tr}) & \textit{ if line not reached}\\
          3 \times min(d_{s}) + 2 \times d_{e}(t) + max(d_{tr}) & \textit{if line reached}\\
          3 \times min(d_{s}) + 2 \times min(d_{e}) + d_{tr}(t) & \textit{if exception thrown}
        \end{array}
      \right.
    \end{equation}
    Where $d_{s}(t) \in [0,1]$ indicates how far $t$ is from reaching the target line and is computed using the normalized approach level and branch distance: $d_{s}(t) = \Vert approachLevel_{s}(t)  + \Vert branchDistance_{s}(t)\Vert\Vert$ ($\Vert$ indicates the normalized value);
    $d_{e}(t) \in \{0,1\}$ shows if the type of the exception thrown by $t$ is the same as the given stack trace ($0$) or not ($1$);
    $d_{tr}(t) \in [0,1]$ measures the stack trace similarity between the given stack trace and the one thrown by $t$. $max(f)$ and $min(f)$ denote the maximum and minimum possible values for a function $f$, respectively.
\end{definition}

In this fitness function, $d_{e}(t)$ and $d_{tr}(t)$ are only considered in the satisfaction of two \textit{constraints}: (i) \textit{exception type coverage} is relevant only 
when we reach the target line and (ii) \textit{stack trace similarity} is important only when we both reach the target line and throw the same type of exception.

As an example, when applying \evocrash on the stack trace from Listing~\ref{lst:stacktrace} with the target frame 3, \WS first checks if the test cases generated by the search process reach the statement pointed to by the target frame (line 413 in class \texttt{BaseClass} in this case). Then, it checks if the generated test can throw a \texttt{ClassCastException} or not. Finally, after fulfilling the first two constraints, it checks the similarity of frames in the stack trace thrown by the generated test case against the given stack trace in Listing~\ref{lst:stacktrace}. 

\evocrash uses \textbf{guided} initialization, mutation and single-point crossover operators to ensure that the target method  (\ie the method appeared in the target frame) is always called by the different tests during the evolution process. 

According to a recent study, \evocrash outperforms other non-search-based crash reproduction approaches in terms of \textit{effectiveness in crash reproduction} and \textit{efficiency}~\cite{Soltani2018a}. This study also shows the helpfulness of tests generated by \evocrash for developers during debugging. 

In this paper, we assess the impact of \bbc as the secondary objective in the \evocrash search process.

\paragraph{\recore.}
This approach utilizes a genetic algorithm guided by a single fitness function, which has been defined according to the core dump and the stack trace produced by the system when the crash happened. To be more precise,  this fitness function is a sum scalarization of three sub-functions: (i) \textbf{TestStackTraceDistance}, which guides the search process according to the given stack trace; (ii) \textbf{ExceptionPenalty}, which indicates whether the same type of exception as the given one is thrown or not (identical to ExceptionCoverage in \evocrash); and (iii) \textbf{StackDumpDistance}, which guides the search process by the given core dump. 
\begin{definition}[\textit{TestStackTraceDistance} \cite{Rossler2013}]
    For a given test case execution $t$, the \textit{TestStackTraceDistance} ($STD$) is defined as follows:
    \begin{equation}
    STD(R,t) = |R| - lcp - (1-StatementDistance(s))
     \end{equation}
\end{definition}
Where $|R|$ is the number of frames in the given stack trace, and $lcp$ is the longest common prefix frames between the given stack trace and the stack trace thrown by $t$. Concretely, $|R| - lcp$ is the number of frames not covered by $t$. Moreover, $StatementDistance(s)$ is calculated using the sum of the approach level and the normalized branch distance to reach the statement $s$, which is pointed to by the first (the utmost) uncovered frame by $t$: $StatementDistance(s) = approachLevel_{s}(t) + \Vert branchDistance_{s}(t) \Vert$.

Since using runtime data (such as core dumps) can cause significant overhead \cite{Chen2015} and leads to privacy issues \cite{nayrolles2015jcharming}, the performance of \recore in crash reproduction was not compared with \evocrash in prior studies \cite{Soltani2018a}, even though two out of three fitness functions in \recore use only the given stack trace to guide the search process. Hence, this paper only considers \textit{TestStackTraceDistance} $+$ \textit{ExceptionPenalty} (called \integ hereafter).

As an example, when applying \recore with \integ on the stack trace in Listing~\ref{lst:stacktrace} with target frame 3, first, \integ determines if the generated test covers the statement at frame 3 (line 413 in class \texttt{BaseClass}). Then, it checks the coverage of frame 2 (line 615 in class \texttt{PropertyClass}). After covering the first two frames by the generated test case, it checks the coverage of the statement pointed to by the deepest frame (line 45 in class \texttt{BaseStringProperty}). For measuring the coverage of each of these statements, \integ uses the approach level and branch distance. After covering all of the frames, this fitness function checks if the the generated test throws a \texttt{ClassCastException} in the deepest frame.

In this study, we perform an empirical evaluation to assess the performance of crash reproduction using \integ with and without \bbc as the secondary objective in terms of \textit{effectiveness in crash reproduction} and \textit{efficiency}.

\section{Basic Block Coverage}
\label{sec:approach}

\subsection{Motivating Example} 

During the search process, the fitness of a test case is evaluated using a fitness function. These fitness functions are different according to the given test criteria. However, one of the main components of these fitness functions is the coverage of specific statements and branches. For instance, one of the main goals in unit test generation is achieving a high structural coverage (\eg line and branch coverage). For this goal, the search process seeks to cover all of the statements and branches in the given CUT. Similarly, the fitness functions used in search-based crash reproduction (either \WS or \integ) require the coverage of specific statements pointed by the given stack trace. 

The distance of the test case from the target statement is calculated using the approach level and branch distance heuristics. As we have discussed in Section~\ref{sec:background:distance}, the approach level and branch distance cannot guide the search process if the execution stops because of implicit branches in the middle of basic blocks (\eg a thrown \texttt{NullPointerException} during the execution of a basic block). As a consequence, these fitness functions may return the same fitness value for two tests, although the tests do not cover the same statements in the block of code where the implicit branching happens.

For instance, assume that one of the objectives of a search process (either for unit test generation or crash reproduction) is covering line \texttt{413} in method \texttt{fromMap} (appeared in Listing \ref{list:fromMap}). This search process generates two test cases T$_1$ and T$_2$ for achieving this objective in a population of solutions. However, T$_1$ stops the execution at line \texttt{404} due to a \texttt{NullPointerException} thrown in method \texttt{getName}, and T$_2$ throws a \texttt{NullPoint\-erException} at line \texttt{405} because it passes a null value input argument to \texttt{map}. Even though T$_2$  covers more lines, the combination of approach level and branch distance returns the same fitness value for both of these test cases: approach level is 2 (nodes \texttt{407-408} and \texttt{410}), and branch distance cannot be helpful in this case as the last covered predicate does not change the execution path away from covering the target line and also the execution stops before covering the next predicate.
This is because these two heuristics assume that each basic block is atomic, and by covering line \texttt{404}, it means that lines \texttt{405} and \texttt{406} are covered, as well.

\subsection{Secondary Objective}

The goal of the Basic Block Coverage (\bbc) secondary objective is to prioritize the test cases with the same fitness value (\ie same approach level and branch distance) according to their coverage within the basic blocks between the closest covered control dependency and the target statement. At each iteration of the search algorithm, test cases with the same fitness value are compared with each other using \bbc. Listing \ref{list:bbc} presents the pseudo-code of the \bbc calculation. Inputs of this algorithm are two test cases T$_1$ and T$_2$, which both have the same approach level and branch distance values (calculated either using crash reproduction or unit test generation fitness functions), as well as line number and method name of the target statement. This algorithm compares the coverage of basic blocks on the path between the last control dependent node executed by both of the given tests and the basic block that contains the target statement (called \textit{effective blocks} hereafter). If T$_1$ and T$_2$ do not cover any control dependency of the target block, \bbc uses the entry point of the CFG of the given method instead as the starting point of the effective blocks' path.
 If \bbc determines there is no preference between these two test cases, it returns 0. Also, it returns a value $<0$ if T$_1$ has higher coverage compared to T$_2$, and vice versa. A higher absolute value of the returned integer indicates a bigger distance between the given test cases.

\begin{lstlisting}[float=t,
    caption={\bbc secondary objective computation algorithm},
    label=list:bbc,
    mathescape=true,
    numbers=left, 
    numberstyle=\tiny,
    keywordstyle=\bfseries\em,
    keywords={,input, output, return, datatype, function, in, if, else, foreach, while, begin, end, }
    numbers=left]
input: test T$_1$, test T$_2$, String method, int line
output: int
begin
    FCB$_1$ $\gets$  fullyCoveredBlocks(T$_1$,method,line);
    FCB$_2$ $\gets$  fullyCoveredBlocks(T$_2$,method,line);
    SCB$_1$ $\gets$  semiCoveredBlocks(T$_1$,method,line);
    SCB$_2$ $\gets$  semiCoveredBlocks(T$_2$,method,line);
    
    if SCB$_1$ = SCB$_2$ $\wedge$ (FCB$_1$ $\subseteq$ FCB$_2$ $\vee$ FCB$_2$ $\subseteq$ FCB$_1$) :
        closestBlock $\gets$ closestSemiCoveredBlocks(SCB$_1$, method, line);
        coveredLines1 $\gets$ getCoveredLines(T$_1$,closestBlock);
        coveredLines2 $\gets$ getCoveredLines(T$_2$,closestBlock);
        return size(coveredLines2) - size(coveredLines1); 
    else if (FCB$_1$ $\subseteq$ FCB$_2$ $\wedge$ SCB$_1$ $\in$ FCB$_2$) $\vee$ (FCB$_2$ $\subseteq$ FCB$_1$ $\wedge$ SCB$_2$ $\in$ FCB$_1$):
        return size(FCB$_2$) - size(FCB$_1$)      
    else:
        return 0;
end  
\end{lstlisting}

In the first step, \bbc detects the effective blocks that are fully covered by each given test case (\ie the test covers all of the statements in the block) and saves them in two sets called \texttt{FCB$_1$} and \texttt{FCB$_2$} (lines 4 and 5 in Listing \ref{list:bbc}). Then, for each of the tests T$_1$ and T$_2$, it detects the closest semi-covered effective block (\ie the closest basic block to the target statement where the test covers the first line but not the last line of the block) and stores them as \texttt{SCB$_1$} and \texttt{SCB$_2$}, respectively (lines~6 and~7). 
The semi-covered blocks indicate the presence of implicit branches. 

\bbc can prioritize given tests in two scenarios: \textbf{Scenario 1}, both tests get stuck in the middle of the same basic block (\ie they both have the same closest semi-covered basic block), or, \textbf{Scenario 2},  one of the tests throws an exception in an effective basic block while the other test fully covers this block.

\paragraph{Scenario 1.} 
Line 9 in Listing \ref{list:bbc} checks if the first scenario is true by determining two conditions. First, \bbc checks if both tests have the same semi-covered basic block. Then, it examines if the fully covered basic blocks of one of the given tests are equal or the subset of the other test. If the second condition is not fulfilled, it means that each of these tests has one covered block that the other one does not cover, and thereby they achieve their semi-covered basic block from different paths. In this case, \bbc cannot find the better test as we do not know which path can lead to covering the target statement. 
If these two conditions are fulfilled, \bbc checks if one of the tests has a higher line coverage in the identified SCB (lines~10 to~13). If this is the case, \bbc will return the number of lines in this block covered only by the winning test case. If the lines covered are the same for T$_1$ and T$_2$ (\ie \texttt{coveredLines1} and \texttt{coveredLines2} have the same size), there is no difference between these two test cases and \bbc returns value 0 (line 13).

\paragraph{Scenario 2.} 
Line 14 in Listing \ref{list:bbc} checks if the effective blocks covered by one test are a subset of the other one. This is true if all of the fully-covered blocks of one test are a subset of fully covered blocks of the other one. Also, the semi-covered block of this test must be among the fully-covered blocks of the test with more coverage (\ie winner test). In this case, \bbc returns the number of blocks that are only fully covered by the winner test case (line 15). If \bbc determines T$_2$ wins over T$_1$, the returned value will be positive, and vice versa. 

Finally, if each of the given tests has a unique covered block in the given method (\ie the tests cover different paths in the method), \bbc cannot determine the winner and returns 0 (lines 16 and 17) because we do not know which path leads to the target block. Even if T$_1$ and T$_2$ reach a particular basic block from different paths in the CFG and both throw exceptions in different lines, \bbc returns 0 and does not select the one with the more coverage in the closest basic block as the winner. The rationale behind this behavior of \bbc is to provide an equal chance for these two tests to evolve as we do not know which path covered by each of these tests has more potential to help the search process to get closer to the target line. If \bbc always selects the test with more coverage in the nearest basic block, even if it covers another path, we are negatively impacting the diversity of the tests chosen for the next generation, thereby reducing the search process's exploration ability.

\paragraph{Example.} 
When giving two tests with the same fitness value (calculated by the primary objective) T$_1$ and T$_2$ from our motivation example to \bbc with target method \texttt{fromMap} and line number 413, this algorithm compares their fully and semi-covered blocks with each other. 
In this example, both T$_1$ and T$_2$ cover the same basic blocks: the fully covered block is \texttt{403} and the semi-covered block is \texttt{404-406}. So, here the conditions in \textbf{Scenario 1} are fulfilled. Hence, \bbc checks the number of lines covered by T$_1$ and T$_2$ in block \texttt{404-406}. Since T$_1$ stopped its execution at line 404, the number of lines covered by this test is 1. In contrast, T$_2$ managed to execute two lines (404 and 405). Hence, \bbc returns $size(coveredLines2) - size(coveredLines1) = 1$. The positive return value indicates that T$_2$ is closer to the target statement, and therefore, it should have a higher chance of being selected for the next generation.

\paragraph{Branchless Methods.} 
\bbc can also be helpful for \textit{branchless methods}. These methods do not contain any branching statement (\eg if conditions or for loops), and thereby theoretically, covering the first line in these methods leads to covering all of the other lines, as well. In other words, by ignoring the \texttt{Entry} and \texttt{Exit} nodes, CFGs of branchless methods contain only one node (\ie basic block) without any edges.
For instance, methods from frames 1 and 2 in Listing \ref{lst:stacktrace} are branchless. The absence of branches in these methods means that there are no control dependent nodes in them, and thereby approach level and branch distance cannot guide the search process in these cases if the generated tests throw implicit exceptions in the middle of these methods. However, in contrast with these two heuristics, \bbc can guide the search process toward covering the most in-depth statement in these cases. As an example, if tests  T$_1$ and T$_2$ both throws implicit branches in the middle of the only basic block ($b_0$) of branchless method $m()$, \bbc enters the Scenario 1 ($FCB_1 = FCB_2  = \emptyset$ and $SCB_1 = SCB_2 = \{b_0\}$) and examines if one of the tests has more lines covered in $b_0$.

\subsection{Application of \bbc}
\label{subsec:bbcapplication}

The time complexity of \bbc is $\mathcal{O}(N \times E  \times log\, V)$ where $E$ and $V$ are the numbers of edges and vertices of the CFG of the given method, respectively; and $N$ is the number of semi-covered basic blocks calculated by \texttt{semiCoveredBlocks} method at lines \texttt{6} and \texttt{7} of Listing \ref{list:bbc}. This complexity stems from the computation of the closest semi-covered basic blocks in Line \texttt{12} of Listing \ref{list:bbc}. In this procedure, \bbc measures the shortest path between each semi-covered basic block and the target basic block (\ie the block containing the given target line) using \textit{Dijkstra}'s shortest path algorithm, which has a time complexity of $\mathcal{O}(E \times log\, V)$.

Given the complexity of \bbc, applying this secondary objective for any generated tests with the same approach level and branch distance may negatively impact the search process's efficiency. In the following paragraphs, we discuss this potential negative impact on search-based crash reproduction and unit test generation.

\subsubsection{Search-Based Crash Reporduction}

The crash reproduction search process can be guided by either \WS or \integ. As discussed in Section~\ref{sec:background:crash}, both of these fitness functions heavily rely on approach level and branch distance. Hence, \bbc can be helpful in the crash reproduction search process.
Since the crash reproduction search process's goal is to cover a specific path in the control dependent graph,  which is indicated by the given stack trace, we apply \bbc without any limitation on any case that includes two test cases with the same (and nonzero) approach level and branch distance.

\subsubsection{Search-Based Unit Test Generation}
\label{sec:approach:application:unit}

In contrast with crash reproduction, the unit test generation search process has multiple statements and branches to cover simultaneously. In \dynamosa, each line or branch to cover is an objective of the search. Hence, the number of times that \bbc is applied as the secondary objective is higher compared to crash reproduction. Therefore, we should limit the number of times that \bbc is applied in this algorithm. We introduce two parameters to bring this limitation: \sleep and \usage.

\paragraph{\sleep.}
When \dynamosa adds a target to the active search objectives, the target will stay active until the search process covers it. Some of the targets are easy to cover, and thereby, approach level and branch distance can simply cover them without \bbc. However, \bbc can help in harder cases where approach level and branch distance cannot cover them in a certain time. \sleep makes sure that \bbc is only applied for the hard-to-cover search objectives. If we set this parameter to \textit{t} seconds, \dynamosa uses \bbc secondary objective only for search objectives that are active for more than \textit{t} seconds.

\paragraph{\usage.}
Like any other evolutionary-based algorithm, the unit test generation search process needs to maintain a balance between the \textit{exploration} and \textit{exploitation}. The former indicates the diversity in the solutions (\ie generated tests execute new paths in the code); the latter indicates searching the solutions in the existing ones' neighborhood (\ie the search process should generate tests similar to the existing ones). By applying \bbc, we improve the exploitation ability of the search process. However, the over-application of \bbc may negatively impact the exploration ability of the search process. \usage makes sure that \bbc does not hinder this balance. Higher \usage means that there is a higher chance of \bbc application during the search process. Assume we set $p \in [0,1]$ as our \usage. Any time that the search process generates two test cases with the same approach level and branch distance for a hard-to-cover target (\ie target which stays as an active objective in \dynamosa for more than \sleep), \bbc will be used with the probability of $p$.

Moreover, by default, \evosuite has eight types of search objectives \cite{rojas2015combining}: \textit{line coverage}, which aims to cover maximum lines in the given CUT; \textit{branch coverage}, which aims to cover maximum branches in the CUT; \textit{exception coverage}, which aims to maximize the number of exceptions captured by the generated tests; \textit{weak mutation}, which aims to generate tests that kill the maximum number of mutants (in weak mutation, a mutant is considered killed if executing one of the generated tests on the mutant leads to a different state compared to the execution on the given CUT); \textit{output coverage}, that aims for generating tests that drive the most diverse outputs; \textit{method coverage}, which aims to cover all of the methods in the given CUT; \textit{no-exception Method Coverage}, checks if each of the methods in the CUT is called directly by one of the tests and this invocation does not lead to any exception; and \textit{direct branch coverage} that makes sure that each branch in the public methods of CUT is covered by a direct call from one of the generated tests.

Since \bbc aims to help the search process rely on the approach level and branch distance in covering lines and branches that cannot be executed with the tests generated by \dynamosa, this secondary objective is only triggered when two tests have the same fitness value either for a non-covered line coverage or branch coverage objective. Hence, \bbc is not involved in segments of the search process in which two tests are getting the same fitness value for other kinds of objectives such as exception coverage. Thereby, despite the fact that \bbc prioritizes tests without throwing implicit exceptions, since this secondary objective is not triggered for objectives other than line coverage and branch coverage, it does not have any negative impact on covering other search objectives (\eg exception coverage).

\section{Empirical Evaluation}
\label{sec:evaluation}

Before evaluating the impact of \bbc, we want to assess its potential usefulness by answering the following research question: 
\begin{enumerate}
    \item[\textbf{RQ 0}] How frequent are implicit branches in a search-based test case generation process?  
\end{enumerate}
This research question serves as a preliminary analysis before the full evaluation of the impact of \bbc on search-based unit test generation and search-based crash reproduction. To answer it, we consider a special configuration of \dynamosa, currently the best algorithm for unit test generation, where the executions of the \bbc algorithm described in Listing \ref{list:bbc} are monitored. We choose \dynamosa, a many-objectives algorithm, because, unlike search-based crash reproduction, it targets each line and branch of a class under test independently, allowing us to collect more data about the execution of \bbc for the different objectives.

To assess the impact of \bbc on search-based unit test generation, we perform an empirical evaluation to answer the following research questions:
\begin{enumerate}
    \item[\textbf{RQ 1}] What is the impact of \bbc on search-based unit test generation? 
    \begin{enumerate}[RQ {1}.1]
        \item[\textbf{RQ 1.1}] What is the impact of \bbc on the structural coverage effectiveness of the unit tests? 
        \item[\textbf{RQ 1.2}] What is the impact of \bbc on the output and implicit exception coverage of the unit tests? 
        \item[\textbf{RQ 1.3}] What is the impact of \bbc on the fault finding capabilities of the unit tests?
        \item[\textbf{RQ 1.4}] What is the impact of \bbc on the structural coverage efficiency of the unit tests?
    \end{enumerate}
\end{enumerate}
In these RQs, we want to evaluate the effect of \bbc on \dynamosa. As for other algorithms, \dynamosa relies on the approach level and branch distance to evaluate the progress of the search process. Previous research has shown that it outperforms other search-based and guided random approaches \cite{Campos2018, Devroey2020, Kifetew2019a, Molina2018, Panichella2018, Panichella2018a}. We compare \dynamosa for 11 different configurations of \bbc in terms of structural coverage effectiveness (RQ 1.1). Since a change in the structural coverage of a class might impact the data flow, we also study the output coverage (\ie diversity of the values returned by the tested methods~\cite{Alshahwan2014}) and captured implicit exceptions produced by the different tests (RQ 1.2). Then, we look at the fault finding capabilities using weak mutation and real faults from the \defectsforj collection (RQ 1.3). Finally, we study the structural coverage efficiency of \bbc (RQ 1.4).

Similarly, for search-based crash reproduction, we answer the following research questions: 
\begin{itemize}
    \item[\textbf{RQ 2}] What is the impact of \bbc on search-based crash reproduction?
    \begin{itemize}
        \item[\textbf{RQ 2.1}] What is the impact of \bbc on the crash reproduction effectiveness?
        \item[\textbf{RQ 2.2}] What is the impact of \bbc on the crash reproduction efficiency?
    \end{itemize}
\end{itemize}
In these two RQs, we want to evaluate the effect of \bbc on the existing fitness functions, namely \integ and \WS, from two perspectives: the crash reproduction ratio of the different configurations (RQ 2.1) and the time required to reproduce a crash (RQ 2.2).

\medskip
In Sections~\ref{sec:setup:unittest} and~\ref{sec:setup:reproduction} we will detail the experimental setup for respectively the study on unit test generation (RQ 0 and RQ 1) and crash reproduction (RQ~2).

\subsection{Setup for search-based unit test generation (RQ 0 and RQ 1)}
\label{sec:setup:unittest}

\subsubsection{Implementation}

We implemented \bbc as a secondary objective (called \texttt{BB\-CO\-VE\-RA\-GE}) in \evosuite~\cite{Fraser2011}, the state-of-the-art tool for search-based unit test generation.
As discussed in Section \ref{sec:approach:application:unit}, since \bbc impacts the exploration-exploitation trade-off and efficiency of the search process, we also defined two additional parameters for \sleep (\texttt{BBC\_SLEEP} with a default value of 60 seconds) and \usage (\texttt{BBC\_US\-AGE\_PER\-CEN\-TAGE} with a default probability of 0.5). 
Our implementation is openly available in our replication package on Zenodo~\cite{pouria_derakhshanfar_2021_4665874}.

\subsubsection{Classes under test selection}

\begin{table}[t]
    \centering
    \caption{Classes under test used for the evaluation of \bbc for unit testing (\textbf{RQ 0} \textbf{RQ 1}): number of classes under test (\textbf{CUTs}), number of non-commented source statements per class (\textbf{NCSS}), number of methods per class (\textbf{Methods}), weighted methods per class (\textbf{WMC}), and cyclomatic complexity per method (\textbf{CCN}).}
    \resizebox{0.99\textwidth}{!}{%
        \begin{tabular}{ l | r | r c | r c | r c | r c }
\textbf{Project} & \textbf{CUTs} & \multicolumn{2}{c|}{\textbf{NCSS}} & \multicolumn{2}{c|}{\textbf{Methods}} & \multicolumn{2}{c|}{\textbf{WMC}} & \multicolumn{2}{c}{\textbf{CCN}} \\ 
  &   & $\overline{x}(\sigma)$ & range & $\overline{x}(\sigma)$ & range & $\overline{x}(\sigma)$ & range & $\overline{x}(\sigma)$ & range \\ 
\hline 
Chart & 10 & 471.0(570.9) & [5,1917] & 54.1(64.7) & [2,229] & 199.7(245.1) & [2,817] & 3.7(6.7) & [1,110] \\ 
Cli & 14 & 140.9(77.5) & [45,236] & 25.4(13.8) & [8,44] & 66.9(39.5) & [17,111] & 2.6(3.3) & [1,19] \\ 
Closure & 11 & 589.4(389.2) & [88,1276] & 63.6(69.4) & [22,265] & 265.6(188.4) & [52,601] & 4.2(10.4) & [1,230] \\ 
Codec & 11 & 152.2(158.1) & [42,564] & 19.6(14.1) & [4,50] & 72.8(83.5) & [10,304] & 3.7(5.1) & [1,42] \\ 
Collections & 4 & 375.8(440.1) & [49,1021] & 67.5(61.0) & [14,153] & 204.5(234.2) & [23,542] & 3.0(3.8) & [1,28] \\ 
Compress & 10 & 257.3(192.7) & [24,569] & 35.9(24.8) & [4,75] & 117.7(81.1) & [10,226] & 3.3(3.6) & [1,27] \\ 
Csv & 11 & 225.9(127.6) & [53,460] & 36.5(24.2) & [15,78] & 119.7(75.8) & [29,250] & 3.3(6.4) & [1,44] \\ 
Gson & 12 & 319.1(242.0) & [56,933] & 38.0(21.9) & [10,80] & 160.9(121.8) & [32,464] & 4.2(6.0) & [1,64] \\ 
JacksonCore & 13 & 852.2(645.4) & [125,2121] & 67.5(22.0) & [32,109] & 386.8(312.7) & [56,1012] & 5.7(7.7) & [1,71] \\ 
JacksonDatabind & 32 & 214.1(196.0) & [19,911] & 34.0(29.7) & [1,126] & 106.8(103.7) & [8,446] & 3.1(4.2) & [1,62] \\ 
JacksonXml & 6 & 290.7(166.0) & [104,526] & 36.7(23.8) & [11,68] & 126.0(66.5) & [49,214] & 3.4(5.0) & [1,40] \\ 
Jsoup & 18 & 273.1(338.6) & [5,1348] & 38.8(37.4) & [2,125] & 116.4(143.5) & [2,583] & 3.0(7.7) & [1,176] \\ 
JxPath & 14 & 239.9(194.4) & [22,488] & 26.0(18.5) & [3,45] & 131.7(108.5) & [9,291] & 5.1(7.1) & [1,61] \\ 
Lang & 10 & 274.1(190.0) & [29,455] & 34.6(25.2) & [2,75] & 153.7(121.4) & [10,329] & 4.4(9.9) & [1,76] \\ 
Math & 18 & 195.7(182.1) & [29,579] & 23.0(17.7) & [4,54] & 78.5(69.3) & [13,198] & 3.4(4.6) & [1,49] \\ 
Mockito & 13 & 68.7(65.8) & [10,220] & 18.5(24.0) & [2,74] & 39.8(45.5) & [3,151] & 2.1(2.5) & [1,31] \\ 
Time & 12 & 273.2(130.0) & [71,442] & 51.8(25.3) & [18,103] & 123.9(53.8) & [45,195] & 2.4(3.0) & [1,28] \\ 
\end{tabular} 
    } 
    \label{tab:cuts}
\end{table}

We selected classes under test from the latest version of \defectsforj (v.2.0.0)~\cite{Just2014b}, a collection of reproducible failures coming from open source projects with the identification of the corresponding faulty classes. \defectsforj has been used in other studies to assess the coverage and the effectiveness of unit-level test case generation \cite{ma2015grt, Panichella2018, Shamshiri2016}, program repair \cite{Smith2015, Martinez2016}, fault localization \cite{pearson2017evaluating, b2016learning}, and regression testing \cite{noor2015similarity, lu2016does}. 
We selected the ten most recent bugs from the 17 available projects for a total of 225 faulty classes, used as classes under test in our evaluation. This offers a good balance between the number of repetitions (\ie statistical power) of each configuration and number of cases (\ie generalization) \cite{Arcuri2014}.

Since \evosuite may face inevitable challenges for generating tests for some particular classes \cite{xiao2011precise, McMinn2011, Fraser2014b}, we performed a trial with default parameters, on all of the classes to filter out the ones for which \evosuite cannot generate any test, as recommended by related work~\cite{Campos2018, Molina2018, Panichella2018}. We filtered out six classes according to our trial experiment results. In three of these classes, \evosuite could not finish the class instrumentation. For the other two, \dynamosa could not find any search objective. Finally, \evosuite failed to generate tests for a class because of missing classes. By filtering these classes, we performed our main experiment on the \ncuts remaining cases.
Table \ref{tab:cuts} provides more information about the classes selected for the evaluation.

\subsubsection{Parameter settings}

To evaluate the impact of \bbc secondary objective on search-based unit test generation, first, we should set values for \sleep and \usage (explained in Section~\ref{sec:approach:application:unit}). To find the optimum \sleep, we performed a pre-analysis on a subset of subjects. We have randomly selected 45 classes (20\% of our subjects) for this pre-analysis. We ran \dynamosa on each of the sampled classes for 30 times and collected the time required by the search process for covering each objective. These collected results indicate that \dynamosa can cover more than 85\% of the objectives in 60 seconds. For this reason, we have set \sleep to 60 seconds for our experiments.

For our pre-analysis (\textbf{RQ 0}), we have enabled \bbc (\usage$ = 1.0$) after 60 seconds (with an additional setting to record the execution results of \bbc) to evaluate the number of implicit branches occurring during the search and the number of times \bbc could help overcoming those implicit branches. 
Furthermore, to draw a comparison between setting different \usage, we have used ten different values of this parameter in our main experiment (\textbf{RQ 1}): 
$\usage \in \{0.1,0.2,0.3,0.4,0.5,0.6,0.7,0.8,0.9,1.0\}$.

Hence, for the main experiment, we have executed \dynamosa and one plus ten configurations of \bbc on \ncuts classes for 30 rounds of execution with a search budget of 10 minutes. Also, we have executed \dynamosa on 45 classes with the same number of repetitions and search budget for finding the optimum \sleep. In total, we ran 80,190 independent executions to answer \textbf{RQ 0} and \textbf{RQ 1}. These executions took about 12 days overall.

\subsubsection{Data collection}

To evaluate the potential impact of \bbc (\textbf{RQ 0}), we collected for each line and branch objective: the number of times its \textit{fitness} has been evaluated, and the number of times \bbc has been \textit{called}, \textit{activated} (\ie the call effectively led to an evaluation of the \bbc, line 13 or 15 in Listing \ref{list:bbc}), and \textit{useful} (\ie the call to \bbc has returned a non-zero value). When \bbc is useful, it indicates that at one or both of the test throw an implicit exception in the middle of a basic block in the method of search objective (\ie line or branch coverage objective).

We compare \bbc to \dynamosa using \textit{branch coverage} for \textbf{RQ 1.1} and \textbf{RQ 1.4} for \nruns rounds of execution. Branch coverage provides an indication on the structural coverage by looking at the percentage of branches covered by the executions of the test cases in the class under test. We recorded the value of the branch coverage every ten seconds to see how it evolves over time and answer~\textbf{RQ~1.4}.

For \textbf{RQ 1.2}, we consider \textit{output coverage} and \textit{implicit exceptions}. Output coverage \cite{Alshahwan2014} denotes the diversity of the outputs of the different methods of the class under test. It provides information about the data output coverage of the generated tests by looking at how many pre-defined abstract values (\ie partitions of the output domain) are returned by the methods of the class under test. We used the method from Rojas \etal \cite{rojas2015combining} available in \evosuite to compute the output coverage. For instance, a method returning integer value has to return negative, zero, and positive values (when the tests are executed) to satisfy the output coverage criterion.
In addition to (expected) outputs, we consider \textit{implicit exceptions} by looking at the number ($e$) of top-level methods in the class under test throwing an undeclared (\ie runtime) exception implicitly (\ie without any \texttt{throw new} instruction). For one execution, we compute the \textit{implicit exception coverage} as the ratio between $e$ and the highest value of $e$ among the all the executions of the different \bbc configurations for that class. 
Since \bbc addresses the challenge of handling implicit branches for search-based unit test generation, we expect it to impact both the output coverage and the number of methods throwing an implicit exception.

We rely on \textit{weak mutation} and \textit{real faults} to assess the fault finding capabilities of the generated tests (\textbf{RQ 1.3}). Weak mutation score \cite{Howden1982, Papadakis2011} gives the percentage of mutants (\ie artificially injected faults) for which at least one test triggers a different program state, compared to the original program, directly after the execution of the mutated statement. Weak mutation is a viable and cheaper alternative to strong mutation, which requires an additional propagation of the erroneous state to the output of the program \cite{Offutt1994}. For our evaluation, weak mutation allows us to assess the diversity of runtime states, allowing to catch more faults, when using \bbc. We use the default set of weak mutation operators available in \evosuite \cite{Fraser2015a}: delete call, delete field, insert unary operator, replace arithmetic operator, replace bitwise operator, replace comparison operator, replace constant, and replace variable.

Additionally, we use real faults from the \defectsforj benchmark to compare the effective fault finding capabilities of tests generated using \bbc. We executed all of the 11 configurations on the buggy versions of the software, and next, we check if the tests generated by each configuration can throw the same exception as the bug exposing stack traces, which are indicated by \defectsforj. The rationale behind running all of the configurations only on the buggy versions, and not the fixed versions, is to have a realistic scenario. In a realistic scenario, developers are neither aware of the bug, nor have access to the fixed version. In this scenario, an automated test generation tool can help developers if it generates tests that throw an exception revealing the bug. Since \evosuite can detect the assertion-based failures only by running it on the fixed version \cite{fraser20151600}, we limited our comparison for fault detection only to the 92 faults that a non-assertion error can expose.

\subsubsection{Data analysis}

For each class under test, we use the Vargha-Delaney \^A$_{12}$ statistic \cite{vargha}, a non-parametric effect size measure, to examine the effect size of differences between using and not using \bbc for branch, output, and implicit exception coverage, and weak mutation score (\textbf{RQs 1.1-1.4}).
For a pair of factors $(A,B)$ a value of \^A$_{12} > 0.5$ indicates that $A$ is more likely to achieve a higher coverage or mutation score, while a value of \^A$_{12} < 0.5$ shows the opposite. Also, \^A$_{12}= 0.5$ means that there is no difference between the factors. 
We used the standard thresholds \cite{vargha} for interpreting the \^A$_{12}$ magnitude: 0.56 (small), 0.64 (medium), and 0.71 (large). 
To assess the significance of effect sizes (\^A$_{12}$), we apply the non-parametric Wilcoxon Rank Sum test, with $\alpha = 0.01$ for the Type~I error ($H_0$: there is no difference between using and not using \bbc for $x$ on a class under test $c$, where $x$ is the branch, output, or implicit exception coverage, or weak mutation score). 

We also rank the different configurations of \bbc, based on their coverage and weak mutation score, using Friedman's non-parametric test for repeated measurements with a significance level $\alpha = 0.05$~\cite{Garcia2009} (\textbf{RQs 1.1-1.3}). This test is used to test  the significance of the  differences between groups (treatments) over the dependent variable (here, coverage and weak mutation score). 
We further complement the test for significance with Nemenyi's post-hoc procedure~\cite{Japkowicz2011,Panichella2021}: two configurations are significantly different if their corresponding average ranks differ by at least the given Critical Distance (CD).

Finally, since fault coverage (\textbf{RQ 1.3}) has a dichotomic distribution (\ie a generated test exposes the fault or not), for each fault, we use the Odds Ratio ($OR$) to measure the impact of each \bbc configuration on the \textit{real faults coverage}. A value $OR > 1$ in a comparison between a pair of factors $(A,B)$ indicates that the application of factor A increases the fault coverage, while $OR < 1$ indicates the opposite. Also, a value of $OR = 1$ indicates that both of the factors have the same performance.
We apply Fisher's exact test, with $\alpha = 0.01$ for the Type I error, to assess the significance of the results ($H_0$: there is no difference between using and not using \bbc in reproduction ratio of the fault). 

\subsection{Setup for search-based crash reproduction (RQ 2)}
\label{sec:setup:reproduction}

\subsubsection{Implementation}

Since \recore and \evocrash are not openly available, we implement \bbc in \botsing \cite{derakhshanfar2020botsing}, an extensible, well-tested, and open-source search-based crash reproduction framework already implementing the \WS fitness function and the guided initialization, mutation, and crossover operators. We also implement \integ (\recore fitness function) in this tool.
\botsing relies on \evosuite for code instrumentation and test case generation by using \texttt{evosuite-client} as a dependency. 
We also implement the \integ fitness function used as baseline in this paper.

\subsubsection{Crash selection}

We select crashes from \crashpack \cite{Derakhshanfar2019}, a benchmark containing hard-to-re\-pro\-duce Java crashes.
We apply the two fitness functions with and without using \bbc as a secondary objective to 124 crashes, which have also been used in a recent study~\cite{Derakhshanfar2020}. 
These crashes stem from six open-source projects: \textrm{JFreeChart}, \textrm{Commons-lang}, \textrm{Commons-math}, \textrm{Mockito}, \textrm{Joda-time}, and \textrm{XWiki}.
For each crash, we apply each configuration on each frame of the crash stack traces. We repeat each execution 30 times to take randomness into account, for a total of 114,120 independent executions.
We run the evaluation on two servers with 40 CPU-cores, 128 GB memory, and 6 TB hard drive. In total, these executions took about 5 days.

\subsubsection{Parameter settings}

We run each search process with five minutes time budget and set the population size to 50 individuals, as suggested by previous studies on search-based test generation~\cite{Panichella2018}. Moreover, as recommended in prior studies on search-based crash reproduction~\cite{Soltani2018a}, we use the \textit{guided mutation} with a probability $p_m=1/n$ ($n = $ length of the generated test case), and the \textit{guided crossover} with a probability $p_c=0.8$ to evolve test cases. We do note that prior studies do not investigate the sensitivity of the crash reproduction to these probabilities. Tuning these parameters should be undertaken as future work.

\subsubsection{Data collection}

To evaluate the crash reproduction ratio (\ie the ratio of success in crash reproduction in 30 rounds of runs) of different assessed configurations (\textbf{RQ 2.1}), we follow the same procedure as previous studies \cite{Derakhshanfar2020, Soltani2018b}: for each crash $C$, we detect the highest frame that can be reproduced by at least one of the configurations ($r_{max}$). 
We examine the crash reproduction ratio of each configuration for crash $C$ targeting frame $r_{max}$.

To evaluate the efficiency of different configurations (\textbf{RQ 2.2}), we analyze the time spent by each configuration on generating a crash reproducing test case.
We do note that the extra pre-analysis and basic block coverage in \bbc is considered in the spent time.
Since measuring efficiency is only possible for the reproduced crashes, we compare the efficiency of algorithms on the crashes that are reproduced at least once by one of the algorithms. 
We assume that the algorithm reached the maximum allowed budget (5 minutes) in case it failed to reproduce a crash.

\subsubsection{Data analysis}

As for real fault coverage (\textbf{RQ 1.3}), crash reproduction data (\textbf{RQ 2.1}) has a dichotomic distribution (\ie an algorithm reproduces a crash $C$ from its $r_{max}$ or not), for each crash, we use the Odds Ratio ($OR$) to measure the impact of each algorithm on the crash reproduction ratio for each crash. 
We apply Fisher's exact test, with $\alpha = 0.01$ for the Type I error, to assess the significance of the results ($H_0$: there is no difference between using and not using \bbc in the reproduction ratio of the crash). 

For \textbf{RQ 2.2}, for each crash, we use the non-parametric Vargha-Delaney \^A$_{12}$ statistic \cite{vargha} with the non-parametric Wilcoxon Rank Sum test to examine differences between using and not using \bbc for efficiency ($H_0$: there is no difference between using and not using \bbc in the reproduction efficiency of the crash).

\subsection{Replicability}

We enable the replicability of our results by providing replication packages on Zenodo (\url{https://zenodo.org}) for \textbf{RQ 0} and \textbf{RQ 1} \cite{pouria_derakhshanfar_2021_4665874} and \textbf{RQ 2} \cite{derakhshanfar_pouria_2020_3953519}.
Those replication packages include the classes under test and crashes used for the evaluation, the evaluation infrastructure (including documentation and scripts to re-run the evaluation), and the data analysis procedure used to produce the graphs, tables, and numbers reported in this paper.

\section{Results}
\label{sec:results}

\subsection{Potential impact of \bbc (\textbf{RQ 0})}

\begin{table}[t]
    \centering
    \caption{Statistics about the number of objectives (\textbf{Obj.}), fitness evaluations (\textbf{Fitness eval.}), calls to \bbc evaluations (\textbf{BBC calls}), calls effectively leading to an evaluation of the \bbc (\textbf{BBC active}), and evaluations returning a non-zero value (\textbf{BBC useful}).}
    \resizebox{0.99\textwidth}{!}{%
\begin{tabular}{ l | r | r r | r r | r r | r r }
\textbf{Project} & \textbf{Obj.} & \multicolumn{2}{c|}{\textbf{Fitness eval.}} & \multicolumn{2}{c|}{\textbf{BBC calls}} & \multicolumn{2}{c|}{\textbf{BBC active}} & \multicolumn{2}{c}{\textbf{BBC useful}} \\ 
  &   & $\overline{count}$ & $\sigma$ & $\overline{count}$ & $\sigma$ & $\overline{count}$ & $\sigma$ & $\overline{count}$ & $\sigma$ \\ 
\hline 
Chart & 3492 & 17,522.59 & 69,478.23 & 15,769.44 & 108,896.01 & 2,267.34 & 15,409.45 & 133.66 & 4,980.58 \\ 
Cli & 963 & 46,395.51 & 144,057.34 & 39,927.05 & 179,255.85 & 5,300.18 & 37,189.29 & 2.74 & 43.16 \\ 
Closure & 4779 & 23,864.65 & 33,537.30 & 34,880.69 & 59,787.43 & 8,716.67 & 28,410.00 & 446.23 & 6,556.16 \\ 
Codec & 526 & 85,859.14 & 138,087.36 & 118,522.38 & 249,495.43 & 49,434.50 & 161,610.84 & 0.00 & 0.07 \\ 
Collections & 915 & 41,404.66 & 40,811.89 & 78,162.33 & 80,603.58 & 2,391.87 & 13,382.00 & 713.11 & 6,706.91 \\ 
Compress & 1602 & 27,870.01 & 56,441.02 & 25,610.46 & 58,955.84 & 10,477.92 & 35,881.90 & 0.06 & 2.13 \\ 
Csv & 1279 & 21,797.16 & 66,812.74 & 21,892.09 & 89,951.27 & 1,617.00 & 16,831.40 & 51.66 & 561.60 \\ 
Gson & 2272 & 50,307.24 & 105,668.55 & 47,428.92 & 143,743.06 & 12,515.74 & 69,460.49 & 972.59 & 22,547.11 \\ 
JacksonCore & 8108 & 16,546.99 & 32,507.93 & 16,406.25 & 49,033.41 & 10,233.04 & 34,686.78 & 240.63 & 5,202.30 \\ 
JacksonDatabind & 4932 & 19,779.36 & 44,533.60 & 26,837.34 & 72,399.41 & 6,323.01 & 21,387.24 & 436.74 & 6,523.39 \\ 
JacksonXml & 1130 & 30,898.57 & 29,490.09 & 55,675.38 & 64,763.15 & 35,723.20 & 47,364.58 & 195.75 & 1,210.80 \\ 
Jsoup & 2458 & 58,216.14 & 117,964.61 & 82,136.18 & 168,880.68 & 2,080.66 & 17,089.75 & 87.03 & 3,178.65 \\ 
JxPath & 2348 & 51,578.30 & 103,321.87 & 29,519.46 & 104,762.47 & 7,402.75 & 42,828.18 & 4.72 & 64.64 \\ 
Lang & 1749 & 37,868.96 & 93,794.17 & 20,247.58 & 60,978.13 & 1,338.74 & 12,510.84 & 2.91 & 38.91 \\ 
Math & 1309 & 27,917.29 & 48,262.64 & 49,197.32 & 84,697.47 & 21,353.59 & 45,462.28 & 2,710.62 & 19,146.62 \\ 
Mockito & 584 & 91,840.19 & 113,787.23 & 156,256.50 & 216,605.91 & 42,901.56 & 95,736.14 & 608.66 & 4,312.73 \\ 
Time & 1891 & 19,180.13 & 45,616.90 & 21,628.31 & 68,101.74 & 1,331.23 & 11,072.58 & 90.19 & 2,319.25 \\ 
\hline 
(all) & 40337 & 30,111.81 & 71,396.34 & 34,988.58 & 100,703.53 & 9,472.14 & 40,567.40 & 354.12 & 7,913.20 \\ 
\end{tabular} 
    }
    \label{tab:preanalysis}
\end{table} 

Table \ref{tab:preanalysis} provides the general statistics of the preliminary analysis answering \textbf{RQ 0} per project. The number of branch and line objectives ranges from 526 for \texttt{Codec} to 8,108 for \texttt{JacksonCore}. In total, the number of fitness evaluations per objective ranges between 1 and 1,143,620 with an average of 30,111.81 evaluations. \bbc has been called between 1 and 1,681,329 times per objective with an average of 34,988.58 calls. It is interesting to note that, since the evaluation of an objective may require to compare multiple test cases, \bbc can be called multiple times for each fitness evaluation. \bbc has been effectively activated up to 1,365,526 (average of 9,472.140) times per objective, and has been useful up to 798,005 (average of 354) times per objective. 

\begin{figure}[t]
    \centering
    \includegraphics[width=\textwidth]{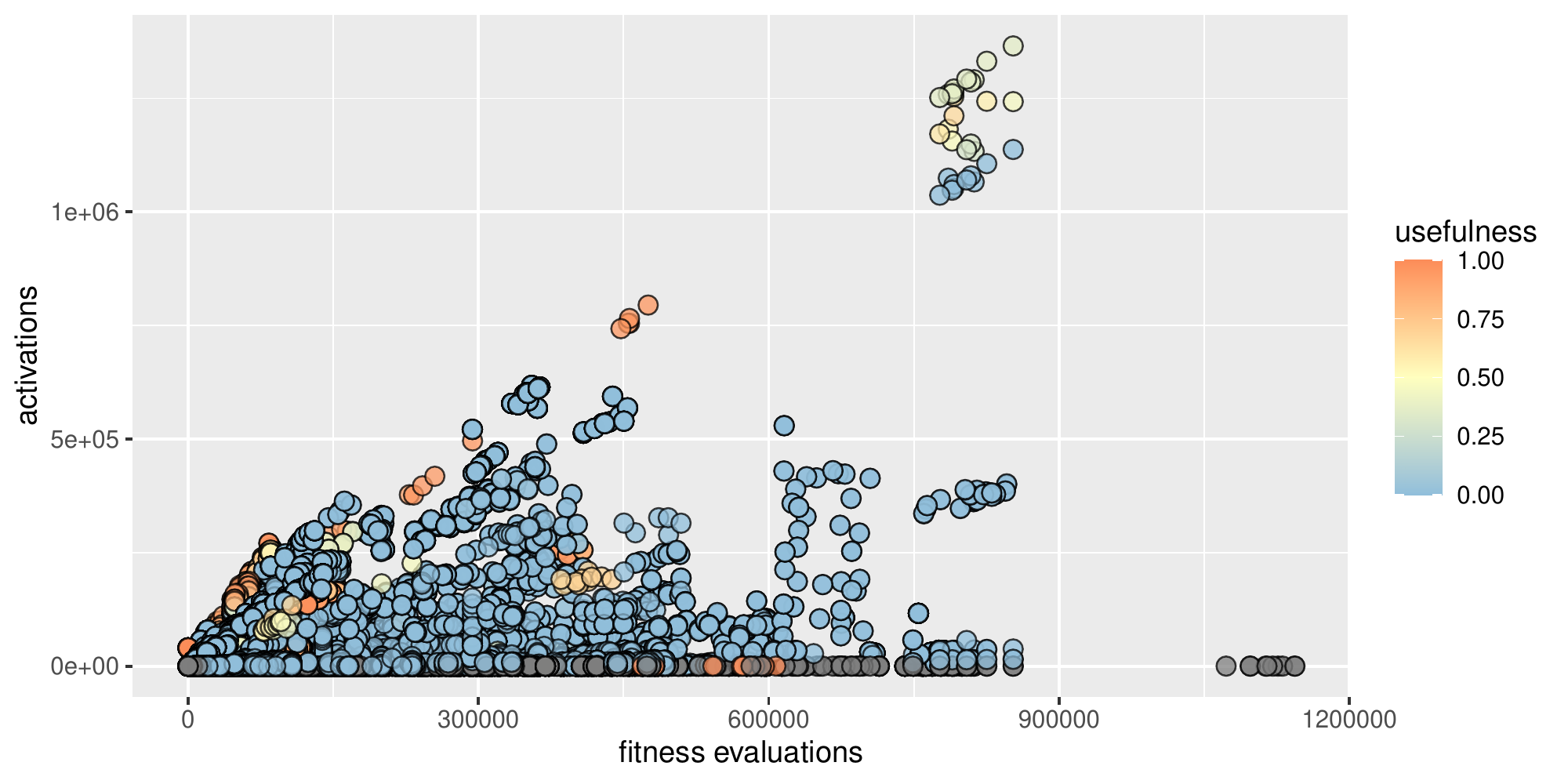}
    \caption{Distribution of the usefulness of \bbc activations per fitness evaluations. The usefulness is defined as the number of \bbc evaluations returning a non-zero value divided by the number of activations. Grey points denote fitness evaluations without any \bbc activation.}
    \label{fig:result:usefulness} 
\end{figure}

Figure \ref{fig:result:usefulness} provides a summary of the usefulness of \bbc. Each data point corresponds to the percentage of useful calls to \bbc per fitness evaluation, measured for one objective and one execution out of \nruns. On average, \bbc has been useful 2.5 times ($\sigma=3.17$ times) per fitness evaluation, with a maximum of 4,0145 times for a single fitness evaluation (which happens when multiple test cases have to be compared). 

\paragraph{\textbf{Summary (RQ 0).}}
Implicit branches are quite common. Our results show that on average, \bbc has been activated (\ie the call to \bbc effectively led to an evaluation) 9,472.140 times with a standard deviation $\sigma=40,567.40$, denoting big variations of the activation among the different objectives. The usefulness rate per activation is 2.39\% on average ($\sigma = 12.09\%$), confirming that not all activations can effectively lead to a distinction between two test cases \wrt to their partial coverage of basic blocks. Those results tend to confirm our design choice to parameterize the activation of \bbc using an activation probability. 

\subsection{Search-based unit test generation (\textbf{RQ 1})}

We first discuss the results of applying \bbc as a secondary objective for unit test generation using \dynamosa. Contrarily to crash reproduction, which seeks to cover only a small number of branches, unit test generation targets all the branches in a class under test. 

\begin{figure}[t]
    \subfloat[Branch coverage.\label{fig:result:branch:coverage}]{%
        \includegraphics[height=43.5mm]{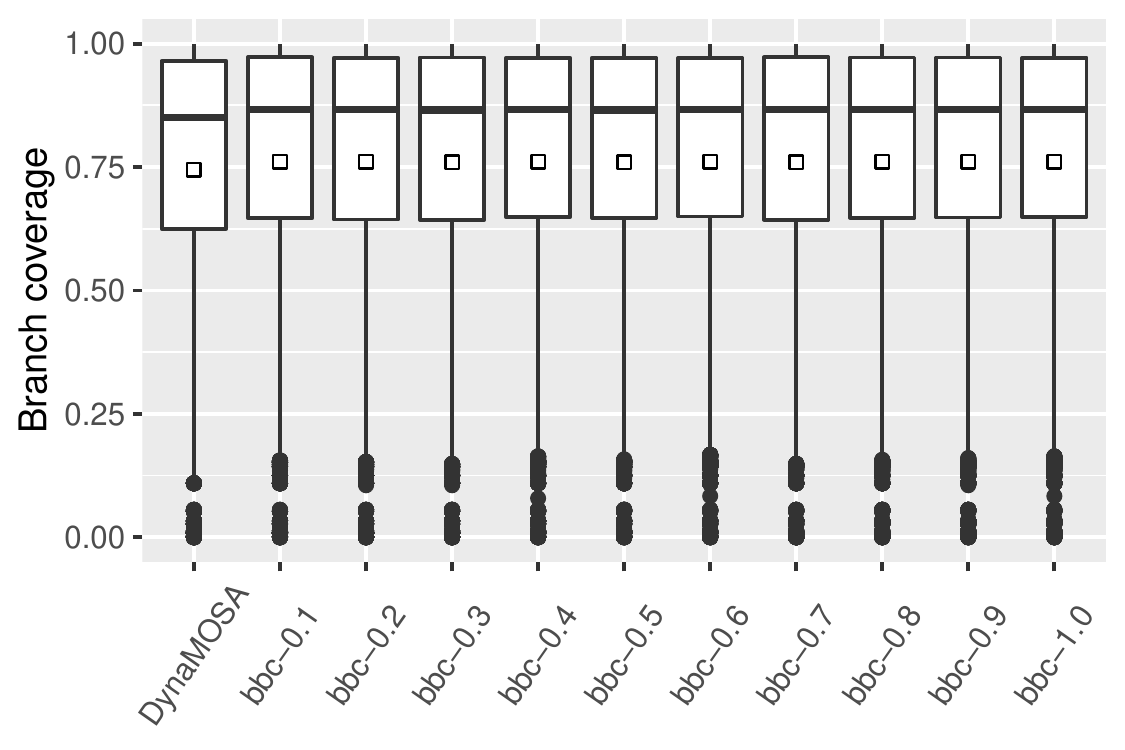}%
    }\hfil
    \subfloat[\^A$_{12}(\bbc_{Pr},\dynamosa)$ magnitudes with a positive (count $>0$) and negative (count $<0$) effect and a $p-value < 0.01$\label{fig:result:branch:vd}]{%
        \includegraphics[height=43.5mm]{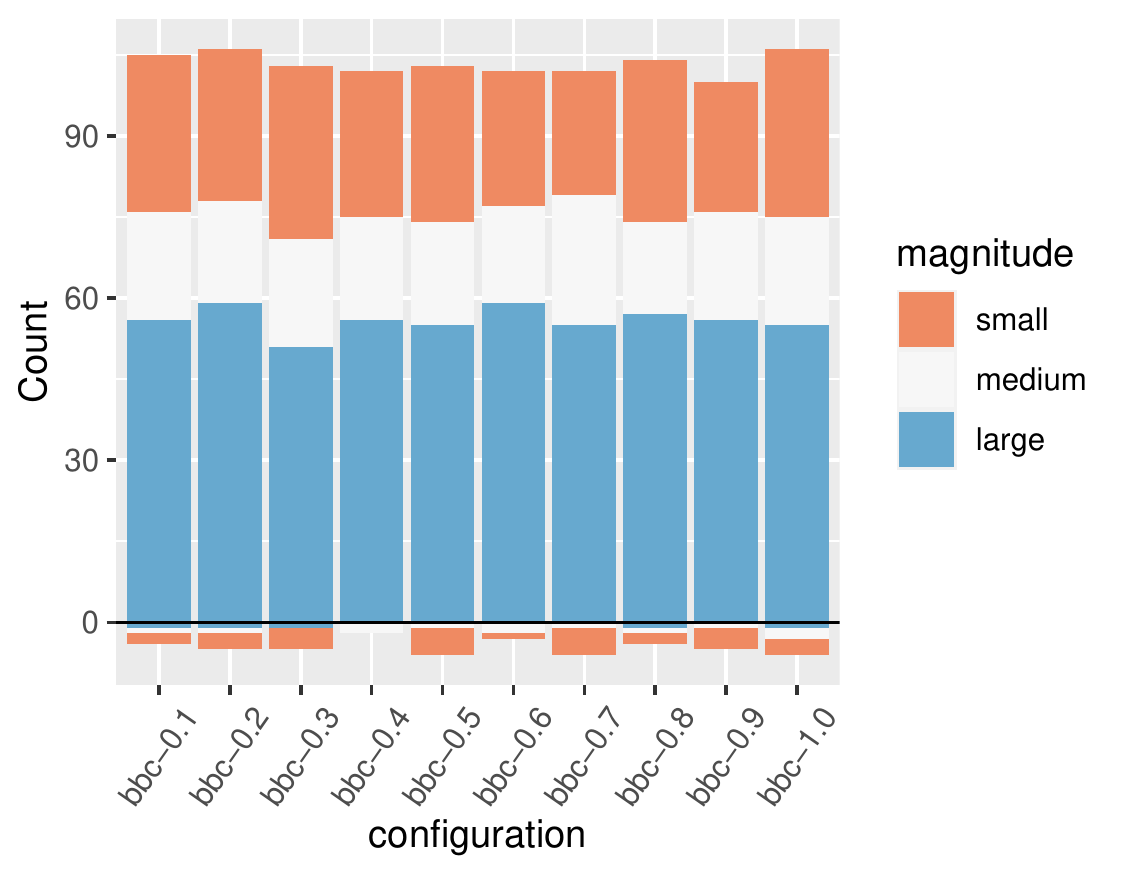}%
    }\hfil    
    \caption{Branch coverage of the tests generated for the \ncuts classes under test (out of 30 executions) for different configurations of  \bbc. The square ($\square$) denotes the arithmetic mean, the bold line (---) is the median.}
    \label{fig:result:branch}
\end{figure}

\paragraph{Branch coverage effectiveness (RQ 1.1).}
Figure \ref{fig:result:branch:coverage} reports the branch coverage of the different classes under test for all the \nruns test suites for the different configurations of \bbc. Generally, the average branch coverage slightly improves when activating \bbc as a secondary objective, from 74.5\% ($\sigma = 28\%$) for \dynamosa up to 76.1\% ($\sigma = 27.5\%$) for \bbc 0.2, 0.4, 0.6, and 1.0. Although small, this improvement is systematic across all \bbc configurations according to the effect sizes reported in Figure \ref{fig:result:branch:vd}. \bbc 0.6 gives the best results with a \textit{large} positive (\^A$_{12}>0.5$) effect size for 59 classes under test (against 0 \textit{large} negative, \^A$_{12}<0.5$, effect size), followed by \bbc 0.2 with 59 classes (against 1 classes), and \bbc 0.8 with 57 classes (against 1 class). 

\begin{figure}[t]
    \centering
    \includegraphics[width=80mm]{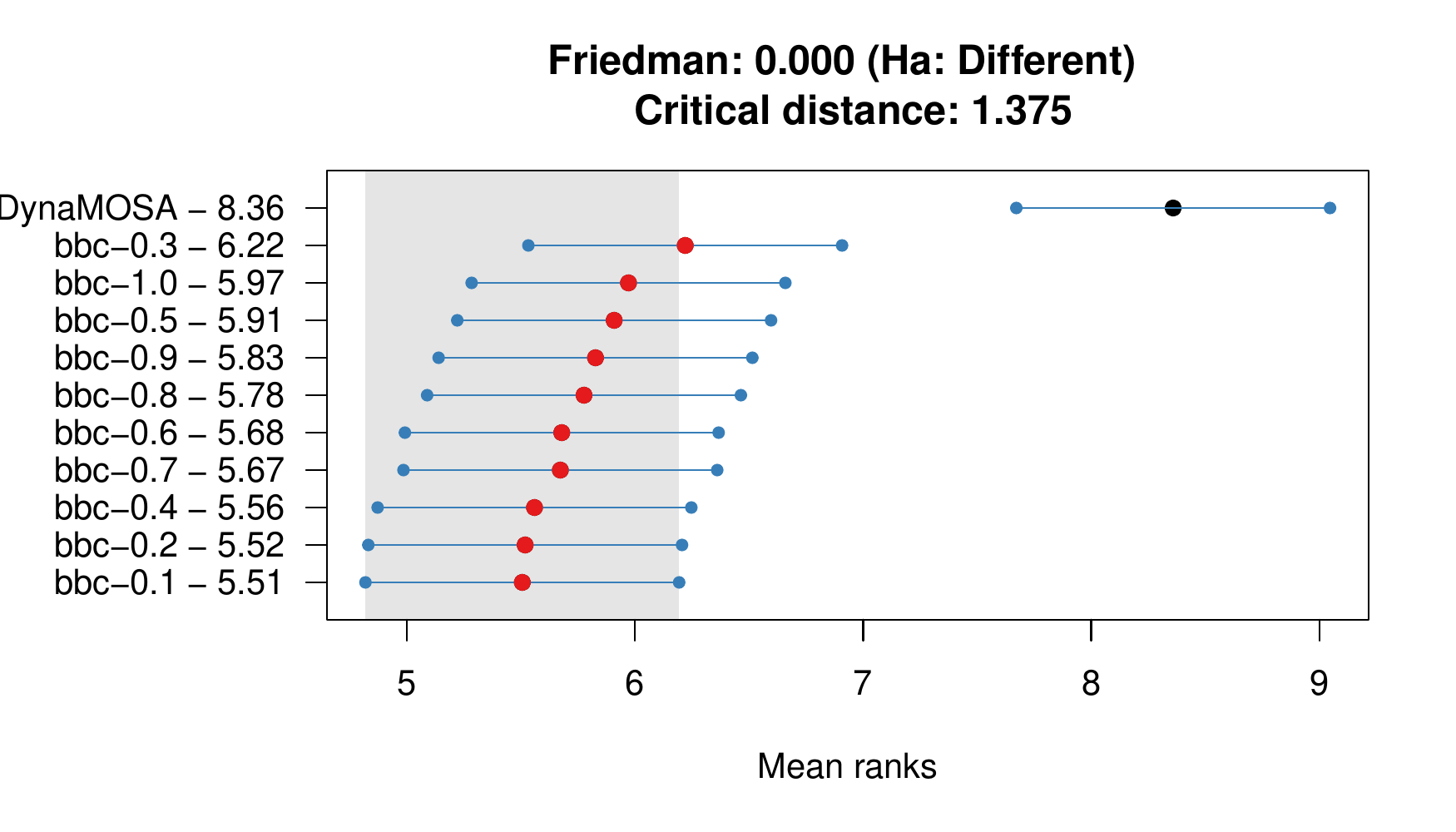}
    \caption{Non-parametric multiple comparisons of the branch coverage using Friedman's test with Nemenyi's post-hoc procedure.}
    \label{fig:result:branch:friedman}
\end{figure}

Figure \ref{fig:result:branch:friedman} provides a graphical representation of the ranking (\ie mean ranks with confidence interval) of the different \bbc configurations. According to Friedman's test, the different treatments \bbc 0.1 to 1.0 achieve significantly different branch coverage (p-values $ <0.01$) compared to \dynamosa. 
Furthermore, the differences between the average ranks of \bbc 0.1 to 1.0 and the average rank of the baseline are larger than the critical distance $CD=1.375$ determined by Nemenyi's post-hoc procedure (denoted by red dots in Figure \ref{fig:result:branch:friedman}). This indicates that \bbc 0.1 to 1.0 achieves a significantly higher branch coverage than \dynamosa.

We analyzed the correlation between the effect sizes (\^A$_{12}$) of the best performing \bbc configuration (according to Friedman's test with Nemenyi's post-hoc procedure) and \bbc usefulness (presented in RQ 0). The result of this analysis indicates that there is a positive correlation between the number of times that \bbc could be useful (\ie select a winner between two given tests with the same approach level and branch distance) and the effect that this secondary objective has on branch coverage improvement (Spearman's $\rho=0.4$ with a p-value $< 0.6e-10$). Hence, in any case that \bbc exposes that one generated test is closer to the target line than another test with the same approach level and branch distance (due to the implicit branch occurrence), there is a considerable chance that it helps the search-based test generation process to generate tests with higher branch coverage.

To confirm if this observed correlation stems from the connection between the potential implicit branches in the middle of basic blocks and improvement in the branch coverage, we manually analyzed some cases in which \bbc application leads to statistically significant improvement in branch coverage achie\-ved by the generated test. In this manual analysis, we identified multiple potential implicit exceptions before the target lines and branches, which are only covered by tests generated by utilizing \bbc as a secondary objective.

\begin{lstlisting}[float=t,
    caption={method \texttt{nextToken()} from \texttt{JacksonDatabind-106}},
    label=list:bbc:example:method,
    numbers=left, 
    language=Java,
    numberstyle=\tiny]
public JsonToken nextToken(){
    [...]
    if (_startContainer) {
        _startContainer = false;
        [...]
        _nodeCursor = _nodeCursor.iterateChildren();
        _currToken = _nodeCursor.nextToken();
        if ([...]) {
            _startContainer = true;
        }
            return _currToken;
    }
    [...]
}
\end{lstlisting}

\begin{lstlisting}[float=t,
    caption={method \texttt{iterateChildren()} in \texttt{JacksonDatabind-106}},
    label=list:bbc:example:called,
    numbers=left, 
    language=Java,
    numberstyle=\tiny]
public final NodeCursor iterateChildren() {
    [...]
    if (n == null){
        throw new IllegalStateException("No current node");
    } 
    if (n.isArray()) { 
        [...]
    }
    if (n.isObject()) {
        [...]
    }
    throw new IllegalStateException([...]);
}
\end{lstlisting}

For instance, for the class under test \texttt{com.fasterxml.jackson.data\-bind.no\-de.Tree\-Tra\-ver\-sing\-Parser} in \texttt{JacksonDatabind-106}, we see that tests generated by \bbc configurations achieve a higher structural coverage against \dynamosa.  In the majority of runs, the tests generated by \bbc managed to cover Lines 6 to 11 in method \texttt{nextToken()} (Listing \ref{list:bbc:example:method}), while \dynamosa is not successful in covering these lines. By looking at method \texttt{nodeCursor.iterateChildren()} (Listing \ref{list:bbc:example:called}), which is called by \texttt{nextToken()} in line 6 of Listing \ref{list:bbc:example:method}, we see that this method may throw an \texttt{IllegalStateException} at lines 4 and 12. Since \dynamosa does not have any information about the branches in the other classes other than the class under test, it cannot guide the search process to execute the method \texttt{iterateChildren()} without raising an exception.

\begin{figure}[t]
    \subfloat[Output coverage.\label{fig:result:output:coverage}]{%
        \includegraphics[height=43.5mm]{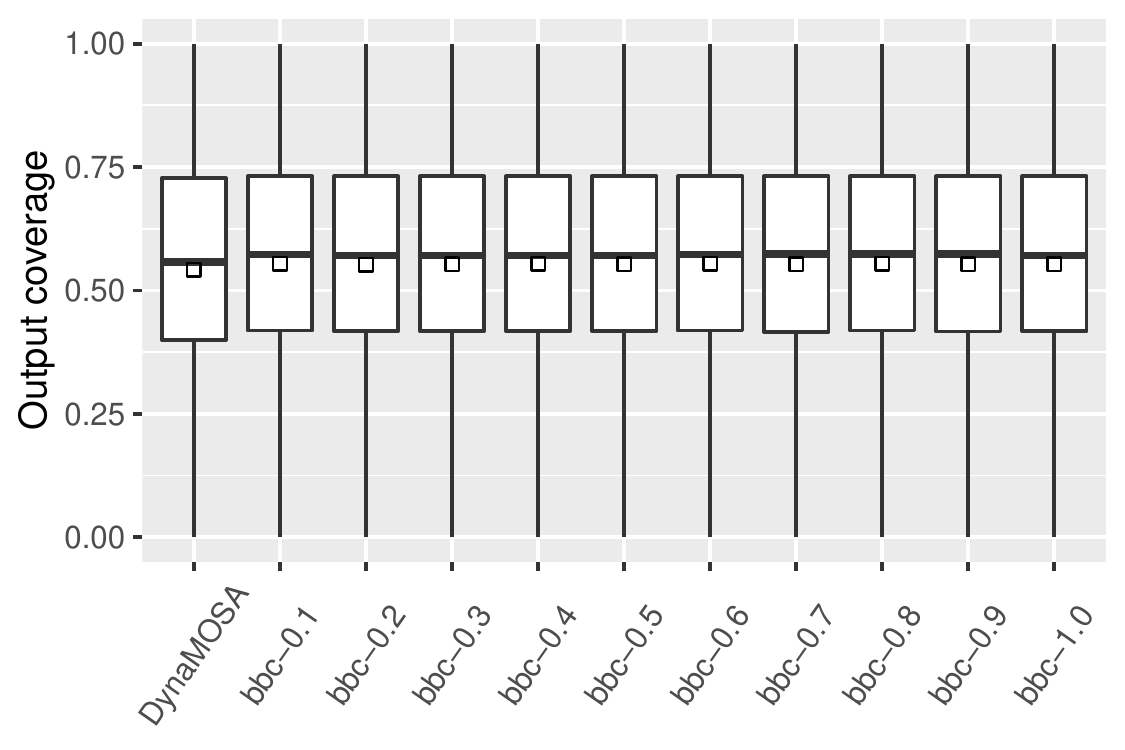}%
    }\hfil
    \subfloat[\^A$_{12}(\bbc_{Pr},\dynamosa)$ magnitudes with a positive (count $>0$) and negative (count $<0$) effect and a $p-value < 0.01$\label{fig:result:output:vd}]{%
        \includegraphics[height=43.5mm]{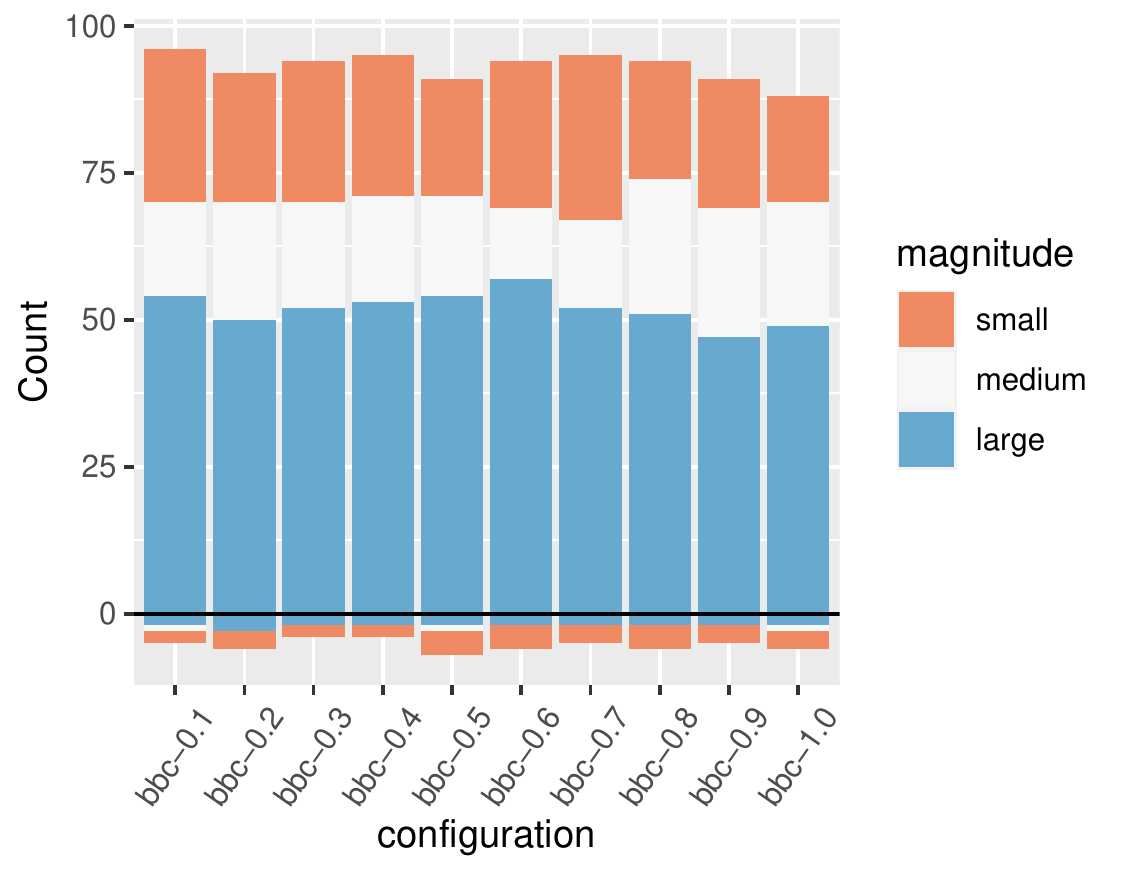}%
    }\hfil    
    \caption{Output coverage of the tests generated for the \ncuts classes under test (out of 30 executions) for different configurations of  \bbc. The square ($\square$) denotes the arithmetic mean, the bold line (---) is the median.}
    \label{fig:result:output}
\end{figure}

\paragraph{Output coverage and implicit exception coverage (RQ 1.2).}
The improvement of branch coverage also leads to more output diversity, reported in Figure~\ref{fig:result:output:coverage}: from 54.2\% ($\sigma = 26.6\%$) for \dynamosa up to 55.5\% ($\sigma = 26.2\%$) for \bbc 0.8. This improvement is also systematic across all \bbc configurations according to the effect sizes reported in Figure \ref{fig:result:output:vd}. \bbc 0.6 give the best results with a \textit{large} positive (\^A$_{12}>0.5$) effect size for 57 classes under test each (against 2 \textit{large} negative, \^A$_{12}<0.5$, effect sizes each), followed by \bbc 0.1 and 0.5 with 54 classes (against 2 classes), and \bbc 0.4 with 53 classes (against 2 classes).  

The two target classes with \textit{large} negative effect sizes on the output coverage are the same classes for the different configurations of \bbc: \ie different versions of the class \texttt{org.a\-pa\-che.com\-mons.cli.Help\-Format\-ter} in \texttt{Cli-31} and \texttt{Cli-32}. Interestingly, all \bbc configurations achieve a statistically significant higher implicit  runtime exception coverage (\ie undeclared runtime exceptions not explicitly thrown by a \texttt{throw new} instruction) with a large effect size for the same class on the same buggy versions of \texttt{Cli}, indicating that for this particular class, the loss of coverage of output values is compensated by a higher number of methods throwing implicit runtime exceptions.

This could be explained by the fact that \bbc favors test cases with a higher coverage of basic blocks, but that are not able to reach the return statements of the methods under test (\eg if the values used by the test cause implicit runtime exceptions). 
There is however no general correlation between the output coverage and the implicit exception coverage (Spearman's $\rho=-0.008$ with a p-value $< 0.001$).

Same as RQ 1.1, we evaluated the correlation between the improvement of \bbc in terms of output coverage and \bbc usefulness (presented in RQ 0). This analysis shows a positive correlation between these two metrics (Spearman's $\rho=0.3$ with a p-value $< 0.1e-5$). As we explained, this observation stems from the correlation between branch coverage and the output coverage achieved by each test: covering more lines and branches increases the chance of seeing more diverse output from CUT. To support this hypothesis, we also checked if there is a correlation between branch coverage and output coverage. Our analysis shows that branch coverage and output coverage are strongly correlated (Spearman's $\rho=0.6$ with a p-value $< 0.3e-16$).

\begin{figure}[t]
    \subfloat[Exception coverage.\label{fig:result:exception:coverage}]{%
        \includegraphics[height=43.5mm]{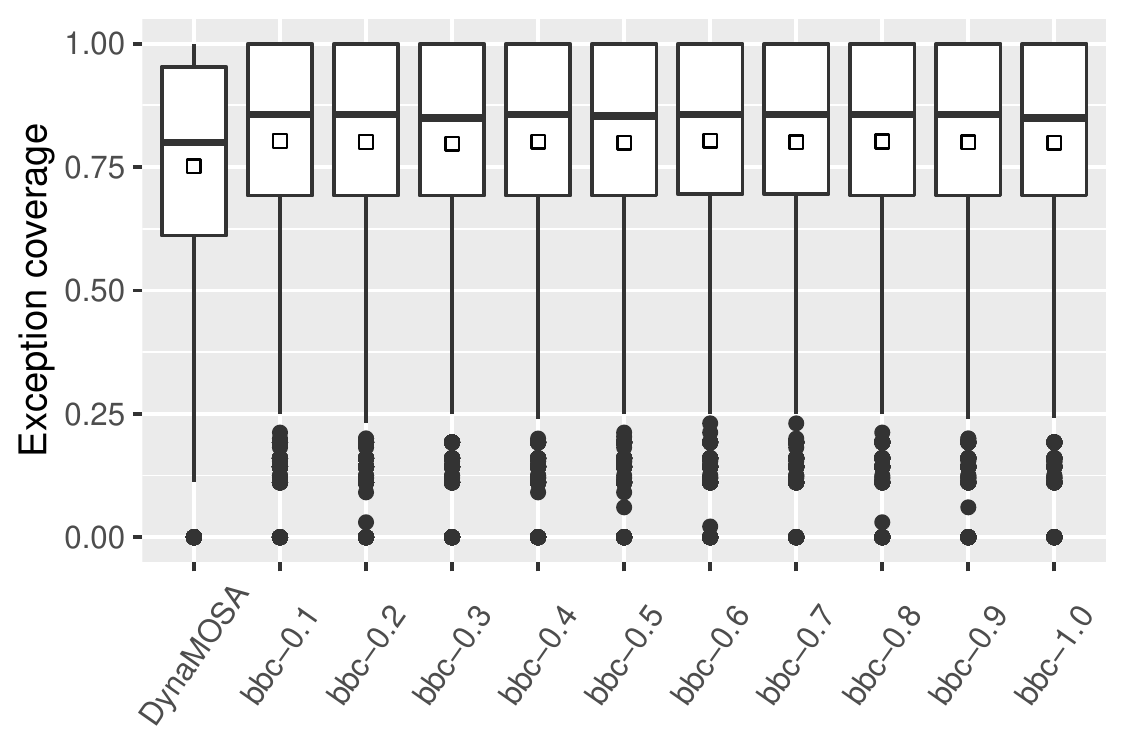}%
    }\hfil
    \subfloat[\^A$_{12}(\bbc_{Pr},\dynamosa)$ magnitudes with a positive (count $>0$) and negative (count $<0$) effect and a $p-value < 0.01$\label{fig:result:exception:vd}]{%
        \includegraphics[height=43.5mm]{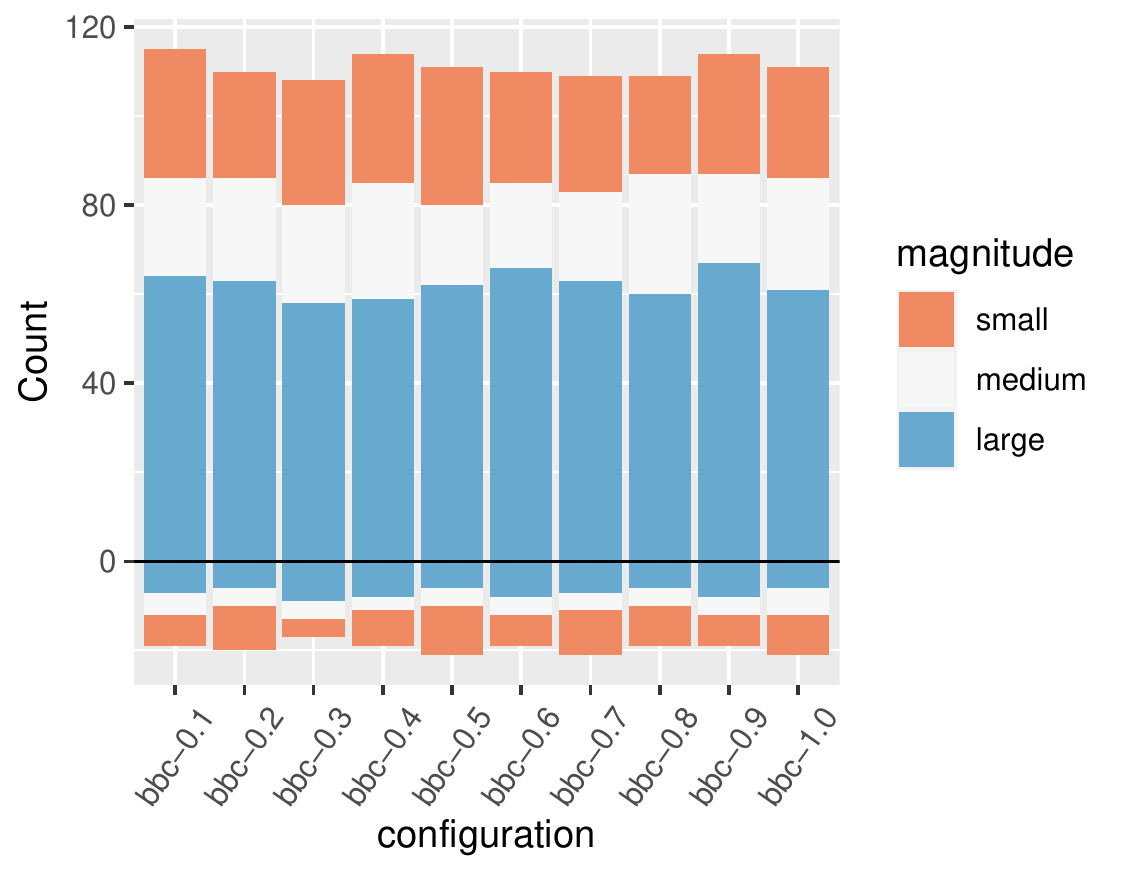}%
    }\hfil    
    \caption{Exception coverage of the tests generated for the \ncuts classes under test (out of 30 executions) for different configurations of  \bbc. The square ($\square$) denotes the arithmetic mean, the bold line (---) is the median.}
    \label{fig:result:exception}
\end{figure}

Figure \ref{fig:result:exception:coverage} reports the implicit runtime exception coverage of the generated tests. Implicit exceptions are not declared in the method under test and are triggered when implicit branches are executed. Results show that the average exception coverage increases when using \bbc as a secondary objective: from 75.1\% ($\sigma = 22.8\%$) when using \dynamosa up to 80.3\% for \bbc 0.1 ($\sigma = 21.2\%$) and 0.6 ($\sigma = 21\%$). \bbc 0.9 gives the best results with a \textit{large} positive (\^A$_{12}>0.5$) effect size for 67 classes under test (against 8 \textit{large} negative, \^A$_{12}<0.5$, effect size), followed by \bbc 0.6 with 66 classes (against 8 classes), and \bbc 0.1 with 64 classes (against 7 classes).

\begin{figure}[t]
    \subfloat[Output coverage.\label{fig:result:output:friedman}]{%
        \includegraphics[width=0.49\textwidth]{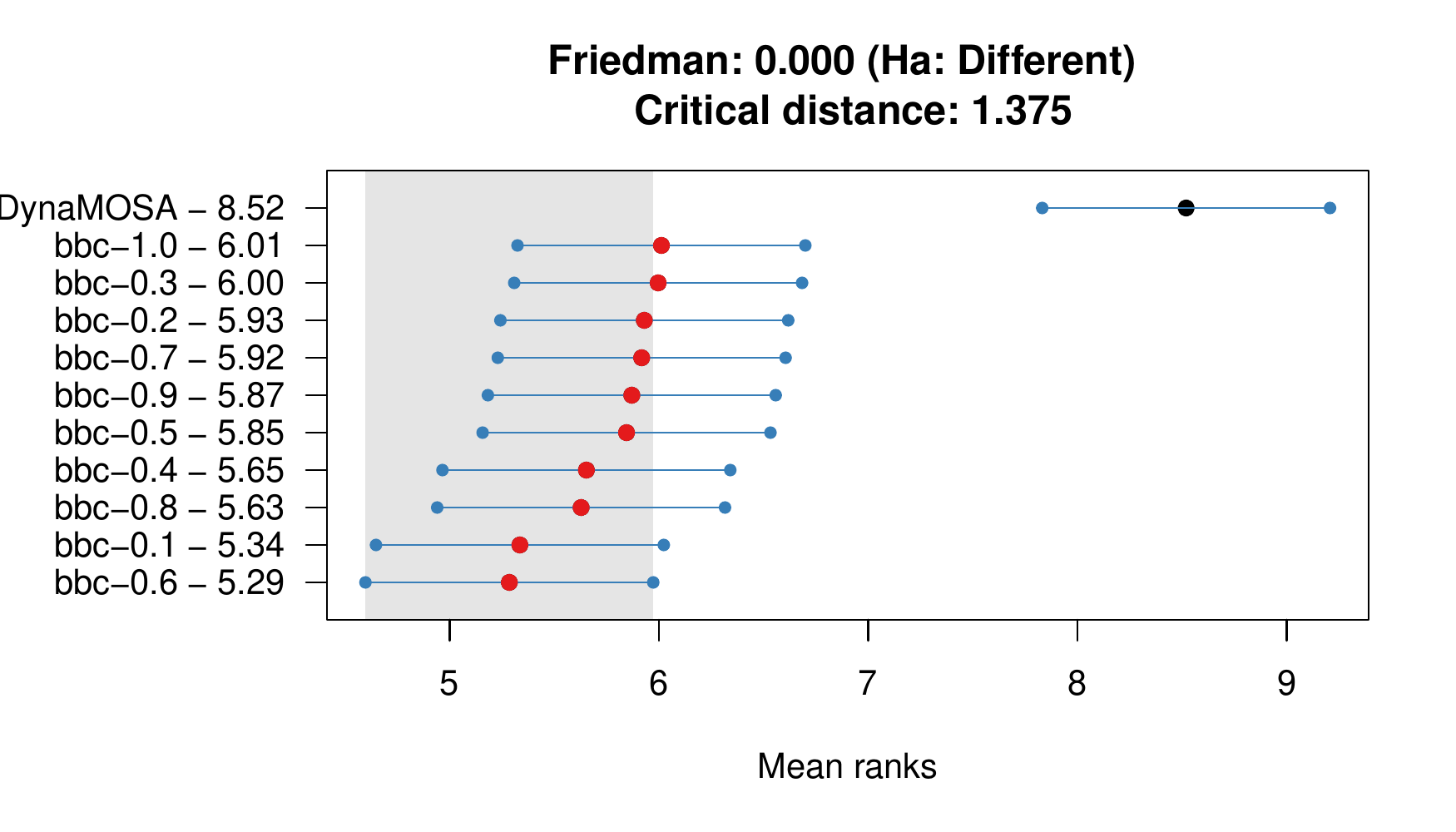}%
    }\hfil
    \subfloat[Exception coverage.\label{fig:result:exception:friedman}]{%
        \includegraphics[width=0.49\textwidth]{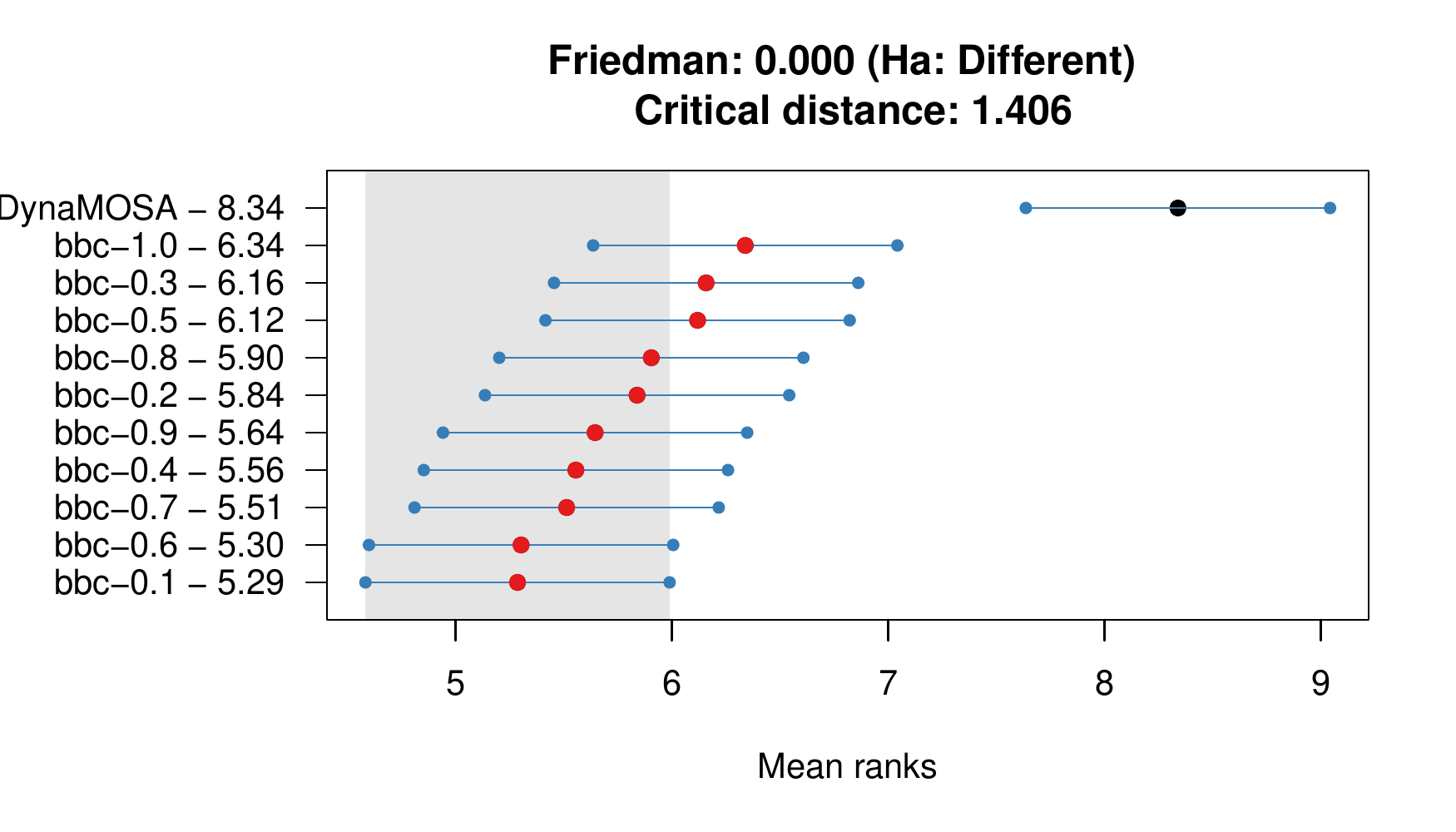}%
    }\hfil    
    \caption{Non-parametric multiple comparisons of the coverage using Friedman's test with Nemenyi's post-hoc procedure.}
    \label{fig:result:exception:output:friedman}
\end{figure}

The rankings in Figure \ref{fig:result:exception:output:friedman} indicate that \bbc 0.1 to 1.0 perform well, with an average rank much smaller than the baseline, both for output and exception coverage. The configurations' average ranks differences with the average rank of the baseline are larger than the critical distance $CD=1.375$ determined by Nemenyi's post-hoc procedure.

In contrast with branch coverage and output coverage, Spearman's test does not show any general correlation between \bbc usefulness and implicit exception coverage (Spearman's $\rho=0.04$ with a p-value $= 0.5$). This result supports our discussion in Section \ref{sec:approach}: since \bbc is only triggered when \dynamosa compares tests regarding a line or branch coverage search objective, it does not have any negative impact on other search objectives, including the implicit exception coverage of the generated tests. We also analyzed some of the exceptions that are only thrown by the tests generated using \bbc. The remainder of this section explains one of these instances.

\begin{lstlisting}[float=t,
    caption={An implicit exception in \texttt{MATH-3} which is thrown significantly more often by tests generated by the search process utilizing \bbc secondary objective.},
    label=list:bbc:example:exception,
    numbers=left, 
    language=Java,
    numberstyle=\tiny]
java.lang.ArrayIndexOutOfBoundsException: 1
    at [...].MathArrays.linearCombination([...]:846)
\end{lstlisting}

\begin{lstlisting}[float=t,
    firstnumber=834,
    caption={method \texttt{linearCombination} from \texttt{Apache Commons MATH}},
    label=list:bbc:example:exception:code,
    numbers=left, 
    language=Java,
    numberstyle=\tiny]
public static double linearCombination(final double[] a, final double[] b)
{
    [...]
    if (len != b.length) {
            throw new DimensionMismatchException(len, b.length);
    }

    for (int i = 0; i < len; i++) {
        [...]
    }

    final double prodHighCur = prodHigh[0];
    double prodHighNext = prodHigh[1]; // target line
    [...]
}
\end{lstlisting}

Listing \ref{list:bbc:example:exception} shows an example of an implicit exception that is thrown significantly more often when using \bbc. \dynamosa managed to capture this exception in 9 our of \nruns runs, while \bbc 0.5 captured it in 23 out of \nruns runs.
This exception occurs in line 846 of method \texttt{linearCombination} (Listing \ref{list:bbc:example:exception:code}).
This exception can be triggered only in one specific case where the input arrays (\texttt{a} and \texttt{b}) both contain only one element.
If these two parameters do not have the same size, this method throws an \textit{explicit} exception at line 838 (\ie this line is formatted as \texttt{throw new [...]}).
Since \evosuite can recognize explicit exception throws in the CUT and convert them to explicit branches while generating the control flow graphs, approach level and branch distance can guide the search process to cover other lines after 839 by prioritizing tests that pass two arrays with the same size to method \texttt{linearCombination}. 
However, since the explicit branch was the only control-dependent branch for the target line (line 846), the search process does not have any guidance to cover the following lines (including the target line).
Assume that test \texttt{T1} generates input parameters \texttt{a} and \texttt{b} with size 0. Then, this method throws \texttt{Ar\-ray\-In\-dex\-Out\-Of\-Bounds\-Exception} in one line before the target line (line~845). This implicit branch will be hidden from the approach level and branch distance heuristics. By adding \bbc, the search process can differentiate these two tests and help the search process to generate tests that can cover the following lines more often. By having more tests that can cover the target line, the search process has a higher opportunity to execute the target line, and thereby find the exception in this line. 

\begin{figure}[t]
    \subfloat[Weak mutation score.\label{fig:result:mutation:score}]{%
        \includegraphics[height=43.5mm]{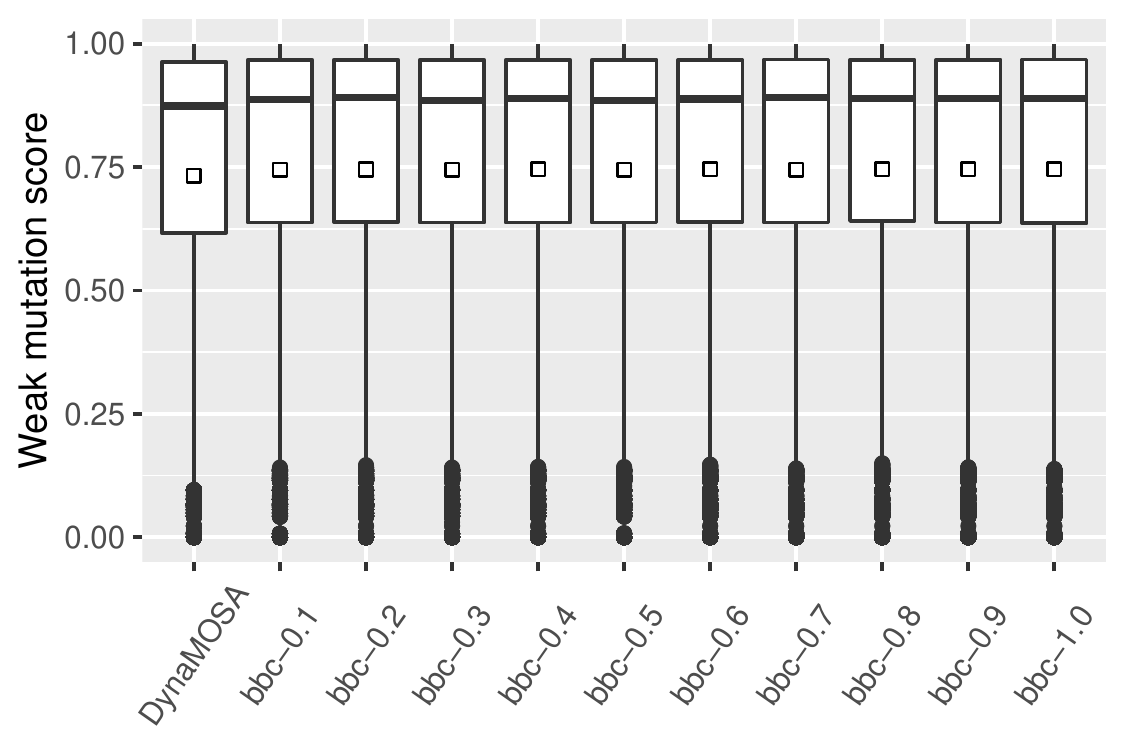}%
    }\hfil
    \subfloat[\^A$_{12}(\bbc_{Pr},\dynamosa)$ magnitudes with a positive (count $>0$) and negative (count $<0$) effect and a $p-value < 0.01$\label{fig:result:mutation:vd}]{%
        \includegraphics[height=43.5mm]{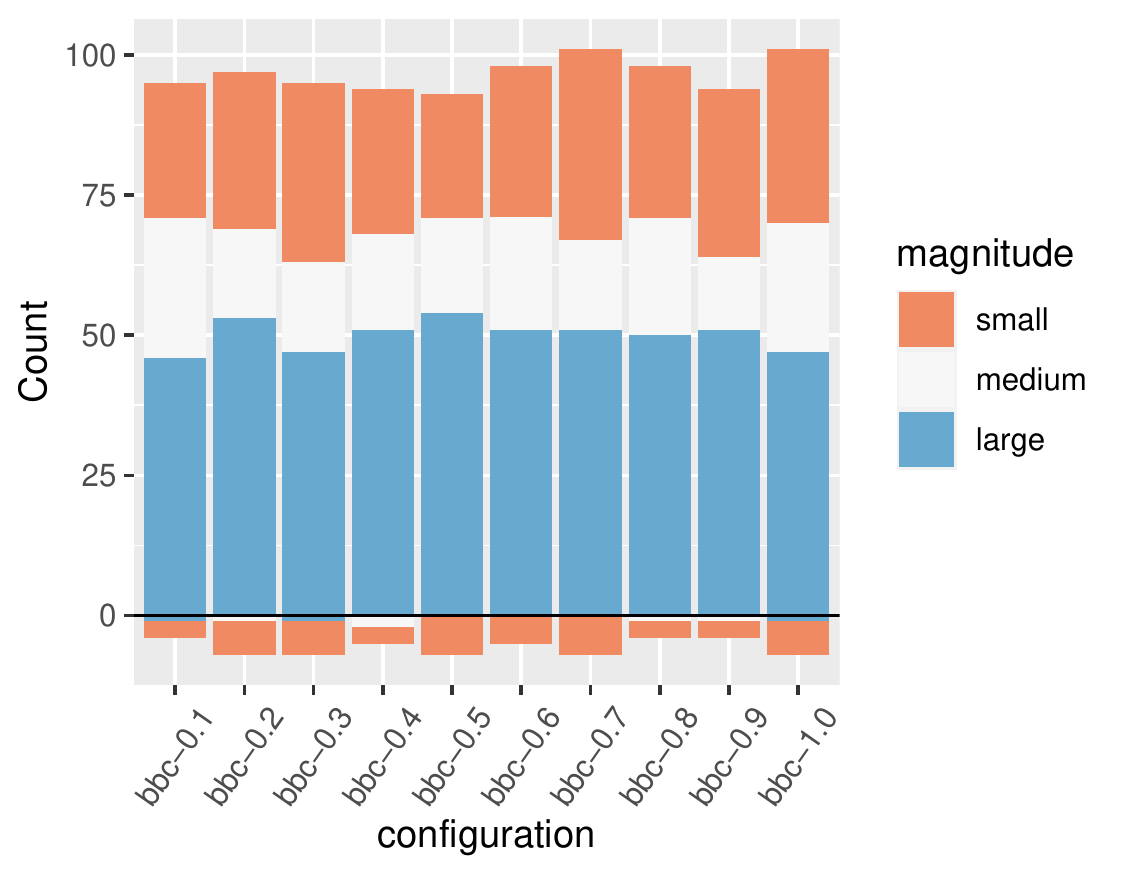}%
    }\hfil 
    \caption{Weak mutation score of the tests generated for the \ncuts classes under test (out of 30 executions) for different configurations of \bbc. The square ($\square$) denotes the arithmetic mean, the bold line (---) is the median.}
    \label{fig:result:mutation}
\end{figure}

\paragraph{Weak mutation score and real faults (RQ 1.3)}
As for branch and output coverage, activating \bbc slightly improves the weak mutation score of the generated tests (reported in Figure \ref{fig:result:mutation:score}). \bbc 0.4, 0.6 and 0.8 achieve the higher average mutation score with 74.6\% ($\sigma = 29.6\%$), compared to 73.2\% ($\sigma = 30.1\%$) for the baseline (\dynamosa). That improvement is also systematic across the different configurations of \bbc according to the effect sizes reported in Figure \ref{fig:result:mutation:vd}. \bbc 0.5 gives the best results with a \textit{large} positive (\^A$_{12}>0.5$) effect size for 54 classes under test (against 0 \textit{large} negative, \^A$_{12}<0.5$, effect size), followed by \bbc 0.2 with 53 classes (against 0 class), and \bbc 0.4, 0.6, 0.7 and 0.9 with 51 classes each (against 0 class). 

\begin{figure}[t]
    \centering
    \includegraphics[width=80mm]{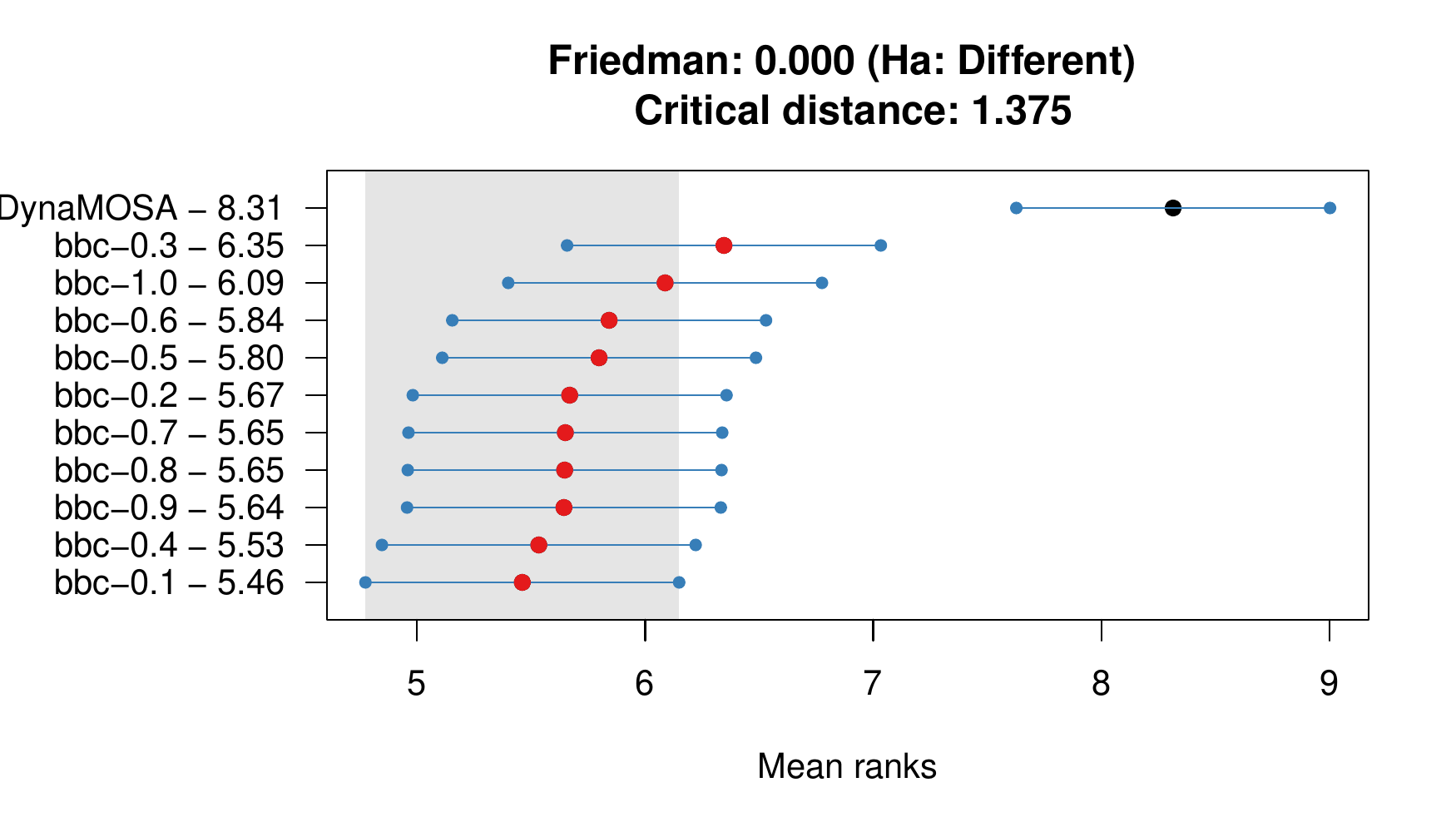}
    \caption{Non-parametric multiple comparisons of the weak mutation score using Friedman's test with Nemenyi's post-hoc procedure.}
    \label{fig:result:mutation:friedman}
\end{figure}

Looking at the ranking reported in Figure \ref{fig:result:mutation:friedman}, \bbc 0.1 to 1.0 have an average rank much smaller than the baseline. Those differences are larger than the critical distance $CD=1.375$ determined by Nemenyi's post-hoc procedure.

Moreover, we checked if we could find any correlation between the weak mutation score and \bbc usefulness (presented in RQ 0). This analysis shows a moderate correlation between these two metrics (Spearman's $\rho=0.37$ with a p-value $< 0.3e-8$). One reason for this correlation could be the strong correlation between weak mutation score and branch coverage (Spearman's $\rho=0.91$ with a p-value $< 0.3e-16$). Thanks to \bbc secondary objective, the search-based test generation process can cover more lines and branches, thereby killing the mutants in these newly covered lines.
 
\begin{table}[t]
    \centering
    \caption{Real faults coverage of the different configurations with the number of faults covered at least once in \nruns runs ($\#$) out of 92 faults, the average coverage frequency ($\overline{freq.}$, $\sigma$), and the number of time a configuration performed better ($>1$) of worse ($<1$) than \dynamosa with a significance level of 0.01. }
\begin{tabular}{ l | c c c | c c c }
\textbf{Config.} & \multicolumn{3}{c|}{\textbf{Faults coverage}} & \multicolumn{3}{c}{\textbf{Odds ratio}} \\ 
  & $\#$ & $\overline{freq.}$ & $\sigma$ & $>1$ & $=1$ & $<1$ \\ 
\hline 
bbc-0.1 & 26 & $22.25\%$ & $38.84\%$ & 1 & - & - \\ 
bbc-0.2 & 27 & $22.79\%$ & $39.18\%$ & 2 & - & - \\ 
bbc-0.3 & 26 & $22.28\%$ & $39.02\%$ & 2 & - & 1 \\ 
bbc-0.4 & 26 & $22.28\%$ & $38.66\%$ & 3 & - & - \\ 
bbc-0.5 & 25 & $22.68\%$ & $39.36\%$ & 3 & - & - \\ 
bbc-0.6 & 27 & $22.46\%$ & $38.86\%$ & 3 & - & - \\ 
bbc-0.7 & 26 & $23.04\%$ & $39.68\%$ & 2 & - & - \\ 
bbc-0.8 & 28 & $22.39\%$ & $38.75\%$ & 3 & - & - \\ 
bbc-0.9 & 25 & $22.57\%$ & $38.96\%$ & 2 & - & - \\ 
bbc-1.0 & 27 & $22.25\%$ & $38.97\%$ & 3 & - & - \\ 
DynaMOSA & 26 & $21.49\%$ & $38.37\%$ & - & - & - \\ 
\end{tabular}
    \label{tab:faults}
\end{table} 

Finally, we compare the fault revealing capabilities of the generated tests using \defectsforj. Table \ref{tab:faults} presents the results for the different configurations of \bbc and the baseline (\dynamosa). In general, the tests reveal between 25 and 28 faults at least once in \nruns rounds of executions out of the 92 faults considered (the selection procedure is detailed in Section \ref{sec:setup:unittest}). For the faults that are revealed in at least one round, the average coverage frequency (for \nruns rounds of execution) varies between 22.25\% (for \bbc 0.1 and 1.0) and 23.04\% (for \bbc 0.7). The table also reports the number of faults for which a configuration performed better (odds ratio above 1) or worse (odds ratio below 1) than the \dynamosa baseline with a significance level of 0.01. The best configurations are \bbc 0.4, 0.5, 0.6, 0.8, and 1.0 with 3 faults (against 0).

\begin{lstlisting}[float=t,
    caption={The fault in \texttt{CHART-4} which is captured significantly more often by tests generated by the search process utilizing \bbc secondary objective.},
    label=list:bbc:example:fault,
    numbers=left, 
    firstnumber=0,
    language=Java,
    numberstyle=\tiny]
java.lang.NullPointerException
	at org.jfree.chart.plot.XYPlot.getDataRange(XYPlot.java:4493) 
	at org.jfree.chart.axis.NumberAxis.autoAdjustRange(NumberAxis.java:434)
	at org.jfree.chart.axis.NumberAxis.configure(NumberAxis.java:417)
	at org.jfree.chart.plot.XYPlot.configureDomainAxes(XYPlot.java:972)
	at org.jfree.chart.plot.XYPlot.setRenderer(XYPlot.java:1644)
	at org.jfree.chart.plot.XYPlot.setRenderer(XYPlot.java:1620)
	at org.jfree.chart.plot.XYPlot.setRenderer(XYPlot.java:1607)
\end{lstlisting}

\begin{lstlisting}[float=t,
    firstnumber=4464,
    caption={method \texttt{getDataRange} from \texttt{JFreeChart}},
    label=list:bbc:example:fault:code1,
    numbers=left, 
    language=Java,
    numberstyle=\tiny]
public Range getDataRange(ValueAxis axis) {
    [...]
    // iterate through the datasets that map to the axis and get the union
    // of the ranges.
    Iterator iterator = mappedDatasets.iterator();
    while (iterator.hasNext()) 
    {
        XYDataset d = (XYDataset) iterator.next();
        if (d != null) {
            XYItemRenderer r = getRendererForDataset(d);
            if (isDomainAxis) {
                if (r != null) {
                    result = Range.combine(result, r.findDomainBounds(d));
                }
                else {
                    result = Range.combine(result,
                            DatasetUtilities.findDomainBounds(d));
                }
            }
            else {
                if (r != null) {
                    result = Range.combine(result, r.findRangeBounds(d));
                }
                else {
                    result = Range.combine(result,
                            DatasetUtilities.findRangeBounds(d));
                }
            }
            
            Collection c = r.getAnnotations(); // target line
            [...]
        }
    }
```
}
\end{lstlisting}

We manually analyzed the three faults that are captured significantly more often by \bbc. In all of them, we identified potential implicit branches before covering the target line (\ie the line in which the fault happens) that can prevent the classical and approach level from guiding the search process towards covering these failures.

For instance, Listing \ref{list:bbc:example:fault} presents the stack trace that reveals a fault in \texttt{J\-Free\-Chart}.\footnote{See case \texttt{CHART-4} in \defectsforj at \url{https://github.com/rjust/defects4j/blob/master/framework/projects/Chart/trigger_tests/4}} When selecting the \texttt{XYPlot} class as class under test, \bbc configurations can throw this exception significantly more often than tests generated by \dynamosa. This stack trace has five frames that are pointing to a method in the target class (\texttt{XYPlot}): Lines 1, 4, 5, 6, and 7 in Listing \ref{list:bbc:example:fault}. By analyzing the methods in these frames, we can see that majority of them are simple methods with one line except the first frame in Line 1 of Listing \ref{list:bbc:example:fault}, which points to method \texttt{getDataRange} that has about 100 lines of codes. 

As we can see in Listing \ref{list:bbc:example:fault:code1}, the target line, in which the \texttt{NullPointerException} occurs (Line 4493), is in an \texttt{if} condition which starts at Line 4472. 
The target line is directly control-dependent on this condition. Hence, when a test fulfills the condition in line 4472, the approach level and branch distance heuristics assume that the generated test eventually will cover the target line (Line 4494), and thereby these two heuristics do not provide any guidance for the test generation search process afterward. However, by taking a closer look, we can see that even after entering the \texttt{if} condition, a test needs to, first, call the \texttt{combine} method (in one of the Lines 4476, 4479, 4485, or 4488) and also call either \texttt{findDomainBounds} (in Lines 4476 or 4479) or \texttt{findRangeBounds} (in Lines 4485 or 4488) before it can reach the target line. Each of these methods can throw explicit exceptions. Since these methods are not part of the class under test, the search process is unaware of those exceptions. Also, each of these methods calls multiple methods that can also throw exceptions.

\bbc can guide the test generation search process to execute these lines without any exception and cover the target line. By covering the target line, the search process has the opportunity to generate a test that throws a \texttt{Null\-Pointer\-Exce\-ption} in this target line, and thereby captures the fault.
 
\begin{figure}[t]
    \subfloat[Data distribution and smoothed conditional means. \label{fig:result:branch:coverage:evolution}]{%
        \includegraphics[width=0.98\textwidth]{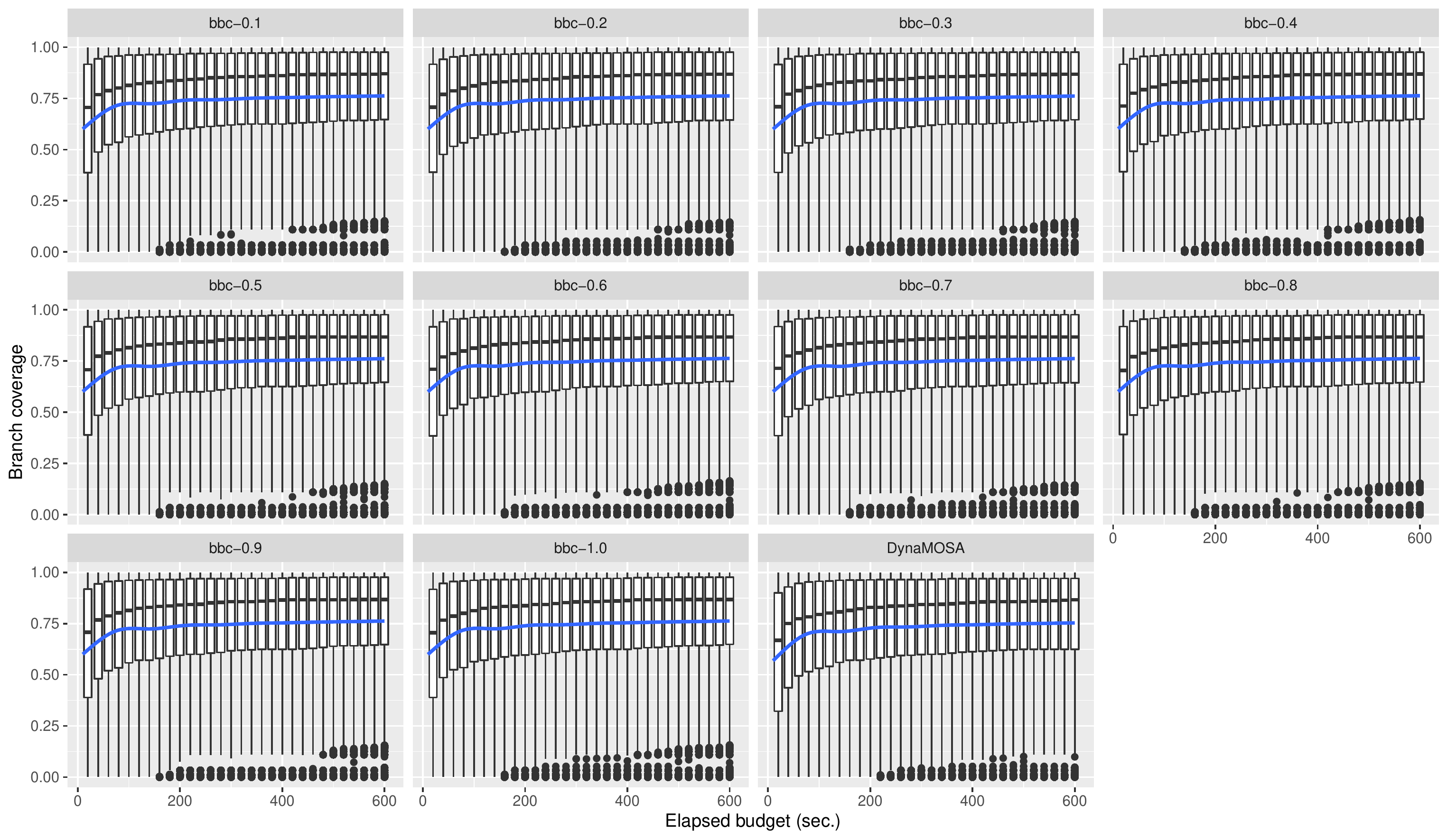}%
    }\hfil
    \subfloat[\^A$_{12}(\bbc_{Pr},\dynamosa)$ magnitudes evolution with a positive (count $>0$) and negative (count $<0$) effect and a $p-value < 0.01$.\label{fig:result:branch:vd:evolution}]{%
        \includegraphics[width=0.98\textwidth]{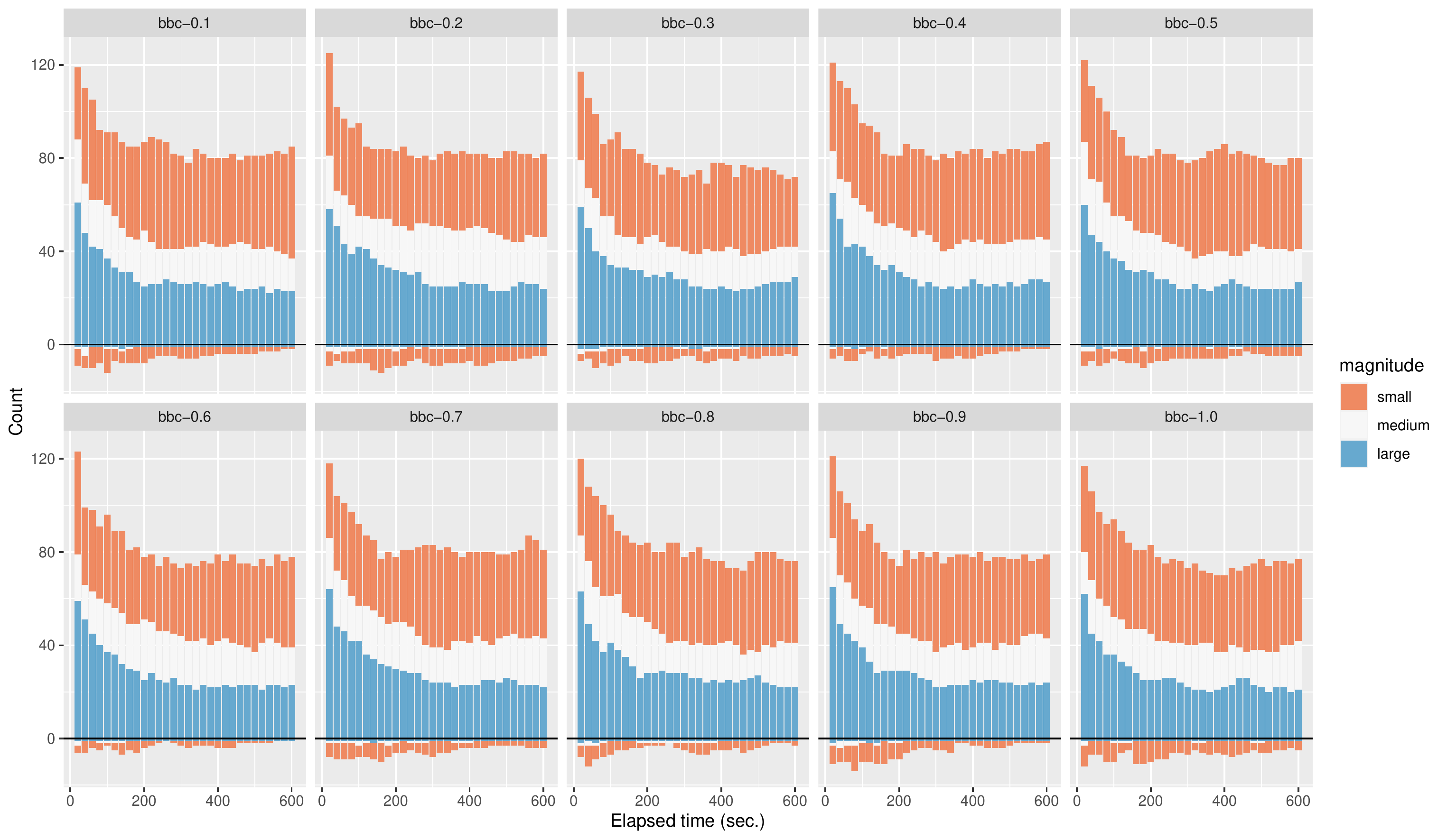}%
    }\hfil    
    \caption{Evolution of the branch coverage of the tests generated for the \ncuts classes under test (out of 30 executions) for different configurations of \bbc.}
    \label{fig:result:branch:evolution}
\end{figure}

\paragraph{Branch coverage efficiency (RQ 1.4).}
Figure \ref{fig:result:branch:coverage:evolution} presents the tendency of the branch coverage over time using the smoothed conditional means. Overall, \bbc~0.5 tends to achieve a higher branch coverage. This is confirmed by the number of classes for which we observe a significant difference (with $\alpha = 0.01$) in the coverage achieved, reported in Figure \ref{fig:result:branch:vd:evolution} and grouped by effect size (\^A$_{12}$) magnitude. Counts above (resp. below) 0 denote the number of classes for which we observe a positive (resp. negative) effect. After three minutes, \bbc 0.4 achieves a large (resp. medium) positive effect size for 34 (resp. 18) classes under test against 1 (resp. 0) large (resp. medium) negative effect sizes. Those numbers slightly decrease over time with 27 (resp. 18) classes under test with a large (resp. medium) effect size after exhaustion of the ten minutes search budget, for 1 (resp. 0) large (resp. medium) classes with a negative effect size.
 
\paragraph{\textbf{Summary (RQ 1).}}
We see an improvement of the branch coverage of the generated tests when activating \bbc as a secondary objective in \dynamosa. This improvement in branch coverage also leads to an increase of the output and exception coverage, and of the diversity of runtime states (denoted by an increase of the weak mutation score). 
Among the different configurations, \bbc 0.5 gives the best results and those results remain stable over time. It also leads to the coverage of three additional faults in \defectsforj without any loss compared to the baseline. 
Giving our results, we can recommend using \bbc 0.5 as a secondary objective for unit test generation. 

\subsection{Search-based crash reproduction (\textbf{RQ 2})}

\begin{figure}[t]
    \subfloat[\integ\label{fig:result:overall:reproduction:integ}]{%
        \includegraphics[width=0.45\textwidth]{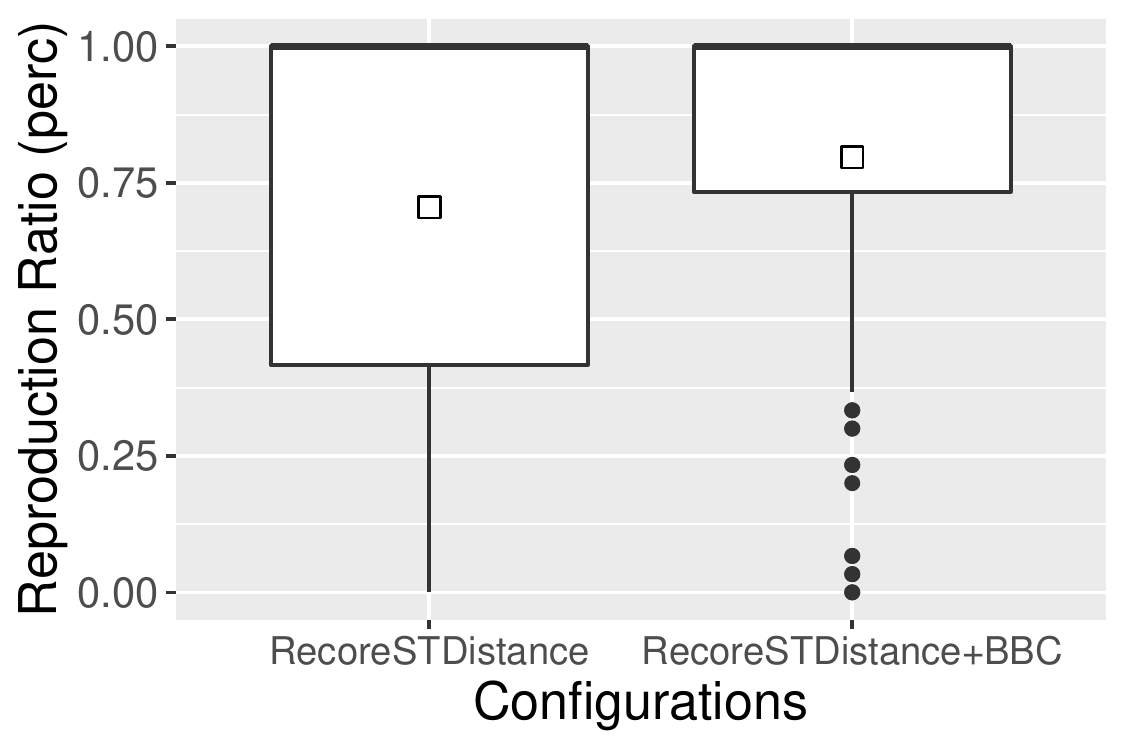}%
    }\hfil
    \subfloat[\WS\label{fig:result:overall:reproduction:ws}]{%
        \includegraphics[width=0.45\textwidth]{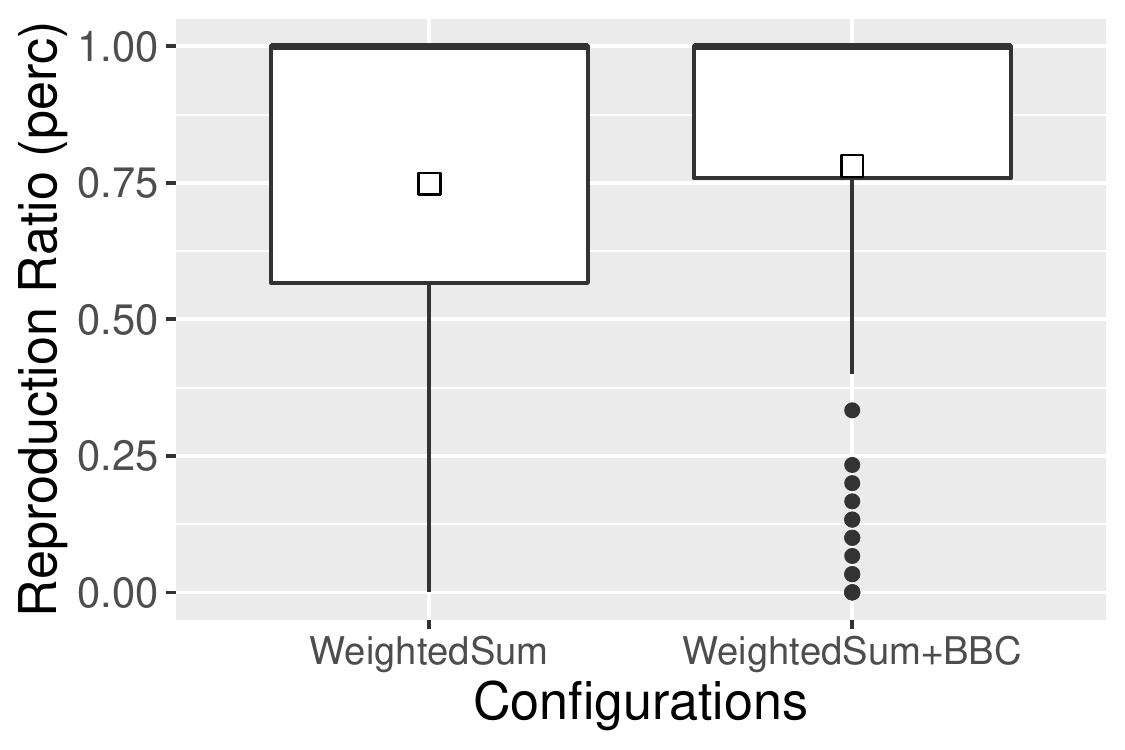}%
    }\hfil    
    \caption{Crash reproduction ratio (out of 30 executions) of fitness functions with and without \bbc. The square ($\square$) denotes the arithmetic mean and the bold line (---) is the median.}
    \label{fig:result:overall:reproduction}
\end{figure}

\begin{figure}[t]
    \subfloat[\integ\label{fig:result:significance:reproduction:integ}]{%
        \includegraphics[width=0.45\textwidth]{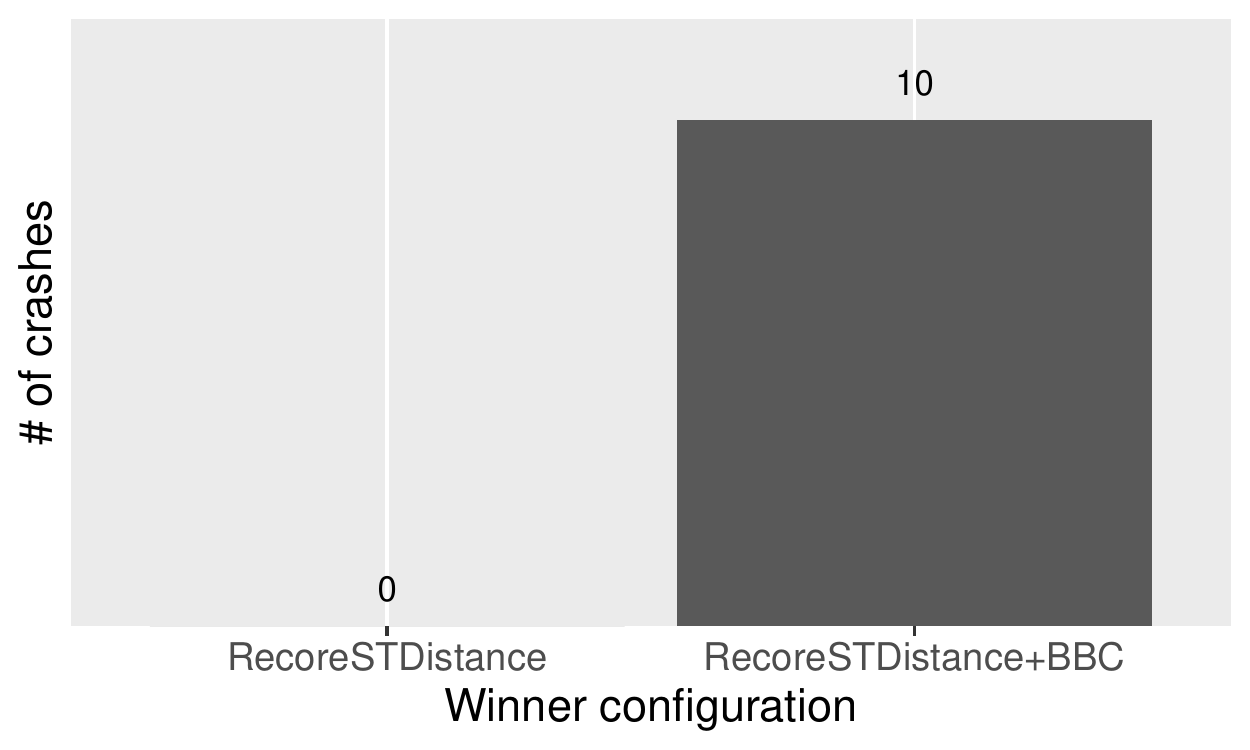}%
    }\hfil
    \subfloat[\WS\label{fig:result:significance:reproduction:ws}]{%
        \includegraphics[width=0.45\textwidth]{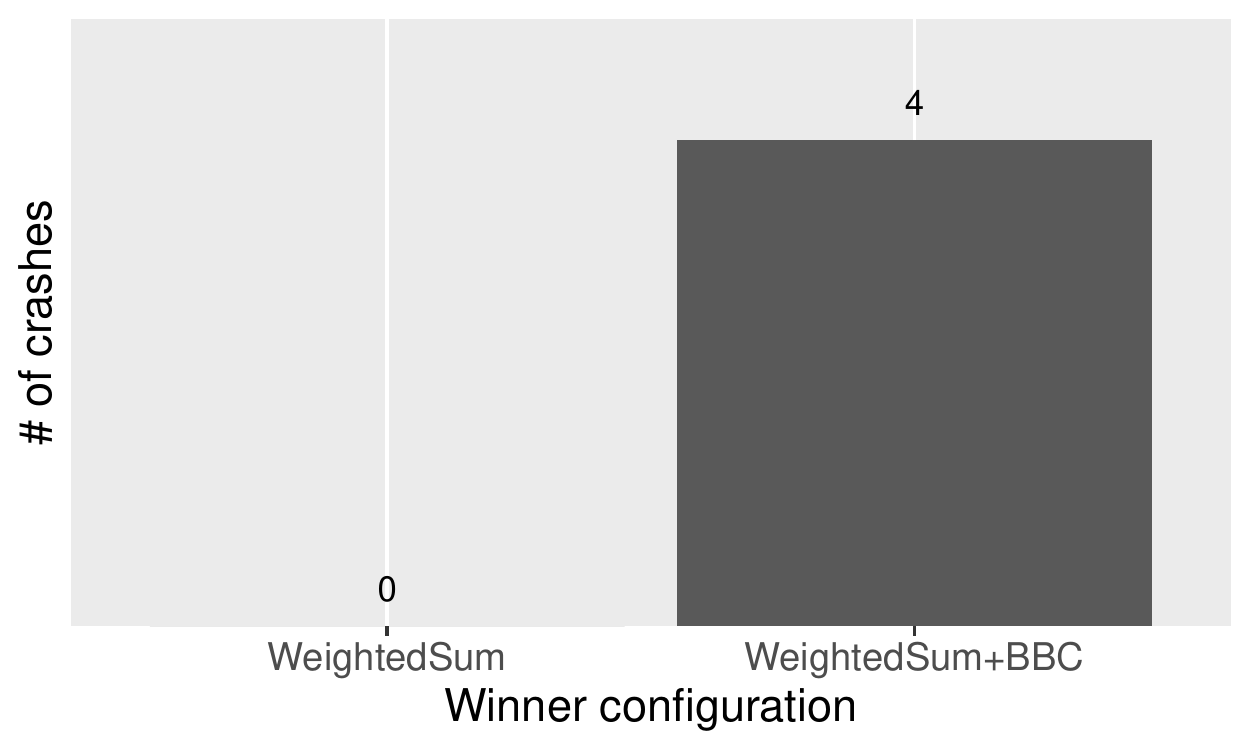}%
    }\hfil    
    \caption{Pairwise comparison of impact of \bbc on each fitness function in terms of crash reproduction ratio with a statistical significance $<0.01$.}
    \label{fig:result:significance:reproduction}
\end{figure}

\paragraph{Crash reproduction effectiveness (RQ 2.1).}
Figure \ref{fig:result:overall:reproduction} presents the crash reproduction ratio of the search processes guided by \integ  (Figure \ref{fig:result:overall:reproduction:integ}) and \WS (Figure \ref{fig:result:overall:reproduction:ws}), with and without \bbc as a secondary objective. This figure shows that, on average, the crash reproduction ratio of \WS improves 3.3\% when using \bbc: the average crash reproduction ratio of \WS is 74.8\% (with standard deviation 38.1\%) while the average crash reproduction of \WS + \bbc is increased to 78.1\% (with standard deviation 36.1\%).
This improvement is higher for crash reproduction using \integ. On average, the crash reproduction ratio achieved by \integ + \bbc is 9.2\% higher than \integ without \bbc: \integ achieves 70.5\% (with standard deviation 38.1\%) average crash reproduction ratio, while the average crash reproduction ratio of \integ + \bbc is 79.7\% (with standard deviation 37.3\%).
Higher improvement in \integ was expected as this fitness function relies more on the approach level and branch distance heuristics for covering each of the frames in the given stack trace. Also, in both of the fitness functions, the lower quartile of crash reproduction ratio has been improved by utilizing \bbc. These improvements in crash reproduction ratio for \WS and \integ are 19.1\% and 31.7\%, respectively.

\begin{table}[t]
    \centering
    \caption{Comparing the crash reproduction ratio between crash reproduction using \WSA and \WSOBB, for cases where one of the configurations has a significantly higher crash reproduction ratio (p-value $<$  0.01)}.
    \begin{tabular}{ l | c c | c c}
\hline 
\textbf{Crash} & \multicolumn{2}{c|}{\textbf{Reproduction ratio}} & \textbf{OR} & \textbf{p-value} \\ 
& WeightedSum & WeightedSum+BBC & & \\ 
\hline 
LANG-54b & 19 & 29 & 0.1 & 2.4659e-03 \\ 
XCOMMONS-1057 & 17 & 27 & 0.2 & 7.4098e-03 \\ 
XWIKI-12889 & 17 & 27 & 0.2 & 7.4098e-03 \\ 
XWIKI-14556 & 11 & 24 & 0.2 & 1.4306e-03 \\ 
\end{tabular}
    \label{tab:rq2.1:OR:ws}
\end{table}

\begin{table}[t]
    \centering
    \caption{Comparing the crash reproduction ratio between crash reproduction using \integA and \integOBB, for cases where one of the configurations has a significantly higher crash reproduction ratio (p-value $<$  0.01)}.
    \begin{tabular}{ l | c c | c c}
\hline 
\textbf{Crash} & \multicolumn{2}{c|}{\textbf{Reproduction ratio}} & \textbf{OR} & \textbf{p-value} \\ 
& RecoreSTDistance & RecoreSTDistance+BBC & & \\ 
\hline 
LANG-54b & 20 & 29 & 0.1 & 5.5791e-03 \\ 
MATH-78b & 10 & 21 & 0.2 & 9.2060e-03 \\ 
TIME-7b & 1 & 12 & 0.1 & 1.0508e-03 \\ 
XWIKI-12667 & 16 & 30 & 0.0 & 1.6767e-05 \\ 
XWIKI-13141 & 13 & 27 & 0.1 & 2.5073e-04 \\ 
XWIKI-13196 & 19 & 30 & 0.0 & 3.1881e-04 \\ 
XWIKI-13316 & 17 & 29 & 0.0 & 4.3102e-04 \\ 
XWIKI-13916 & 19 & 30 & 0.0 & 3.1881e-04 \\ 
XWIKI-14152 & 3 & 18 & 0.1 & 9.4143e-05 \\ 
XWIKI-14556 & 0 & 24 & 0.0 & 3.2940e-11 \\ 
\end{tabular}
    \label{tab:rq2.1:OR:recore}
\end{table}

To make our observations in Figure~\ref{fig:result:overall:reproduction} more robust, we performed an additional statistical analysis. Figure~\ref{fig:result:significance:reproduction} depicts the number of crashes, for which \bbc has a significant impact on the effectiveness of crash reproduction guided by \integ (Figure~\ref{fig:result:significance:reproduction:integ}) and \WS (Figure~\ref{fig:result:significance:reproduction:ws}). 
\bbc significantly improves the crash reproduction ratio in 10 and 4 crashes for fitness functions \integ and \WS, respectively. Notably, the application of this secondary objective does not have any significant adverse effect on crash reproduction. 
Tables \ref{tab:rq2.1:OR:ws} and \ref{tab:rq2.1:OR:recore} present the odds ratio and p-value in cases that \bbc leads to a significant improvement in crash reproduction ratios of \WS and \integ, respectively. As we can see in these tables, the odds ratio values in all cases are lower or equal to 0.2, indicating the high impact of \bbc.
Finally, we observed that \bbc helps each of the  \integ and \WS to reproduce 3 new crashes that could not be reproduced without this secondary objective.

\begin{figure}[t]
    \subfloat[\integ\label{fig:result:significance:efficiency:integ}]{%
        \includegraphics[width=0.45\textwidth]{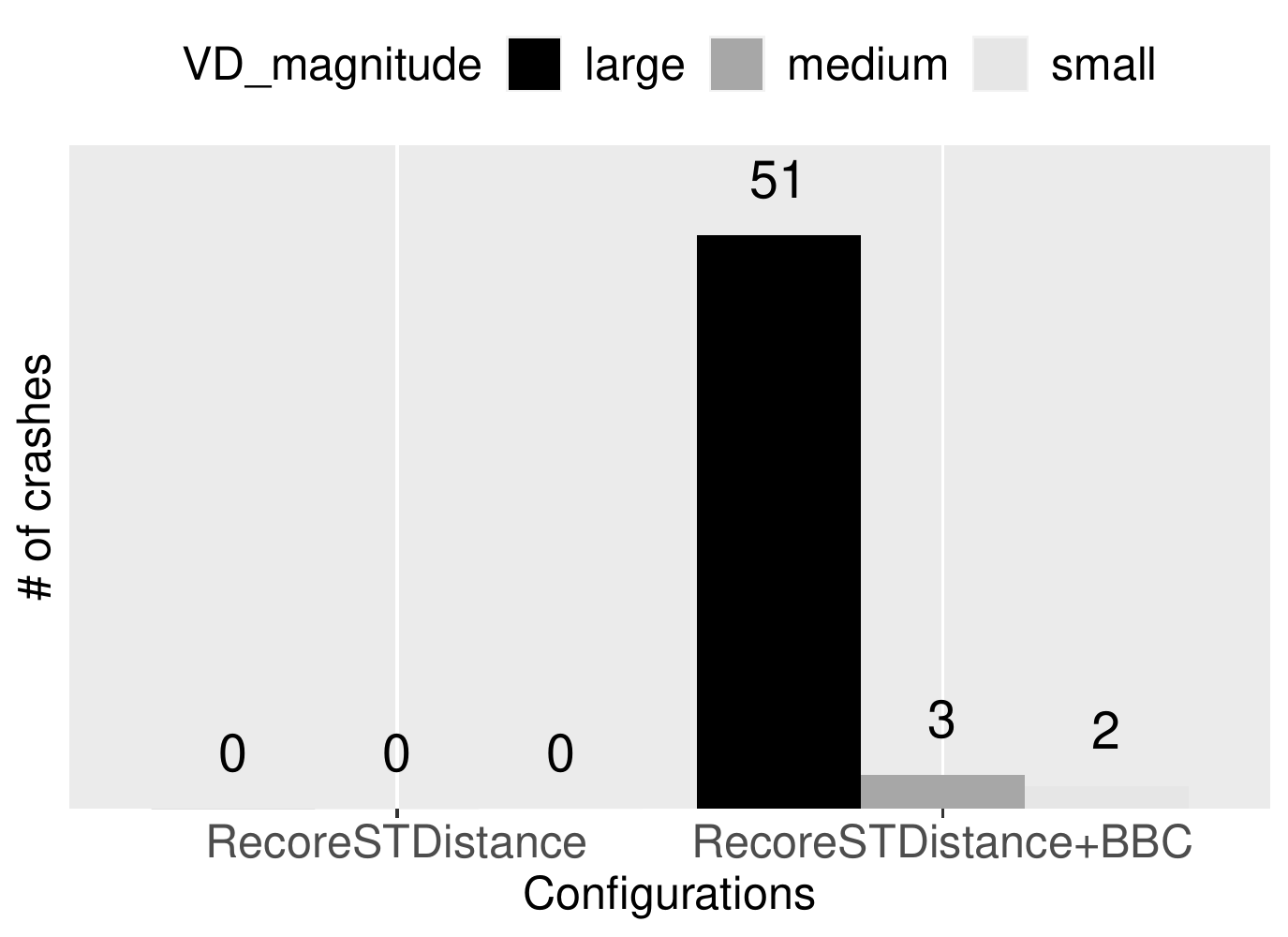}%
    }\hfil 
    \subfloat[\WS\label{fig:result:significance:efficiency:ws}]{%
        \includegraphics[width=0.45\textwidth]{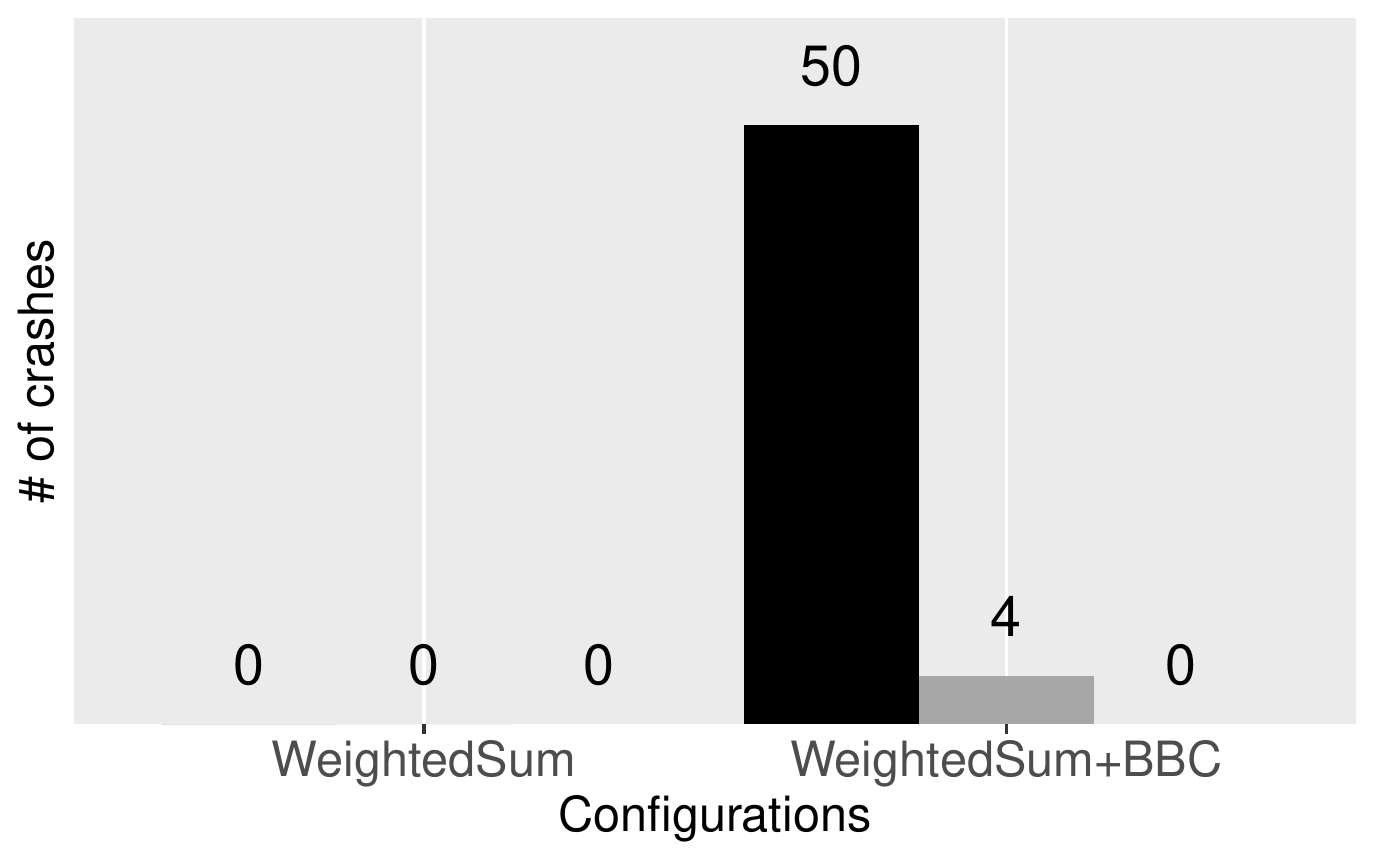}%
    }\hfil
    \caption{Pairwise comparison of impact of \bbc on each fitness function in terms of efficiency with a small, medium, and large effect size $\textit{\^{A}}_{12} < 0.5$ and a statistical significance $<0.01$.}
    \label{fig:result:significance:efficiency}
    \vspace{-1.5em}
\end{figure}

\paragraph{Crash reproduction efficiency (RQ 2.2).}
Figure \ref{fig:result:significance:efficiency} illustrates the number of cra\-shes, in which \bbc significantly affects the time consumed by the crash reproduction search process. As Figure \ref{fig:result:significance:efficiency:ws} shows, \bbc significantly improves the speed of crash reproduction guided by \WS in 54 crashes (43.5\% of cases), while it does not lose efficiency in the reproduction of any crash. 
Similarly,  Figure \ref{fig:result:significance:efficiency:integ} shows that \bbc has a higher positive impact on the efficiency of the search process guided by \integ. It significantly reduces the time consumed by the search process in 56 crashes (45.1\% of cases), while it had no adverse impact on the reproduction efficiency of any crash. 

\begin{figure}[t]
    \subfloat[\integ\label{fig:result:improvement:efficiency:integ}]{%
        \includegraphics[width=0.5\textwidth]{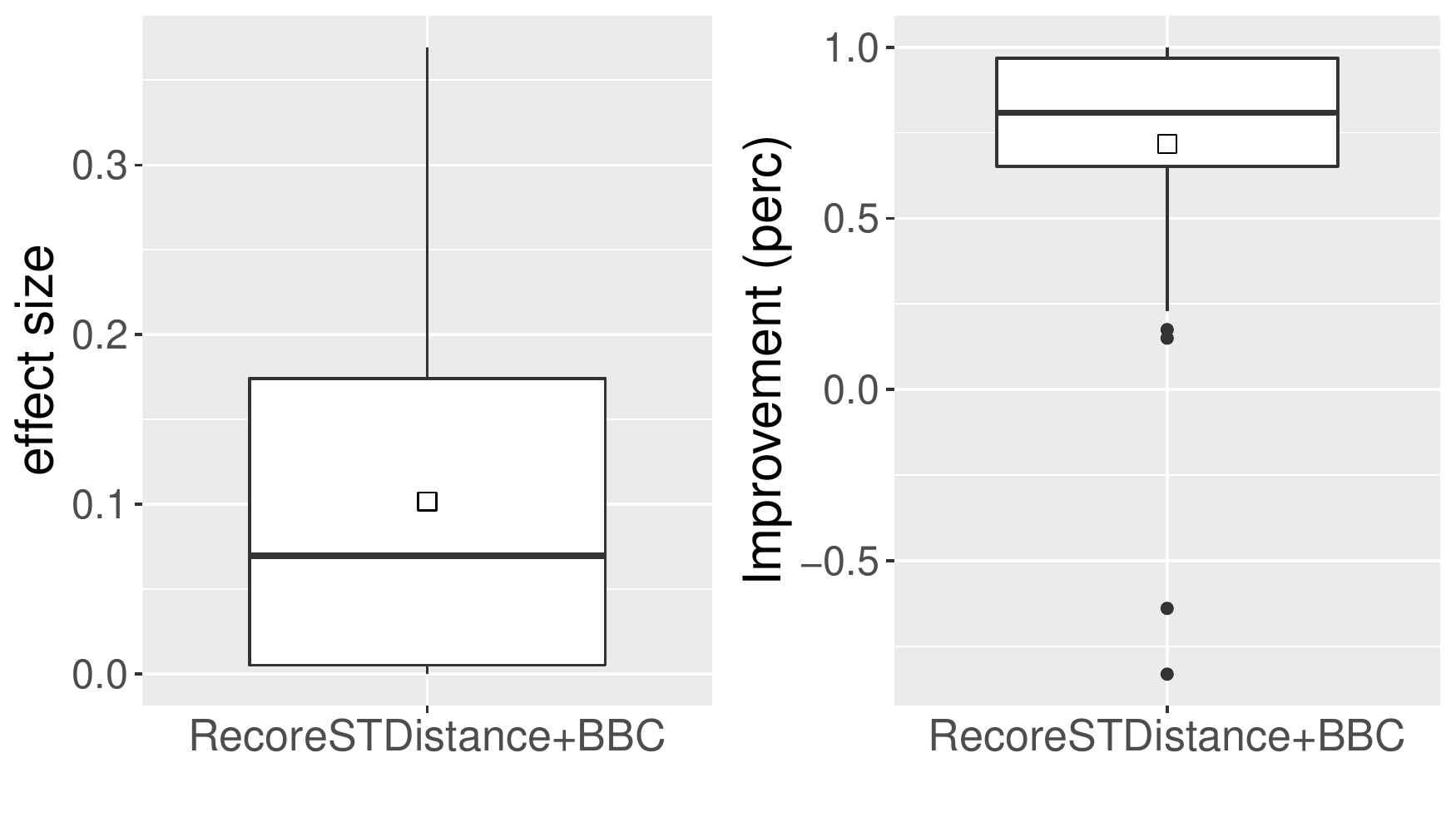}%
    }\hfil 
    \subfloat[\WS\label{fig:result:improvement:efficiency:ws}]{%
        \includegraphics[width=0.5\textwidth]{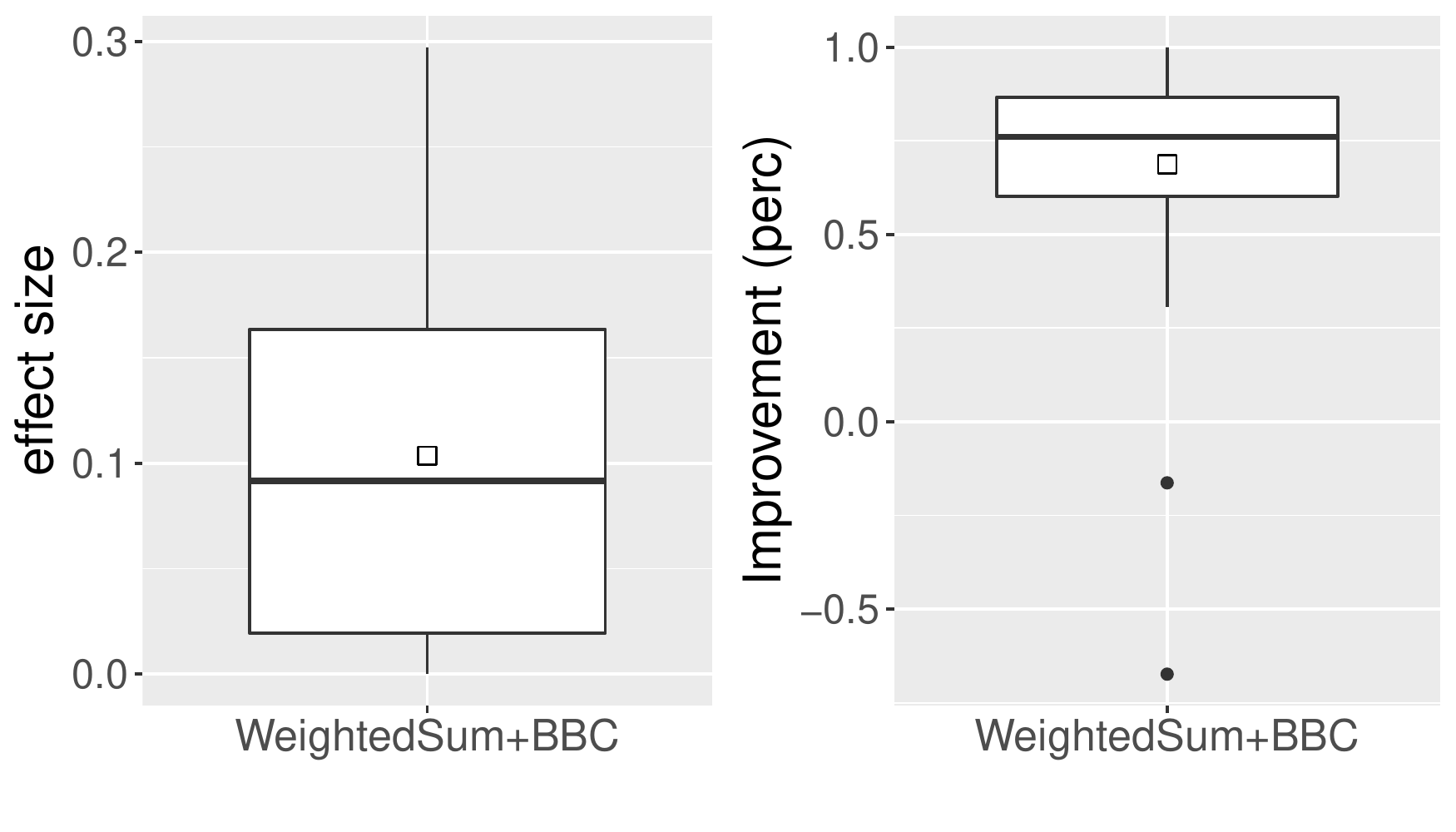}%
    }\hfil
    \caption{The effect size and the average improvement achieved by \bbc on each of the fitness functions in cases that \bbc makes a significant difference in terms of efficiency.}
    \label{fig:result:improvement:efficiency}
    \vspace{-2em}
\end{figure} 

Figure \ref{fig:result:improvement:efficiency} depicts the average improvements in the efficiency and effect sizes for crashes where the difference in the consumed budget when using \bbc or not was significant.
According to the right-side plot in Figure \ref{fig:result:improvement:efficiency:integ}, \bbc reduces the time consumed by the search process guided by \integ up to 98\% (being 71.7\% on average).
Also, the left-side plot indicates that the average effect size of differences between \integ and \integ+\bbc (calculated by Vargha-Delaney) is 0.102 (lower than 0.5 indicates that \bbc improved the efficiency).
Figure \ref{fig:result:improvement:efficiency:ws} shows that the average improvement (right-side plot) achieved by using \bbc as the second objective of \WS is 68.7\%, and the average effect size (left-side plot), in terms of the crash reproduction efficiency, is 0.104.

\paragraph{\textbf{Summary (RQ 2).}} 
\bbc improves the crash reproduction ratio for both of the \WS and \integ fitness functions. This improvement is higher for \integ as this fitness function relies more on approach level and branch distance. Moreover,
\bbc improves the efficiency of the search process with both of the crash reproduction fitness functions.

\section{Discussion}
\label{sec:discussion}

\subsection{\bbc for unit test generation}

\paragraph{Increase in program state and return value diversity.}
Using \bbc as a secondary objective leads to better branch coverage. Although small on average, the improvement is systematic, as demonstrated by the effect sizes. More interestingly, \bbc also leads to a better output and implicit exception coverage. This is particularly interesting in a unit testing context because it allows to capture more diverse returned values (including implicit exceptions) from the methods under test. We observe the same trends for weak mutation, denoting more diverse program states. Although the evaluation of the quality of the generated tests is outside of the scope of this study, we believe that diverse return values and program states can have a positive impact on the quality of the generated assertions, which is one of the known current limitations preventing the large industrial adoption of search-based unit test generation \cite{Almasi2017}. 

\paragraph{Adaptive secondary objectives.}
As explained in Section \ref{subsec:bbcapplication}, applying \bbc can be expensive ($\mathcal{O}(N \times E \times log\, V)$), compared to classical secondary objectives (linear time). Therefore, \bbc should be activated only when it can effectively contribute to decide between two test cases with the same fitness value. As shown by our preliminary analysis, this is especially relevant in the context of unit test generation, where each branch should be covered, which could trigger a high number of \bbc evaluations. In our implementation of \bbc for unit testing (described in Section~\ref{sec:approach:application:unit}), we limit the number of activations of \bbc, based on the activation time of an objective (\sleep) and a user-defined probability (\usage). This approach might however not be optimal. For instance, for classes under test with a high number of implicit branches, activating \bbc sooner and more often might improve the search process. In our future work, we will explore how the secondary objective can be dynamically adapted during the search, for instance, based on the evolution of the fitness values of the different objectives in \dynamosa.

\subsection{\bbc for crash reproduction}

Generally, using \bbc as secondary objective leads to a better crash reproduction ratio and higher efficiency in search-based crash reproduction. This improvement is achieved thanks to the additional ability to guide the search process when facing implicit branches during the search. 
Combining \bbc with \integ shows an important improvement compared to the combination of \bbc with \WS. This result was expected, since only one (out of three) component in \WS is allocated to line coverage, and thereby most parts of the fitness function do not use the approach level and branch distance heuristics. In contrast, \integ uses the approach level and branch distance to cover each of the frames in the given stack trace incrementally.

Our results show that \bbc helps the crash reproduction process to reproduce new crashes. For instance, the crash that we used in this study (XWIKI-13377) can be reproduced only by \integ + \bbc.

\section{Threats to validity}
\label{sec:threats}

\paragraph{Internal validity.}
We cannot guarantee that our implementation of \bbc in \evosuite and \botsing is bug-free. However, we mitigated this threat by testing our implementations and manually examining some samples of the results. 
Moreover, following the guidelines of the related literature \cite{Arcuri2014}, we executed each configuration \nruns times to take the randomness of the search process into account. 

\paragraph{External validity.}
We cannot ensure that our results are generalizable to all cases. However, for both of our experiments for unit test generation and crash reproduction, we have used two earlier established benchmarks: \crashpack \cite{Derakhshanfar2019}, which is a benchmark for crash reproduction containing 124 hard-to-reproduce crashes provoked by real bugs in a variety of open-source applications, and \defectsforj \cite{Just2014b}, a collection of real-world Java projects failures containing 835 bugs.

To increase the external validity while maintaining a good balance between the statistical power and the overall execution, analysis, and reporting time, we choose to consider only the ten most recent bugs from the 17 projects available in \defectsforj. After filtering out classes that cannot be handled by \evosuite, we ran our evaluation on \ncuts classes. Among those \ncuts classes, 44 come from different versions of the same projects. Although involved in different bugs, those classes might be similar and influence our results. To mitigate this threat, we performed a qualitative analysis to confirm the effect of \bbc.

\paragraph{Construct validity.} 
For unit test generation (\textbf{RQ 1}), we left the parameters of \dynamosa to their default values used by \evosuite. Those values are commonly used in the literature and it has been empirically shown that they give good results \cite{Panichella2018, Arcuri2013, Panichella2018a, Fraser2014b}. We can, however, not guarantee that these default values are the best when used with \bbc. Nevertheless, our results show that \bbc can improve search-based unit test generation when using the default parameter values.

For search-based crash reproduction (\textbf{RQ 2}), we used \bbc with two different fitness functions and left other parameters to their default values, used in previous studies \cite{Soltani2018a, Derakhshanfar2020e}. Those studies do not investigate the sensitivity of search-based crash reproduction to these values, and tuning these parameters should be undertaken as future work. However, as for unit test generation, our results show that \bbc can improve search-based crash reproduction with the default parameter values.

\paragraph{Conclusion validity.} 
We based our conclusion on standard statistical analysis for significance \cite{Arcuri2014} with $\alpha = 0.01$. Effects of multiple comparisons are mitigated by adjusting $p-values$ via Nemenyi's post-hoc procedure~\cite{Japkowicz2011,Panichella2021}. Furthermore, we complemented our quantitative analysis with qualitative investigations to confirm the observed effects.

\paragraph{Verifiability.}
Finally, we openly provide all our implementations: \botsing~\cite{derakhshanfar2020botsing}, as an open-source crash reproduction tool, and the implementation of \bbc in \evosuite~\cite{pouria_derakhshanfar_2021_4665874}.
Also, the data and the processing scripts used to present the results are available as two replication packages on Zenodo \cite{derakhshanfar_pouria_2020_3953519, pouria_derakhshanfar_2021_4665874}.

\section{Related work}
\label{sec:relatedwork}

\subsection{Handling implicit branches}

Related to our approach, the Testability Transformations (\tet) technique addresses the problem of implicit branches in unit test generation \cite{li2011bytecode, fraser20151600}. This strategy transforms the code to make implicit branches explicit by adding extra branches for error conditions and brings more guidance for the approach level and branch distance heuristics. For code transformation of each class, \tet needs extra bytecode instrumentation. Since instrumenting some classes can be difficult due to several known issues \cite{fraser2013evosuite}, instrumenting each class, which is coupled with the class under test, may fail. Also, if we limit the testability transformations to the class under test, the search process will not have any extra guidance in cases of facing the implicit branches in the other classes.

In this study, we tried to evaluate \tet in \dynamosa. However, \evosuite failed before starting the search process for all the different classes under test. After a deeper investigation, we found out that \tet is not compatible with \dynamosa, which is the default search algorithm in \evosuite. Moreover, \tet faces extra challenges while it needs extra bytecode instrumentation.
In theory, given the nature of \tet and \bbc, these two techniques can be applied simultaneously to the search process. Hence, these two approaches can complement each other to achieve high structural coverage and detect more faults. Studying the impact of using both \tet and \bbc on search-based test generation calls for further implementation and efforts, and thereby, it is part of our future research agenda.

\subsection{Search-based crash reproduction}

Many previous papers have studied search-based crash reproduction approaches. Two of these papers introduced new fitness functions to guide the search process. EvoCrash~\cite{Soltani2018a} measures the distance of a generated test from a given crash, and R{\"{o}}{\ss}ler \etal~\cite{Rossler2013} have proposed an approach called \recore in order to guide the crash reproduction search process using the given crash and core dump. We have described both  these approaches with their corresponding fitness functions in Section~\ref{sec:background:crash}; we consider them as baselines in our evaluation.

We have previously performed multiple studies on the search process introduced in \evocrash.
One of our recent studies evaluated the crash reproduction ability of \evocrash against 200 real-world crashes~\cite{Derakhshanfar2019}. We have also performed an extensive manual analysis of the \evocrash execution results to identify the challenges in this search process. We have also carried out other studies on other aspects of this search process to address some of the identified challenges. For instance, we have proposed an approach called Behavioral Model Seeding~\cite{Derakhshanfar2020}. In this approach, the usages of objects in the source code of software under test are transformed into transition systems, and these models are later used for generating more realistic solutions (\ie tests) during the search process. Furthermore, in other studies~\cite{Soltani2018b, Derakhshanfar2020d}, we rely on multi-objectivization techniques to improve the diversity of the population during the search.

Each of the aforementioned studies show that the proposed approaches can improve crash reproduction in their respective way. All of these studies use the \evocrash approach as baseline. We also used this approach (\ie \WS) as a baseline, and our results are consistent with those of our prior studies~\cite{Derakhshanfar2020, Derakhshanfar2020d} (\ie the \textit{no seeding} configuration in~\cite{Derakhshanfar2020}, and the \textit{Single} configuration in~\cite{Derakhshanfar2020d}). However, it should be noted that these results can slightly differ from \cite{Soltani2018b} and \cite{Derakhshanfar2019} as the experiments for these studies are performed using the \evocrash tool. We previously re-implemented the \evocrash approach (\ie including the \WS fitness function) in \botsing \cite{derakhshanfar2020botsing}, a framework for search-based crash reproduction. Since \botsing is a well-tested and more mature tool compared to the early versions of \evocrash, it can achieve more stable results.

In this study, we applied \bbc only on \WS. We have not considered other strategies introduced in our previous studies \cite{Derakhshanfar2019, Derakhshanfar2020, Derakhshanfar2020d} because each of these strategies works independently, and thereby can be applied simultaneously on the search process. For instance, model seeding improves the test generation capability of the search process, while \bbc focuses on improving the guidance that the second objectives can provide for the search process. Hence, both of them can be activated during the crash reproduction process. If we wanted to apply BBC for each of these strategies, we would have many configurations to assess and compare. This kind of analysis is out of the scope of this study, which only concentrates on \bbc, and calls for further studies in our future work.

In addition, for the first time, we have also considered the \integA fitness function as one of the baselines \cite{Rossler2013}. As we explained in Section \ref{sec:background:crash}, \integA is part of a main fitness function in \recore. This sub-function measures the distance of a generated test from covering a given crash. Since this study considers that we only have the crash stack trace and do not have any other information like core dumps, we only implemented \integA as an independent fitness function in \botsing.

\section{Conclusion and future work}
\label{sec:conclusion}

Approach level and branch distance are two well-known heuristics, widely used by search-based test generation approaches to guide the search process towards covering target statements and branches. These heuristics measure the distance of a generated test from covering the target using the coverage of control dependencies. However, these two heuristics do not consider implicit branches. For instance, if a test throws an exception during the execution of a non-branch statement, approach level and branch distance cannot guide the search process to tackle this exception. 
In this paper, we extended our previous work on Basic Block Coverage (\bbc{}), a secondary objective addressing this issue. We complemented our previous study into \bbc on search-based crash reproduction with an investigation of \bbc for unit test generation. 

Our results show that \bbc improves the branch coverage for unit tests generated using \dynamosa. Although small ($\sim$1\%), this improvement in the branch coverage is systematic and leads to an increase of the output and implicit runtime exception coverage, and of the diversity of runtime states.  
\bbc also helps \integ and \WS to reproduce 6 and 1 new crashes, respectively.
Finally, \bbc significantly improves the efficiency in 26.6\% and 13.7\% of the crashes using \integ and \WS{}, respectively.

An important implication of our work for future research is that we need to investigate secondary search objectives that can be \emph{dynamically} activated depending on the software under test. 
In this work, we applied the activation mechanism for secondary search objectives (\bbc) based on user-provided (static) meta-parameters. We have seen indications that such a mechanism can both improve the search process and at the same time reduce the computational cost, yet it can be counter-productive in some cases. 
We envision that \bbc and other secondary objectives would benefit from an adaptive activation, depending on the runtime behavior (\eg if the number of implicit runtime exceptions increases) or structure (\eg high coupling or deep inheritance hierarchy) of the classes under test.  

In our future work, we will investigate the application of \bbc for other search-based test generation techniques (such as testability transformations, and system and integration testing), as well as the implications of an increase of the diversity of program states in the generated unit tests (\eg for assertions generation). We will also investigate how \bbc can be dynamically activated using an adaptive secondary objectives approach to reduce the computational overload on the search process.

\begin{acknowledgements}
This research was partially funded by the EU Horizon 2020 ICT-10-2016-RIA ``STAMP'' project (No.731529), the EU Horizon 2020 H2020-ICT-2020-1-RIA ``COSMOS'' project (No.957254), and the Vici ``TestShift'' project (No. VI.C.182.032) from the Dutch Science Foundation NWO.
\end{acknowledgements}

\bibliographystyle{spmpsci}      
\bibliography{bibliography}

\begin{thebibliography}{10}
\providecommand{\url}[1]{{#1}}
\providecommand{\urlprefix}{URL }
\expandafter\ifx\csname urlstyle\endcsname\relax
  \providecommand{\doi}[1]{DOI~\discretionary{}{}{}#1}\else
  \providecommand{\doi}{DOI~\discretionary{}{}{}\begingroup
  \urlstyle{rm}\Url}\fi

\bibitem{Allen:1970:CFA:800028.808479}
Allen, F.E.: Control flow analysis.
\newblock {ACM} {SIGPLAN} Notices \textbf{5}(7), 1--19 (1970).
\newblock \doi{10.1145/390013.808479}

\bibitem{Almasi2017}
Almasi, M.M., Hemmati, H., Fraser, G., Arcuri, A., Benefelds, J.: {An
  Industrial Evaluation of Unit Test Generation: Finding Real Faults in a
  Financial Application}.
\newblock In: 2017 IEEE/ACM 39th International Conference on Software
  Engineering: Software Engineering in Practice Track (ICSE-SEIP), pp.
  263--272. IEEE (2017).
\newblock \doi{10.1109/ICSE-SEIP.2017.27}

\bibitem{Alshahwan2014}
Alshahwan, N., Harman, M.: {Coverage and fault detection of the
  output-uniqueness test selection criteria}.
\newblock In: Proceedings of the 2014 International Symposium on Software
  Testing and Analysis - ISSTA 2014, pp. 181--192. ACM Press (2014).
\newblock \doi{10.1145/2610384.2610413}

\bibitem{arcuri2019restful}
Arcuri, A.: {REST}ful {API} automated test case generation with evomaster.
\newblock ACM Transactions on Software Engineering and Methodology (TOSEM)
  \textbf{28}(1), 1--37 (2019)

\bibitem{Arcuri2014}
Arcuri, A., Briand, L.: {A hitchhiker's guide to statistical tests for
  assessing randomized algorithms in software engineering}.
\newblock Software Testing, Verification and Reliability \textbf{24}(3),
  219--250 (2014).
\newblock \doi{10.1002/stvr.1486}

\bibitem{Arcuri2013}
Arcuri, A., Fraser, G.: {Parameter tuning or default values? An empirical
  investigation in search-based software engineering}.
\newblock Empirical Software Engineering \textbf{18}(3), 594--623 (2013).
\newblock \doi{10.1007/s10664-013-9249-9}

\bibitem{b2016learning}
B~Le, T.D., Lo, D., Le~Goues, C., Grunske, L.: A learning-to-rank based fault
  localization approach using likely invariants.
\newblock In: Proceedings of the 25th International Symposium on Software
  Testing and Analysis, pp. 177--188. ACM (2016)

\bibitem{borba2010testing}
Borba, P., Cavalcanti, A., Sampaio, A., Woodcook, J.: Testing techniques in
  software engineering: Second pernambuco summer school on software
  engineering, PSSE 2007, Recife, Brazil, December 3-7, 2007, Revised Lectures,
  vol. 6153.
\newblock Springer (2010)

\bibitem{Campos2018}
Campos, J., Ge, Y., Albunian, N., Fraser, G., Eler, M., Arcuri, A.: {An
  empirical evaluation of evolutionary algorithms for unit test suite
  generation}.
\newblock Information and Software Technology \textbf{104}(August), 207--235
  (2018).
\newblock \doi{10.1016/j.infsof.2018.08.010}

\bibitem{Chen2015}
Chen, N., Kim, S.: {STAR: Stack trace based automatic crash reproduction via
  symbolic execution}.
\newblock IEEE Trans. on Software Engineering \textbf{41}(2), 198--220 (2015).
\newblock \doi{10.1109/TSE.2014.2363469}

\bibitem{derakhshanfar_pouria_2020_3953519}
Derakhshanfar, P., Devroey, X.: {Replication package of Basic Block Coverage
  for Search-Based Crash Reproduction}.
\newblock \doi{10.5281/zenodo.3953519}.
\newblock \urlprefix\url{https://doi.org/10.5281/zenodo.3953519}

\bibitem{pouria_derakhshanfar_2021_4665874}
Derakhshanfar, P., Devroey, X.: {pderakhshanfar/EMSE-BBC-experiment:
  Replication package for EMSE journal extension} (2021).
\newblock \doi{10.5281/zenodo.4665874}.
\newblock \urlprefix\url{https://doi.org/10.5281/zenodo.4665874}

\bibitem{derakhshanfar2020integ}
Derakhshanfar, P., Devroey, X., Panichella, A., Zaidman, A., van Deursen, A.:
  Towards integration-level test case generation using call site information.
\newblock arXiv preprint arXiv:2001.04221  (2020)

\bibitem{derakhshanfar2020botsing}
Derakhshanfar, P., Devroey, X., Panichella, A., Zaidman, A., Van~Deursen, A.:
  Botsing, a search-based crash reproduction framework for java.
\newblock In: 35th IEEE/ACM International Conference on Automated Software
  Engineering (ASE), pp. 1278--1282. IEEE (2020)

\bibitem{Derakhshanfar2020d}
Derakhshanfar, P., Devroey, X., Panichella, A., Zaidman, A., {Van Deursen}, A.:
  {Botsing, a Search-based Crash Reproduction Framework for Java}.
\newblock In: 35th IEEE/ACM International Conference on Automated Software
  Engineering (ASE '20), September 21–25, 2020, Virtual Event, Australia, pp.
  1278--1282. ACM/IEEE (2020).
\newblock \doi{10.1145/3324884.3415299}

\bibitem{Derakhshanfar2020}
Derakhshanfar, P., Devroey, X., Perrouin, G., Zaidman, A., Deursen, A.:
  {Search-based crash reproduction using behavioural model seeding}.
\newblock STVR \textbf{30}(3), e1733 (2020).
\newblock \doi{10.1002/stvr.1733}

\bibitem{Derakhshanfar2020c}
Derakhshanfar, P., Devroey, X., Zaidman, A.: {It Is Not Only About Control
  Dependent Nodes: Basic Block Coverage for Search-Based Crash Reproduction}.
\newblock In: A.~Aleti, A.~Panichella (eds.) Search-Based Software Engineering
  - 12th International Symposium, SSBSE 2020, pp. 42--57. Springer (2020).
\newblock \doi{10.1007/978-3-030-59762-7_4}

\bibitem{Derakhshanfar2020e}
Derakhshanfar, P., Devroey, X., Zaidman, A., {Van Deursen}, A., Panichella, A.:
  {Good Things Come In Threes: Improving Search-based Crash Reproduction With
  Helper Objectives}.
\newblock In: 35th IEEE/ACM International Conference on Automated Software
  Engineering (ASE '20), pp. 211--223. IEEE / ACM (2020).
\newblock \doi{10.1145/3324884.3416643}

\bibitem{Devroey2020}
Devroey, X., Panichella, S., Gambi, A.: {Java Unit Testing Tool Competition -
  Eighth Round}.
\newblock In: Proceedings of the IEEE/ACM 42nd International Conference on
  Software Engineering Workshops, pp. 545--548. ACM (2020).
\newblock \doi{10.1145/3387940.3392265}

\bibitem{Fraser2011}
Fraser, G., Arcuri, A.: Evosuite: Automatic test suite generation for
  object-oriented software.
\newblock In: Proceedings of the 19th ACM SIGSOFT Symposium and the 13th
  European Conference on Foundations of Software Engineering, ESEC/FSE '11, pp.
  416--419. ACM, New York, NY, USA (2011).
\newblock \doi{10.1145/2025113.2025179}

\bibitem{fraser2013evosuite}
Fraser, G., Arcuri, A.: Evosuite: On the challenges of test case generation in
  the real world.
\newblock In: 2013 IEEE Sixth International Conference on Software Testing,
  Verification and Validation, pp. 362--369. IEEE (2013)

\bibitem{fraser2012whole}
Fraser, G., Arcuri, A.: {Whole test suite generation}.
\newblock IEEE Transactions on Software Engineering \textbf{39}(2), 276--291
  (2013).
\newblock \doi{10.1109/TSE.2012.14}

\bibitem{Fraser2014b}
Fraser, G., Arcuri, A.: {A large-scale evaluation of automated unit test
  generation using EvoSuite}.
\newblock ACM Transactions on Software Engineering and Methodology
  \textbf{24}(2), 1--42 (2014).
\newblock \doi{10.1145/2685612}

\bibitem{fraser20151600}
Fraser, G., Arcuri, A.: 1600 faults in 100 projects: Automatically finding
  faults while achieving high coverage with evosuite.
\newblock Empirical Software Engineering \textbf{20}(3), 611--639 (2015)

\bibitem{Fraser2015a}
Fraser, G., Arcuri, A.: {Achieving scalable mutation-based generation of whole
  test suites}.
\newblock Empirical Software Engineering \textbf{20}(3), 783--812 (2015).
\newblock \doi{10.1007/s10664-013-9299-z}

\bibitem{Garcia2009}
Garc{\'{i}}a, S., Molina, D., Lozano, M., Herrera, F.: {A study on the use of
  non-parametric tests for analyzing the evolutionary algorithms' behaviour: a
  case study on the CEC'2005 Special Session on Real Parameter Optimization}.
\newblock Journal of Heuristics \textbf{15}(6), 617--644 (2009).
\newblock \doi{10.1007/s10732-008-9080-4}

\bibitem{Howden1982}
Howden, W.W.E.: {Weak Mutation Testing and Completeness of Test Sets}.
\newblock IEEE Transactions on Software Engineering \textbf{SE-8}(4), 371--379
  (1982).
\newblock \doi{10.1109/TSE.1982.235571}

\bibitem{Japkowicz2011}
Japkowicz, N., Shah, M.: {Evaluating Learning Algorithms: A Classification
  Perspective}.
\newblock Cambridge University Press (2011)

\bibitem{Just2014b}
Just, R., Jalali, D., Ernst, M.D.: {Defects4J: a database of existing faults to
  enable controlled testing studies for Java programs}.
\newblock In: Proceedings of the 2014 International Symposium on Software
  Testing and Analysis - ISSTA 2014, pp. 437--440. ACM (2014).
\newblock \doi{10.1145/2610384.2628055}

\bibitem{Kifetew2019a}
Kifetew, F., Devroey, X., Rueda, U.: {Java Unit Testing Tool Competition -
  Seventh Round}.
\newblock In: 2019 IEEE/ACM 12th International Workshop on Search-Based
  Software Testing (SBST), pp. 15--20. IEEE (2019).
\newblock \doi{10.1109/SBST.2019.00014}

\bibitem{li2011bytecode}
Li, Y., Fraser, G.: Bytecode testability transformation.
\newblock In: International Symposium on Search Based Software Engineering, pp.
  237--251. Springer (2011)

\bibitem{lu2016does}
Lu, Y., Lou, Y., Cheng, S., Zhang, L., Hao, D., Zhou, Y., Zhang, L.: How does
  regression test prioritization perform in real-world software evolution?
\newblock In: 2016 IEEE/ACM 38th International Conference on Software
  Engineering (ICSE), pp. 535--546. IEEE (2016)

\bibitem{ma2015grt}
Ma, L., Artho, C., Zhang, C., Sato, H., Gmeiner, J., Ramler, R.: Grt:
  Program-analysis-guided random testing (t).
\newblock In: 2015 30th IEEE/ACM International Conference on Automated Software
  Engineering (ASE), pp. 212--223. IEEE (2015)

\bibitem{Martinez2016}
Martinez, M., Monperrus, M.: {ASTOR: a program repair library for Java (demo)}.
\newblock In: Proceedings of the 25th International Symposium on Software
  Testing and Analysis, pp. 441--444. ACM (2016).
\newblock \doi{10.1145/2931037.2948705}

\bibitem{McMinn2004}
McMinn, P.: {Search-based software test data generation: A survey}.
\newblock Software Testing Verification and Reliability \textbf{14}(2),
  105--156 (2004).
\newblock \doi{10.1002/stvr.294}

\bibitem{McMinn2011}
McMinn, P.: Search-based software testing: Past, present and future.
\newblock In: Proceedings of the 2011 IEEE Fourth International Conference on
  Software Testing, Verification and Validation Workshops, ICSTW '11, pp.
  153--163. IEEE Computer Society, Washington, DC, USA (2011).
\newblock \doi{10.1109/ICSTW.2011.100}

\bibitem{Molina2018}
Molina, U.R., Kifetew, F., Panichella, A.: {Java Unit Testing Tool Competition
  - Sixth Round}.
\newblock In: Proceedings of the 11th International Workshop on Search-Based
  Software Testing - SBST '18, i, pp. 22--29. ACM (2018).
\newblock \doi{10.1145/3194718.3194728}

\bibitem{nayrolles2015jcharming}
Nayrolles, M., Hamou-Lhadj, A., Tahar, S., Larsson, A.: {JCHARMING: A bug
  reproduction approach using crash traces and directed model checking}.
\newblock In: 2015 IEEE 22nd International Conference on Software Analysis,
  Evolution, and Reengineering (SANER), pp. 101--110. IEEE (2015).
\newblock \doi{10.1109/SANER.2015.7081820}

\bibitem{noor2015similarity}
Noor, T.B., Hemmati, H.: A similarity-based approach for test case
  prioritization using historical failure data.
\newblock In: 2015 IEEE 26th International Symposium on Software Reliability
  Engineering (ISSRE), pp. 58--68. IEEE (2015)

\bibitem{Offutt1994}
Offutt, A., Lee, S.: {An empirical evaluation of weak mutation}.
\newblock IEEE Transactions on Software Engineering \textbf{20}(5), 337--344
  (1994).
\newblock \doi{10.1109/32.286422}

\bibitem{Panichella2021}
Panichella, A.: {A Systematic Comparison of search-Based approaches for LDA
  hyperparameter tuning}.
\newblock Information and Software Technology \textbf{130}, 106411 (2021).
\newblock \doi{10.1016/j.infsof.2020.106411}

\bibitem{Panichella2018a}
Panichella, A., Kifetew, F.M., Tonella, P.: {A large scale empirical comparison
  of state-of-the-art search-based test case generators}.
\newblock Information and Software Technology \textbf{104}(June), 236--256
  (2018).
\newblock \doi{10.1016/j.infsof.2018.08.009}

\bibitem{Panichella2018}
Panichella, A., Kifetew, F.M., Tonella, P.: {Automated test case generation as
  a many-objective optimisation problem with dynamic selection of the targets}.
\newblock IEEE Transactions on Software Engineering \textbf{44}(2), 122--158
  (2018).
\newblock \doi{10.1109/TSE.2017.2663435}

\bibitem{Papadakis2011}
Papadakis, M., Malevris, N.: {Automatically performing weak mutation with the
  aid of symbolic execution, concolic testing and search-based testing}.
\newblock Software Quality Journal \textbf{19}(4), 691--723 (2011).
\newblock \doi{10.1007/s11219-011-9142-y}

\bibitem{pearson2017evaluating}
Pearson, S., Campos, J., Just, R., Fraser, G., Abreu, R., Ernst, M.D., Pang,
  D., Keller, B.: Evaluating and improving fault localization.
\newblock In: Proceedings of the 39th International Conference on Software
  Engineering, pp. 609--620. IEEE Press (2017)

\bibitem{rojas2015combining}
Rojas, J.M., Campos, J., Vivanti, M., Fraser, G., Arcuri, A.: {Combining
  Multiple Coverage Criteria in Search-Based Unit Test Generation}.
\newblock In: Search-Based Software Engineering (SSBSE 2015), \emph{LNCS}, vol.
  9275, pp. 93--108. Springer (2015).
\newblock \doi{10.1007/978-3-319-22183-0_7}

\bibitem{Rossler2013}
R{\"{o}}{\ss}ler, J., Zeller, A., Fraser, G., Zamfir, C., Candea, G.:
  {Reconstructing core dumps}.
\newblock In: Proc. International Conference on Software Testing, Verification
  and Validation (ICST), pp. 114--123. IEEE (2013).
\newblock \doi{10.1109/ICST.2013.18}

\bibitem{Shamshiri2016}
Shamshiri, S., Just, R., Rojas, J.M., Fraser, G., McMinn, P., Arcuri, A.: {Do
  Automatically Generated Unit Tests Find Real Faults? An Empirical Study of
  Effectiveness and Challenges}.
\newblock In: 2015 30th IEEE/ACM International Conference on Automated Software
  Engineering (ASE), pp. 201--211. IEEE (2015).
\newblock \doi{10.1109/ASE.2015.86}

\bibitem{Smith2015}
Smith, E.K., Barr, E.T., {Le Goues}, C., Brun, Y.: {Is the cure worse than the
  disease? overfitting in automated program repair}.
\newblock In: Proceedings of the 2015 10th Joint Meeting on Foundations of
  Software Engineering, pp. 532--543. ACM (2015).
\newblock \doi{10.1145/2786805.2786825}

\bibitem{Derakhshanfar2019}
Soltani, M., Derakhshanfar, P., Devroey, X., van Deursen, A.: {A
  benchmark-based evaluation of search-based crash reproduction}.
\newblock Empirical Software Engineering \textbf{25}(1), 96--138 (2020).
\newblock \doi{10.1007/s10664-019-09762-1}

\bibitem{Soltani2018b}
Soltani, M., Derakhshanfar, P., Panichella, A., Devroey, X., Zaidman, A., van
  Deursen, A.: {Single-objective Versus Multi-objectivized Optimization for
  Evolutionary Crash Reproduction}.
\newblock In: T.E. Colanzi, P.~McMinn (eds.) Symposium on Search-Based Software
  Engineering. SSBSE 2018., \emph{LNCS}, vol. 11036, pp. 325--340. Springer,
  Montpellier, France (2018).
\newblock \doi{10.1007/978-3-319-99241-9\_18}

\bibitem{Soltani2018a}
Soltani, M., Panichella, A., {Van Deursen}, A.: {Search-Based Crash
  Reproduction and Its Impact on Debugging}.
\newblock IEEE Transactions on Software Engineering \textbf{46}(12), 1294--1317
  (2020).
\newblock \doi{10.1109/TSE.2018.2877664}

\bibitem{vargha}
Vargha, A., Delaney, H.D.: A critique and improvement of the {CL} common
  language effect size statistics of {McGraw} and {Wong}.
\newblock Journal of Educational and Behavioral Statistics \textbf{25}(2),
  101--132 (2000)

\bibitem{xiao2011precise}
Xiao, X., Xie, T., Tillmann, N., De~Halleux, J.: Precise identification of
  problems for structural test generation.
\newblock In: Software Engineering (ICSE), 2011 33rd International Conference
  on, pp. 611--620. IEEE, {ACM}, Waikiki, Honolulu , HI, USA (2011)

\bibitem{Xuan2015}
Xuan, J., Xie, X., Monperrus, M.: {Crash reproduction via test case mutation:
  Let existing test cases help}.
\newblock In: Proceedings of the 2015 10th Joint Meeting on Foundations of
  Software Engineering - ESEC/FSE 2015, pp. 910--913. ACM Press, New York, New
  York, USA (2015).
\newblock \doi{10.1145/2786805.2803206}

\bibitem{Zeller2009}
Zeller, A.: Why Programs Fail, Second Edition: A Guide to Systematic Debugging,
  2nd edn.
\newblock Morgan Kaufmann Publishers Inc., San Francisco, CA, USA (2009)

\end{thebibliography}

\end{document}